\documentclass[a4paper,11pt]{article}
\usepackage{hyperref}
\usepackage{epsfig}
\usepackage{amsmath}
\usepackage{amssymb}
\usepackage{bbm}
\usepackage{bm}
\usepackage{color}
\usepackage{xcolor}
\usepackage{dsfont}
\usepackage[footnotesize]{caption}
\usepackage[makeroom]{cancel}
\usepackage{cite}
\usepackage{graphicx}
\usepackage{multirow}
\usepackage{feynmf}
\usepackage{pdfpages}
\usepackage{lscape}

\newcommand{\eVq}  {\text{eV}^2}

\newcommand{\R}{\mbox{\tiny$R$}}

\newcommand{\T}{\mbox{\tiny$T$}}

\numberwithin{equation}{section}

\begin{document}

\begin{center}
{\Large\bf $A_4$ realization of inverse seesaw: neutrino masses,
$\theta_{13}$ and leptonic non-unitarity}
\\[2mm]
\vskip 2cm

{ Biswajit Karmakar$^{a,}$\footnote{k.biswajit@iitg.ernet.in},
Arunansu Sil$^{a,}$\footnote{asil@iitg.ernet.in}}\\[3mm]
{\it{
$^a$ Department of Physics, Indian Institute of Technology Guwahati, 781039 Assam, India}
}
\end{center}

\vskip 1cm

\begin{abstract}

\noindent We provide an $A_4$ based flavor symmetric scenario to accommodate 
the 
inverse seesaw mechanism for explaining light neutrino masses and mixing.  We 
find that the lepton mixing, in particular the tri-bimaximal mixing pattern 
and its deviation through nonzero $\theta_{13}$, is originated solely from the 
flavor structure of the lepton number violating contribution of the neutral 
lepton mass matrix. Here we discuss in detail how a nonzero value of 
$\theta_{13}$ is correlated with the other parameters in the framework and its 
impact on the Dirac CP phase $\delta$.  We also analyze the non-unitarity 
effects 
on lepton mixing matrix and its implication in terms of the lepton flavor 
violating decays, etc. 
 
\end{abstract}


\section{Introduction}\label{sec1}
Even after the discovery of the Higgs boson at LHC,
understanding the origin of smallness associated with neutrino mass still 
remains an open question. In this respect seesaw mechanism serves as a guiding 
tool hinting toward the existence of new physics beyond the electroweak scale 
($v$).  The conventional type-I seesaw~\cite{Minkowski:1977sc, GellMann:1980vs, 
Mohapatra:1979ia}  tries to explain the smallness of neutrino mass by 
adding three right-handed (RH) neutrinos $N_{R i=1,2,3}$ to the Standard Model (SM). They have Majorana mass 
$M_R$ which is representative of the lepton number violation. With the Yukawa couplings of order 
unity, the left handed neutrinos can be light enough, $m_{\nu} \sim v^2/M_R$, provided the new physics 
scale $M_R$ is sufficiently high $\sim 10^{13}$ GeV or so. Though it suggests an interesting and natural 
explanation of why neutrinos are so light, such a high new physics scale  is beyond the reach of present
and future neutrino experiments. 

Inverse seesaw~\cite{Mohapatra:1986aw,Mohapatra:1986bd} on the other hand turns out to be 
a viable alternate scenario where the new physics scale responsible for neutrino mass generation 
can be brought down near  TeV scale at the expense of involving additional 
fields (SM singlet fermions $S_{i=1,2,3}$). In presence of 
additional symmetry like a global $U(1)_{B-L}$, the corresponding neutral 
lepton $9 \times 9$ mass matrix takes the form 
\begin{eqnarray}\label{CM}
M_{\nu} = \left(
\begin{array}{ccc}
         0 &m_D &0\\
         m_D^T &0     &M\\
         0 &M^T &{0}
\end{array}
\right),
\end{eqnarray}
using the basis  $(\nu_L^c, N_R, S)$. Note that at this level, neutrinos are massless. Once  
the lepton number violating term $ \frac{1}{2} \overline{S^c}\mu S$  is 
introduced with  $\mu<<m_D<M$, the effective $3 \times 3$ light neutrino mass 
matrix  is given by
\begin{equation}\label{mnudef}
 m_{\nu}=m_{D}M^{-1}\mu(M^{T})^{-1}m_D^T=F\mu F^T,
\end{equation}
where $F=m_D M^{-1}$.  
Since the lepton number turns out to be only an approximate symmetry of 
nature, it is perhaps more natural to be broken by a small amount $\mu$ rather than by a 
large mass  $M_R$ as happened in case of type-I seesaw. Also note that 
the other mass scale $M$  (say the new physics scale) in Eq. (\ref{CM}) can be as 
low as TeV  since there exists a double suppression by this new physics 
scale through Eq.(\ref{mnudef}) and smallness of $\mu$ is then justified to 
produce correct amount of light neutrino mass.  

Apart from the smallness associated with the neutrino mass, the origin of lepton mixing 
matrix, being quite different from the quark mixing, needs to be understood. 
The study of underlying principle behind this typical mixing is particularly interesting 
with the recent finding of nonzero $\theta_{13}$ \cite{Abe:2011fz, An:2012eh, 
Ahn:2012nd, Abe:2013hdq}. The present global analysis \cite{Forero:2014bxa,
Capozzi:2013csa, Gonzalez-Garcia:2014bfa} from 
several experimental data \cite{NuExpt} can be summarized as
\begin{center}
  $\Delta m^2_{21}=(7.11 - 8.18)\times10^{-5}\hspace{.1cm} \eVq$, \hspace{.5cm} $|\Delta m^2_{31}|=(2.30 - 2.65)\times10^{-3}\hspace{.1cm} \eVq$,\\
  $\sin^2\theta_{12}=0.278-0.375$, \hspace{.2cm} $\sin^2\theta_{23}=0.392-0.643$, \hspace{.2cm} $\sin^2\theta_{13}=0.0177-0.0294$.
\end{center}
In this regard, a particular pattern yielding $\tan^2 \theta_{23} = 1, \tan^2 
\theta_{12} = 1/2$ and $\theta_{13}=0$ is called the tri-bimaximal mixing 
(TBM)~\cite{Harrison:1999cf}. It has received a lot of attention as such a 
pattern can be elegantly generated using flavor symmetries. Use of non-Abelian 
discrete symmetries (for a review see~\cite{King:2013eh}) like $A_4, S_4$ etc. 
is very well known~\cite{Ma:2001dn,Altarelli:2005yx} in this context. However a 
deformation from TBM mixing 
becomes essential after the precise measurement of $\theta_{13}$. The details of such 
deformation are studied for type-I~\cite{ Barry:2010zk,Karmakar:2014dva} and type-II seesaw~
\cite{Karmakar:2015jza} in the context of $A_4$.
In general, extra flavon fields (SM singlet scalar 
fields transforming non-trivially under the flavor symmetry) are employed\footnote{ Deviations from TBM mixing 
can also be realized by perturbing the vev alignments of the $A_4$ scalars involved~\cite{ Barry:2010zk}.}. for this approach
~\cite{Shimizu:2011xg,King:2011zj}. 
Once these fields get their vacuum expectation values (vev), the requisite flavor 
structure is generated. 

In this work we aim to study the lepton mixing matrix in the inverse seesaw 
framework based on an $A_4$ flavor symmetry. In its minimal form, ref 
\cite{Hirsch:2009mx} discusses 
how a TBM pattern can be incorporated in an $A_4$ symmetric inverse seesaw scenario. 
They have shown (among one of the few possibilities discussed there) that if 
$m_D,  M$ and 
$\mu$ matrices all posses the following structure: 
\begin{eqnarray}\label{m}
M_0 = \left(
\begin{array}{ccc}
         X   &0     &0\\
         0   &Y     &Z\\
         0   &Z     &Y
\end{array}
\right),
\end{eqnarray}
the light neutrino mass matrix obtains a typical form,
\cite{Altarelli:2010gt}
\begin{eqnarray}\label{mtbm}
m_\nu = \left(
\begin{array}{ccc}
         A   &B      &B\\
         B   &A+D    &B-D\\
         B   &B-D    &A+D
\end{array}
\right).
\end{eqnarray}
The diagonalizing matrix of  the above form of $m_{\nu}$ is representative 
of the TBM mixing in 
the basis where charged lepton mass matrix is diagonal. In 
\cite{Dorame:2012zv}, authors have shown that in a $S_4$ 
based inverse seesaw, nonzero $\theta_{13}$ can be generated from the 
correction in the charged lepton sector. Few earlier attempts in realizing 
inverse seesaw in the framework of discrete flavor symmetries can be found 
in~\cite{Abbas:2015zna}.  Here the construction is such that the charged lepton 
mass matrix becomes diagonal. 
Now with a simpler form for $m_D$ and $M$  (where 
$X=Z$ and $Y=0$ in $M_0$), $F$ in Eq. (\ref{mnudef}) becomes proportional to 
identity matrix and hence the structure of $\mu$ matrix coincides with 
that of $m_{\nu}$. This means that $\mu$ matrix (and hence $m_{\nu}$ matrix 
also) of the form similar to 
$M_0$ would generate the TBM pattern of lepton mixing matrix.  Therefore 
we finally adopt a  $\mu$ matrix different from $M_0$ structure so as to 
accommodate the observed value of $\theta_{13}$. It is interesting to note 
that in the inverse seesaw, $\mu$ matrix (the coefficient matrix of the $\overline{S^c}S$ term) 
being different from zero is the source of violation of the lepton number as stated 
earlier and now it also turns out that the same $\mu$ is also the source of non-zero 
$\theta_{13}$ as well as other mixings (the charged lepton mass matrix is diagonal) 
in our scenario. This is a salient feature of our model. 
We have then discussed the possible 
correlation between the Dirac CP phase ($\delta$) with $\theta_{13}$ and other 
parameters involved. We have tried to address the smallness associated with the 
$\mu$ term by considering its origin from a higher dimensional operator. The 
$A_4$ symmetry along with other non-Abelian discrete symmetries like 
$Z_{4}\times Z_3$ play important role. We have estimated the effective neutrino 
mass parameter associated with neutrinoless 
double beta decay~\cite{Asakura:2014lma, Albert:2014awa} and studied the correlation 
with $\delta$ as well.

Furthermore $m_D$ being close to $M$, in general the inverse seesaw framework 
allows non-negligible mixing between the light and heavy neutrino states 
resulting  non-unitarity contributions to the lepton flavor mixing. Since the 
flavor structure is completely known in our framework, 
we are then able to study the non-unitarity involved in the set-up and in turn constrain some 
of the parameters. Lepton flavor violating (LFV) decays also result from this 
non-unitarity effect. However it turns out in our scenario that branching 
ratio of those LFV decays are vanishingly small due to exact cancellation of 
elements involved followed from the particular flavor structure we have 
considered.

This paper is organized as follows. In the Section \ref{sec2} below, we describe the 
construction of the model based on the symmetries of the framework. The 
detailed phenomenology constraining the parameters of the model from the 
available data of neutrino experiments takes place in Section \ref{sec3} and \ref{sec:analysis}.  Section 
\ref{sec5} is devoted in studying the non-unitarity effect and we comment on lepton 
flavor violating decays and additional contribution to neutrinoless double beta 
decay. Finally we conclude in Section \ref{sec6}.  

\section{The Model}\label{sec2}
In order to realize the usual inverse seesaw mechanism for the generation of  
light neutrino masses, we extend he SM particle content by introducing three 
RH neutrinos, $N_{R_{i=1,2,3}}$, and three other singlet fermions, 
$S_{i=1,2,3}$ as already mentioned. In addition few flavons ($\phi_{S}$, $\phi_{T}$,  $\xi$, $\xi'$, $\rho$)
are included to understand the flavor structure  of the lepton mixing.  An additional global $U(1)_{B-L}$ 
symmetry is considered along with the flavor symmetry  $A_4 \times Z_4 \times 
Z_3$. The field content of the model and their charges under the symmetry of 
the model (appropriate for the discussion) are mentioned in Table  \ref{t2}. 
Once the flavon fields get vev (along suitable directions), 
the desired structures of the
mass matrices are generated as we will find below. 
\begin{table}[h]
\centering
\resizebox{10cm}{!}{%
\begin{tabular}{|c|ccccccc|ccccc|}
\hline
 Fields & $e_{R}$ & $\mu_{R}$& $\tau_{R}$ &    $L$ & $H$ & $N_{R}$ & $S$ & 
$\phi_{S}$ & $\phi_{T}$ &  $\xi$ & $\xi'$ & $\rho$ \\
\hline
$A_{4}$ & 1 & $1'$  &  $1''$ & 3 & 1 & 3 & 3 & 3 & 3& 1 &$1'$&1\\
\hline
$Z_{4}$ &-$i$ & -$i$ & -$i$ &  -$i$ & 1 &-$i$ & 1 & -1 & 1 &-1&-1  &$i$ 
\\
\hline
$Z_{3}$ &1 &1 &1 &1 &1 &1 &$\omega^2$  &1 & 1 &1 & 1& $\omega$\\
\hline
$B-L$ &  -1 &  -1 & -1 &-1 &0& -1 & 1 & -2 & 0 & -2  &-2&0\\
\hline
\end{tabular}
}\
\caption{\label{t2} {\small Fields content and transformation properties under
the symmetries imposed on the model.}}
\end{table}

The charged lepton Yukawa terms in the Lagrangian are given by\footnote{ In Eq. (\ref{charged:L}), 
one can introduce a contribution like $\bar{L}\phi_T^{\dagger}H e_{\R}$. 
But such a term can be absorbed in the original contribution by a mere redefinition of the coupling.}, 
\begin{equation}\label{charged:L}
 \mathcal{L}_l =  \frac{y_e}{\Lambda}(\bar{L}\phi_{\T})H e_{\R}
+\frac{y_{\mu}}{\Lambda}(\bar{L}\phi_{\T})'H\mu_{\R}+ 
\frac{y_{\tau}}{\Lambda}(\bar{L}\phi_{\T})''H\tau_{\R},
\end{equation}
to the leading order, where $\Lambda$ represents the cut-off scale of the 
theory and $y_e, y_{\mu}$ and $ y_{\tau}$ are the respective coupling constants.
Terms within the first parenthesis describe the product of two $A_4$ triplets, which 
further contracts with $A_4$ singlets $1$, $1''$ and $1'$ corresponding to 
$e_{\R}, \mu_{\R}$ and $\tau_{\R}$ fields respectively to constitute a true $A_4$ 
singlet. $A_4$ multiplication rules can be summarized as: $1'\times 1'=1''$, $1'\times 1''=1, 1''\times 1''= 1'$ 
and $3\times 3=1+1'+1''+3_A+3_S$. Further details about $A_{4}$ group can be found in~\cite{Altarelli:2010gt}.  
Now we choose the vev of $\phi_T$ as $\langle \phi_T \rangle=v_T(1,0,0)$ \cite{Altarelli:2005yx} so that the 
charged lepton mass matrix turns out to be diagonal in the leading order and 
can be written as   
$M_{l} = v\frac{v_{T}}{\Lambda} {\rm{diag}}\left( y_{e},  y_{\mu}, y_{\tau} \right)$. 

The allowed terms in the neutrino sector invariant under the symmetries considered are given by:
\begin{equation}\label{l1}
 \mathcal{L_{\nu}} = y_1\bar{L}\widetilde{H}N_R+ y_2 \overline{N_R^c}S\rho
+(\mu_1 \xi \rho^2/\Lambda^2+\mu_2\phi_S \rho^2/\Lambda^2) \overline{S^c}S
 +\mu_3 \overline{S^c}S\xi'\rho^2/\Lambda^2, 
\end{equation}
where $y_i, \mu_i$ are the  respective  couplings. 
To construct the flavor structures we consider the flavons acquire vevs along $
 \langle \phi_S \rangle=v_S(1,1,1),
 \langle \xi    \rangle=v_{\xi},
 \langle \xi'   \rangle=v_{\xi'}\hspace{.05cm}{\rm and}\hspace{.05cm}
 \langle \rho \rangle=v_\rho.  
$
In appendix \ref{apa}, we have written the complete scalar potential invariant under 
$A_4 \times Z_4 \times Z_3$ and the additional global $U(1)_{B-L}$ symmetry. There we have argued 
that such choices of vev alignments are indeed possible. 
With such vev alignment Eq. (\ref{l1}) yields the following  $9 \times 9$ mass matrix $M_{\nu}$ in the 
basis
$(\nu_L^c, N_R, S)$ 
\begin{eqnarray}\label{CMmu}
M_{\nu} = \left(
\begin{array}{ccc}
         0 &m_D &0\\
         m_D^T &0     &M\\
         0 &M^T &{\mu}
\end{array}
\right).
\end{eqnarray}
The $3\times 3$ mass matrices present in Eq. (\ref{CMmu}) are 
\begin{eqnarray}\label{mdmn}
m_D=y_1v \left(
\begin{array}{ccc}\label{mD-M}
      1 &0 &0\\
       0 &0 &1\\
       0 &1 &0
\end{array}
\right); 
M =y_2 v_\rho \left(
\begin{array}{ccc}
      1  &0    &0\\
      0  &0    &1\\
      0  &1    &0
\end{array}
\right) \hspace{.5cm} {\rm and }
\end{eqnarray}

\begin{eqnarray}\label{mumatrix}
{\mu}=\left(
\begin{array}{ccc}
 a-2b/3 & b/3   & b/3 \\
 b/3    & -2b/3 & a+b/3\\
 b/3    & a+b/3 & -2b/3
\end{array}
\right)
+
 \left(
\begin{array}{ccc}
      0  &0    &d\\
      0  &d    &0\\
      d  &0    &0
\end{array}
\right),
\end{eqnarray}
with $a=2 \mu_1v_\xi v_\rho^2/\Lambda^2$, $b=-2 \mu_2v_S v_\rho^2/\Lambda^2$ and $d=2 
\mu_3v_\xi' v_\rho^2/\Lambda^2$. Note that $\mu$ term follows from a higher dimensional 
contribution and hence is expected to be naturally small compared to $m_D$ and $M$. 

\section{Neutrino masses and Mixings}\label{sec3}

The specific flavor structure of the model ensures that  (as evident from Eq. (\ref{mnudef}) and 
Eq. (\ref{mD-M})) $F = m_D M^{-1}\propto\mathds{I}$. Hence in our set-up, the 
effective light 
neutrino mass matrix becomes 
\begin{eqnarray}\label{mnuc}
m_{\nu}&=& F\mu F^T =\frac{v^2y_1^2}{v_{\rho}^2y_2^2}~\mu.  
\end{eqnarray}
Eq. (\ref{mnuc}) clearly shows that the flavor structure of $m_\nu$  matrix is entirely 
dictated by that of $\mu$. 
Such an interesting feature was also pointed out in \cite{Lindner:2005pk}, calling it 
screening mechanism in the context of double seesaw. Additionally we note here that $\mu$  
serves the purpose of generating non-zero $\theta_{13}$ as well with a modification of its original 
TBM structure (similar to $M_0$ in Eq. (\ref{m})). This makes our model an interesting scenario 
to study, as the source of $\theta_{13}$ is connected with the lepton number violating 
parameter ($\mu$).
Now let us focus our attention to the $\mu$ matrix in Eq. (\ref{mumatrix}). It is well known from the 
very specific structure of the first matrix of right hand side of Eq. 
(\ref{mumatrix}) involving $a, ~b$  
only~\cite{Altarelli:2005yx} that it leads to a TBM pattern of the lepton
mixing matrix (as the charged lepton mass matrix is diagonal), given by 
\begin{eqnarray}\label{utb}
 U_{TB}=\left(
\begin{array}{ccc}
 \sqrt{\frac{2}{3}} & \frac{1}{\sqrt{3}} & 0 \\
 -\frac{1}{\sqrt{6}} & \frac{1}{\sqrt{3}} & -\frac{1}{\sqrt{2}} \\
 -\frac{1}{\sqrt{6}} & \frac{1}{\sqrt{3}} & \frac{1}{\sqrt{2}}
\end{array}
\right),
\end{eqnarray}
resulting $\theta_{13}=0$. The second matrix  in Eq. (\ref{mumatrix}) breaks the TBM pattern and we expect a deviation of
$\theta_{13}$ from zero. To find out the deviation and possible correlations 
between the mixing angles and parameters of the model, we first rotate $m_{\nu}$ from Eq. (\ref{mnuc}) by $U_{TB}$ so
as to get
\begin{eqnarray}\label{mnup}
m'_{\nu}&=&U_{TB}^Tm_{\nu}U_{TB},\\
&=&\frac{v^2y_1^2}{v_\rho^2y_2^2}\left(
\begin{array}{ccc}
 a-b-d/2     & 0   & \sqrt{3}d/2 \\
 0           & a+d & 0 \\
 \sqrt{3}d/2 & 0   & -a-b+d/2\\
\end{array}
\right).
\end{eqnarray}
As evident, a further rotation by $U_1$ (another unitary matrix) in
the 13 plane will diagonalize the light neutrino mass matrix, $i.e. ~m^{diag}_{\nu}=U^T_{1}m'_{\nu}U_1$ . 
The angle $\theta$ and phase $\psi$ associated in $U_1$ are therefore related 
with the parameters $a,b,d$ involved in $m_{\nu}$\footnote{The overall factor $\frac{y_1^2v^2}{y_2^2v_{\rho}^2}$
does not take part in determining the mixing angles and phases. However it would be important in determining the 
exact magnitude of light neutrino masses as we will see later.}.

Let us consider the form of $U_1$ as, 
\begin{eqnarray}\label{u1}
U_{1}=\left(
\begin{array}{ccc}
 \cos\theta               & 0 & \sin\theta{e^{-i\psi}} \\
     0                    & 1 &            0 \\
 -\sin\theta{e^{i\psi}} & 0 &        \cos\theta
\end{array}
\right),
\end{eqnarray}
where $\theta$ and $\psi$ are the angle and phase respectively. The diagonalization of $m_{\nu}$ takes place through 
\begin{eqnarray}\label{diag:m}
(U_{TB}U_1)^Tm_{\nu}U_{TB}U_1 = {\rm diag} (m_1e^{i\gamma_1},
m_2e^{i\gamma_2},m_3e^{i\gamma_3}),
\end{eqnarray}
where $m_{i=1,2,3}$ are the real 
and positive eigenvalues and $\gamma_{i=1,2,3}$ are the phases extracted from  
the corresponding complex eigenvalues. We are now in a position to evaluate the 
effective light neutrino mixing $U_{\nu}$ such that $U_{\nu}^{T}m_{\nu} U_{\nu} 
= {\rm diag} (m_1, m_2, m_3)$. The $U_{\nu}$ then becomes 
$ U_{\nu} =U_{TB}U_1U_m$, where $U_{m}={\rm diag} (1,e^{i\alpha_{21}/2},e^{i\alpha_{31}/2})$ is 
the Majorana phase matrix with $\alpha_{21}=(\gamma_1-\gamma_2)$ and 
$\alpha_{31}=(\gamma_1-\gamma_3)$, one common phase being irrelevant. 
Now this $U_{\nu}$ (charged lepton mass matrix being diagonal) can be compared with 
$U_{PMNS}$ which in its standard parametrization is given by~\cite{Agashe:2014kda},

{
\begin{eqnarray}
U_{PMNS}=\left(
\begin{array}{ccc} 
    c_{12}c_{13}                                &   s_{12}c_{13}
               & s_{13}e^{-i\delta}\\
    -s_{12}c_{23}-c_{12}s_{13}s_{23}e^{i\delta} &
c_{12}c_{23}-s_{12}s_{13}s_{23}e^{i\delta}    &   c_{13}s_{23} \\
     s_{12}s_{23}-c_{12}s_{13}c_{23}e^{i\delta} &
-c_{12}s_{23}-s_{12}s_{13}c_{23}e^{i\delta}    &   c_{13}c_{23}
\end{array}
\right)
U_{m},
\label{upmns}
\end{eqnarray}
}
where $c_{ij}=\cos\theta_{ij}$, $s_{ij}=\sin\theta_{ij}$, 
the angles $\theta_{ij} = [0, \pi/2]$, $\delta=[0, 2\pi]$ is the
CP-violating Dirac phase while $\alpha_{21}$ and $\alpha_{31}$ are the two
CP-violating Majorana phases.

We consider  $a = \vert a \vert e^{i\phi_a}$, $b = \vert b \vert e^{i\phi_b}$ 
and $d = \vert d \vert e^{i\phi_d}$ ($i.e.$ they are in general complex) the 
phases of which are indicated by $\phi_{a,b,d}$. For calculational purpose, 
we define parameters $|\alpha| = |b|/|a|$, $\beta| = |d|/|a|$ and the difference of phases 
by  $\phi_{ba}=\phi_b - \phi_a$ and $\phi_{da}=\phi_d - \phi_a$. 
As $U_1$ diagonalizes the $m'_{\nu}$ matrix in Eq. (\ref{mnup}), $\theta$ and $\psi$ can be expressed 
in terms of $a, b$ and $ d$ as, 
\begin{eqnarray}\label{tpsi}
\tan{2\theta} &=&\frac{ 
                    \sqrt{3}\beta\cos\phi_{da}
                    }
                    {
                    (\beta\cos\phi_{da}-2)\cos\psi+2\alpha\sin\phi_{ba}
                  \sin\psi
                    },\\
\tan\psi&=& \frac{\sin\phi_{da}}
                 {\alpha\cos(\phi_{ba}-\phi_{da})}\label{tpsi2}. 
\end{eqnarray}

Comparing $U_{\nu} = U_{TB}U_1U_m$ with $U_{PMNS}$ as in Eq.(\ref{upmns}), we obtain the 
following expressions for $\theta_{13}$  and Dirac CP phase $\delta$~\cite{Karmakar:2015jza}  
\begin{eqnarray}\label{ang0}
\sin\theta_{13}=\sqrt{\frac{2}{3}}\left|\sin\theta\right|, \hspace{.2cm}
\delta={\rm arg}[(U_1)_{13}]. 
\end{eqnarray}
These correlations are among the usual characteristics of the $A_4$ flavor 
symmetry~\cite{He:2006qd,Albright:2008rp,King:2011zj,Altarelli:2012bn}. 
For $\sin\theta>0$ (depending on the choices of $\alpha, \beta$),  the relation 
$\delta={\rm arg}[(U_1)_{13}]$ implies $\delta=\psi$. Again if $\sin\theta < 0$, 
the relation becomes $\delta=\psi \pm \pi$. Hence for both the cases, we have 
$\tan{\psi} = \tan{\delta}$ and Eq. (\ref{tpsi2}) becomes 
\begin{equation}
 \tan\delta= \frac{\sin\phi_{da}}
                 {\alpha\cos(\phi_{ba}-\phi_{da})}. 
\end{equation}

Using Eq. (\ref{diag:m}), the complex light neutrino mass eigenvalues  are evaluated as 
\begin{eqnarray}
m^c_{1,3}&=&\frac{v^2y_1^2}{v_\rho^2y_2^2}\left[-b \pm \sqrt{a^2-ad+d^2}\right], \\
 m^c_{2}&=&\frac{v^2y_1^2}{v_\rho^2y_2^2}(a+d). 
\end{eqnarray}
The real and positive mass eigenvalues ($m_i$) can be then extracted having the 
following expressions, 
\begin{eqnarray}
m_1&=&k\left[(P- \alpha\cos\phi_{ba})^2+(Q- \alpha\sin\phi_{ba})^2\right]^{1/2}\label{m1},\\
m_{2}&=&k\left[1+\beta^2+2\beta\cos\phi_{da}\right]^{1/2}\label{m2}, \\
m_{3}&=&k\left[(P+\alpha\cos\phi_{ba})^2+(Q+\alpha\sin\phi_{ba})^2\right]^{1/2},
\label{m3}
\label{absmass}
\end{eqnarray}
where $k= |a |{v^2|y_1|^2}/{v_\rho^2|y_2|^2}$ and 
\begin{eqnarray}  
 P&=&\left[\frac{1}{2}(A+\sqrt{A^2+B^2})\right]^{1/2}, ~~Q ~=~\left[\frac{1}{2}(-A+\sqrt{A^2+B^2})\right]^{1/2}, \\    
 A&=&1+\beta^2\cos 2\phi_{da}-\beta\cos \phi_{da}, ~~B~ = ~ \beta^2\sin 2\phi_{da}-\beta\sin 
 \phi_{da}.
\label{pqab} 
\end{eqnarray}
The three phases associated with these mass eigenvalues are $\gamma_i = \phi_i +\phi_{0}$,
where $\phi_0$ is the overall phase for $a y_1^2/y_2^2$ and $\phi_i$ are given 
by 
\begin{eqnarray}\label{majo:eq}
 \phi_{1} &=&\tan^{-1}\left(\frac{Q-\alpha\sin\phi_{ba}}{P-\alpha\cos\phi_{ba}}\right),\nonumber\\
 \phi_2&=&\tan^{-1}\left(\frac{\beta\sin\phi_{da}}{1+\beta\cos\phi_{da}}\right),\\
 \nonumber
\phi_3&=&\tan^{-1}\left(\frac{Q+\alpha\sin\phi_{ba}}{P+\alpha\cos\phi_{ba}}\right).\nonumber
\end{eqnarray}
Note that the Majorana phases $\alpha_{21}$ and $\alpha_{31}$ depend on $\phi_i$ 
only.

\section{Constraining parameters from neutrino data}\label{sec:analysis}
Using Eqs.(\ref{m1}-\ref{m3}) one can define a ratio of solar to the 
atmospheric 
mass-squared differences as
\begin{equation}\label{exp:r}
 r = \frac{\Delta m^2_{\odot}}{\vert \Delta m^2_{atm} \vert}, 
\end{equation}
with  $\Delta m^2_{\odot} \equiv \Delta m^2_{21} = m^2_2 - m^2_1$ and  
$\vert \Delta m^2_{atm} \vert \equiv |\Delta m^2_{31} |= |m^2_3 - m^2_1|$ .
From the expressions above in Section 3, it is clear that neutrino mixing 
angle  $\theta_{13}$, Dirac CP phase $\delta$, Majorana phases ($\alpha_{21} 
= \phi_1 - \phi_2$ and $ \alpha_{31} = \phi_1 - \phi_3$) and ratio $r$ are functions of 
four parameters, $\alpha, \beta, \phi_{ba}$ and $\phi_{da}$. The other two 
angles ($\theta_{23}, \theta_{12}$) are obtained from the comparison between $U_{\nu}$ and $U_{PMNS}$.
Among these, $\theta_{13},\theta_{23},\theta_{12}$ and ratio $r$ are 
precisely known from the neutrino oscillation data.  Since $\delta$ (also the 
Majorana phases) is yet not be known from the experimental data, we perform the 
analysis 
for several choices of $\delta$. Then the four parameters can be constrained using values of
$\theta_{13}, r$ and $\delta$ once we keep one of them fixed. For convenience, we have divided our 
analysis into five cases: 
(i) Case A [$\phi_{ba} = \phi_{da} = 0$], (ii) Case B [$\phi_{ba} = 0$], 
(iii) Case C [$\phi_{da} = 0$], (iv) Case D [$\phi_{ba} = \phi_{da} = 
\phi$] and  (v) the General Case.

Following \cite{Forero:2014bxa}, the best fit values of $\Delta m^2_{\odot} = 7.6\times 10^{-5} $ eV$^2$
and $\vert\Delta m^2_{atm}\vert = 2.48 \times 10^{-3} $ eV$^2$ along with their 3$\sigma$ ranges are used 
for our analysis. We have fixed $r$ at 0.03. Though there exists another 
parameter $k$ (see Eqs. (\ref{m1}-\ref{m3})), 
this cancels out in the expression for $r$. The magnitude of $k$ will be fixed in order to reproduce
the solar or atmospheric mass 
square difference(s). Once this is also obtained, we essentially get the estimate of the absolute neutrino 
masses and Majorana phases. Expression for the effective neutrino mass 
parameter $\vert m_{ee} \vert$  appearing in the neutrinoless double beta decay is given by\cite{Agashe:2014kda}, 
\begin{eqnarray}
\left|m_{ee}\right|=\left|m_1c_{12}^2c_{13}^2+
m_2s_{12}^2c_{13}^2e^{i\alpha_{21}}+
m_3s_{13}^2e^{i(\alpha_{31}-2\delta)}\right|. 
\end{eqnarray}
Hence we have a prediction for $\vert m_{ee} \vert$ for the allowed range of 
parameters. Note that in this analysis we should be able to find out values of 
$\alpha,\beta, k, \phi_{ba}, \phi_{da}$ and $\delta$ which are consistent with 
experimental data. However the scales involved as flavons vev, cut-off scale 
$\Lambda$, order of $\mu$ matrix ($i.e.$ the magnitude of $|a|,|b|, |d|$) can 
not be determined, specifically here in this section. Latter while discussing 
the non-unitarity effects in Section 5, we would be able to set limits on those 
scales.


\subsection{Case A: [$\phi_{ba}=\phi_{da}=0$]} 

In this case, we make the simplest choice for the associated phases as $\phi_{ba}=\phi_{da}=0$.  
Then Eq. (\ref{tpsi} and \ref{ang0}) can be written as 
\begin{equation}\label{dd0}
\tan{2\theta} =\frac{ 
                    \sqrt{3}\beta
                    }
                    {
                    (\beta-2)
                    }~~{\rm and} 
~~\sin\theta_{13}=\sqrt{\frac{2}{3}}|\sin\theta|,
\end{equation}
with $\tan\delta=0$. Hence  we note that $\sin\theta_{13}$ solely depends on $\beta$. 
\begin{figure}[h!]
\centering
\includegraphics[height=4.5cm]{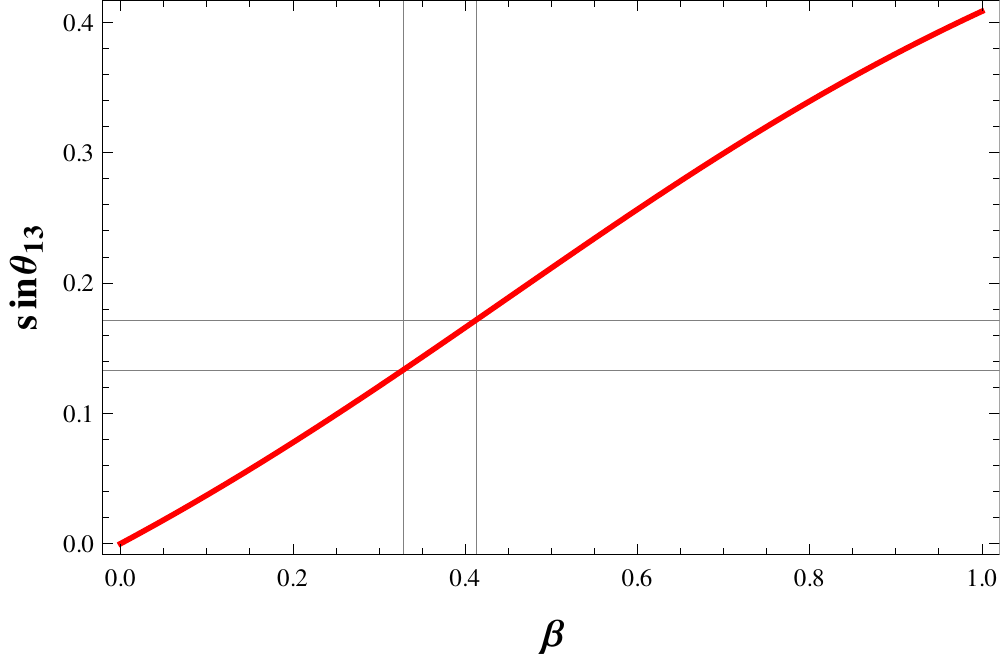}
\includegraphics[height=4.5cm]{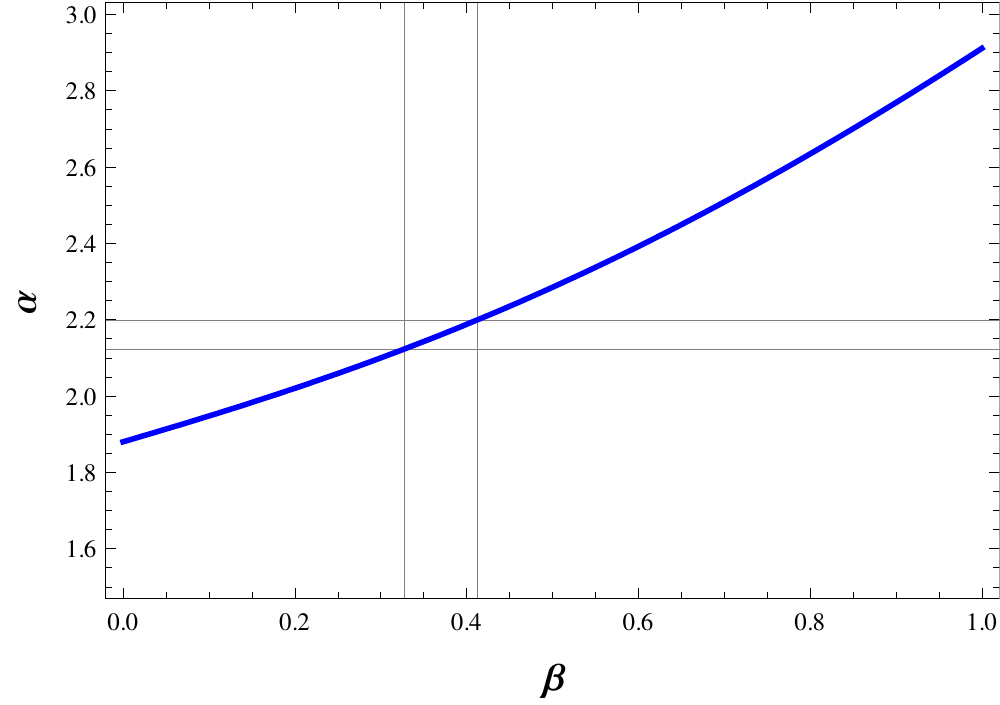}
\caption{ [Left panel] Plot for $\sin\theta_{13}$ vs $\beta$ . Here 3$\sigma$ range for 
$\sin\theta_{13}$ fixes $\beta$ in the range 0.328-0.413. [Right panel] $r=0.03$ contour
in the $\alpha$-$\beta$ plane.}
\label{fig:s13r00}
\end{figure}
With $\beta=0$ we get back the TBM pattern of neutrino mixing matrix. In Fig. \ref{fig:s13r00} 
left panel, we plot the variation of $\sin\theta_{13}$ against $\beta$ 
using Eq. (\ref{dd0}) the 3$\sigma$ 
range of $\sin\theta_{13}$ (between 0.133 and 0.177 as indicated by the two 
horizontal lines) predicts a range of 
$\beta$: $=0.328-0.413$ (denoted by the vertical lines).

With $\phi_{ba}=\phi_{da}=0$, expressions of absolute neutrino masses in Eq. (\ref{m1}-\ref{m3}) simplify into,  
\begin{eqnarray}
m_{1}&=&k\left|\sqrt{1+\beta^2-\beta}-\alpha\right|,\label{m100}\\
 m_{2}&=&k\left[1+\beta\right],\label{m200}\\
m_{3}&=&k\left[\sqrt{1+\beta^2-\beta}+\alpha\right].\label{m300}
\end{eqnarray}
Thereby the  ratio of solar to atmospheric mass-squared differences, $r$ (as 
defined in Eq. (\ref{exp:r})), 
now takes the form 
\begin{eqnarray}
 r=\frac{1}{2}-\frac{\alpha^2-3\beta}{4\alpha\sqrt{1+\beta^2-\beta}}.
\end{eqnarray}
Note that this ratio depends upon both $\alpha$ and $\beta$. To understand this 
 dependence in a better way, we draw the contour plot for $r=0.03$ 
\cite{Agashe:2014kda} in 
$\alpha-\beta$ plane as shown in Fig. \ref{fig:s13r00} (right panel). We find that 
the allowed range of $\beta$ from Fig. \ref{fig:s13r00} (left panel) indicates 
a range of the 
other parameter $\alpha$ to be within (2.12 - 2.18) as seen from Fig. 
\ref{fig:s13r00} 
(right panel). 
Note that contour plot of $r$ provides 
a one to one correspondence between $\alpha$ and $\beta$ values within this range. 
For example, the best fit values of $\sin\theta_{13}$ and $r$ corresponds to   
 $\alpha =2.16 $ and $\beta=0.372$.
So the sets of $(\alpha, \beta)$ values within this allowed range would be used 
for rest of our analysis in Case A. It is observed that $\theta_{12}$ and 
$\theta_{23}$ also fall within their $3\sigma$ value~\cite{Forero:2014bxa} 
for the entire allowed range of $\alpha$ and $\beta$. 

\begin{figure}[h]
$$
\includegraphics[height=4.5cm]{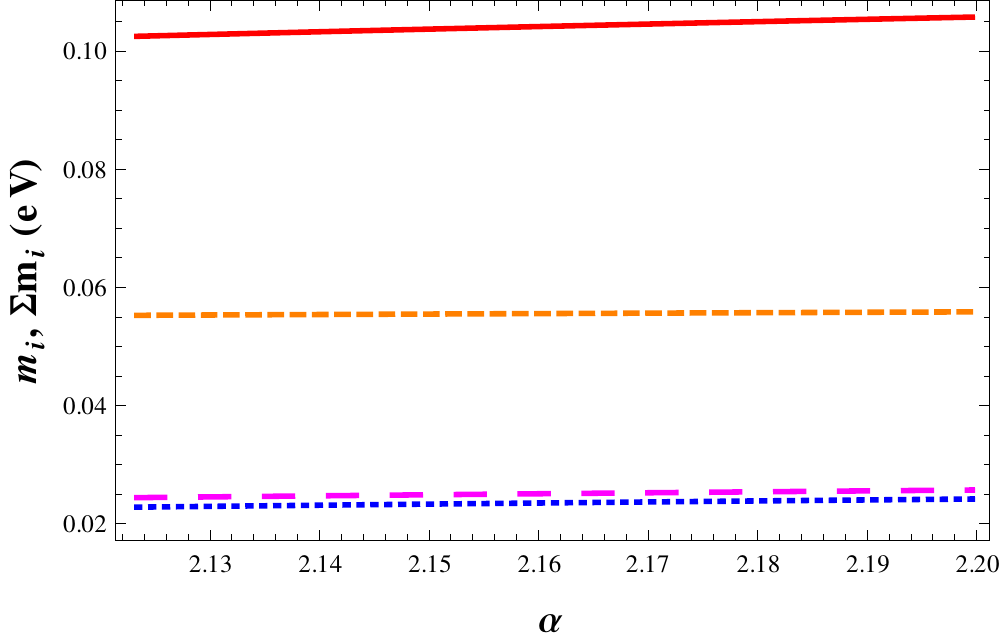}
\includegraphics[height=4.5cm]{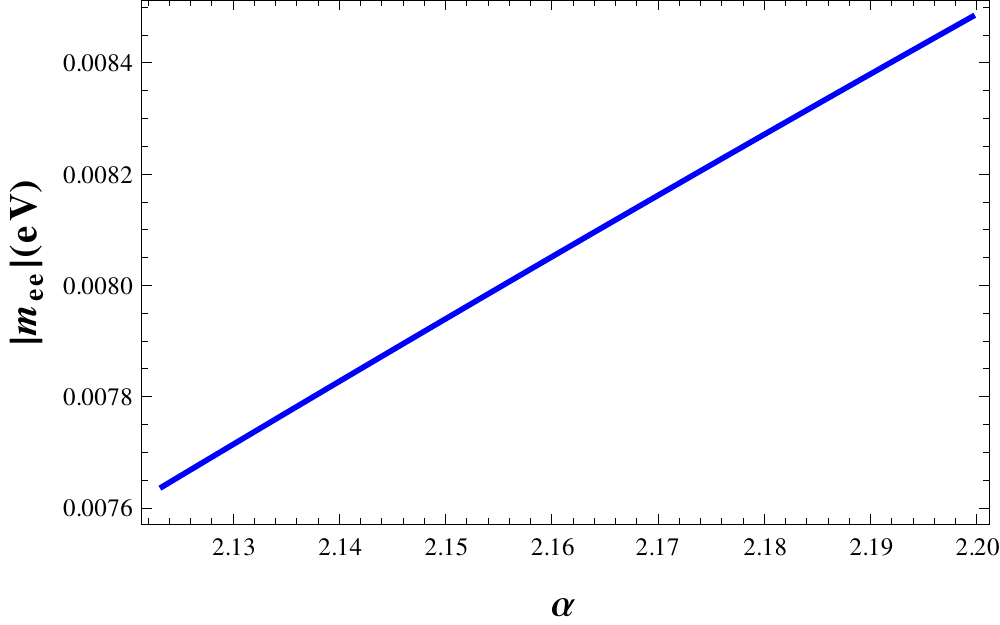}
$$
\caption{ [Left panel] Absolute neutrino masses vs $\alpha$ (blue dotted,  
magenta large dashed, orange dashed  and red continuous lines represent $m_1$, 
$m_2$, $m_3$ and $\sum m_{i}$ respectively); [Right panel] Plot for $|m_{ee}|$ 
vs $\alpha$ [Case A].}
\label{fig:meem00}
\end{figure}

\begin{table}[h]
\centering
\begin{tabular}{c|c}
\hline
Parameters/Observables & Allowed Range\\
\hline
$\beta$         & 0.328-0.412  \\
$\alpha$        & 2.12-2.18     \\ 
$k$    (eV)    & $1.84\times 10^{-2}$ - $1.82\times 10^{-2}$   \\
$\sum m_i$ (eV)  & 0.102462 - 0.105713\\
$|m_{ee}|$   (eV)  & 0.0076-0.0085\\
\hline\hline
\end{tabular}\caption{\label{tab:000} {\small Range of $\beta, \alpha, 
k, \sum m_i$ and $|m_{ee}|$ for 3$\sigma$ variation of $\sin\theta_{13}$ [Case 
A].}}			
\end{table}

From Eq. (\ref{m100}-\ref{m300}) it is evident that along with $\alpha$ and 
$\beta$, individual absolute light neutrino masses   depend also  upon another 
parameter $k  ( = |a |{v^2|y_1|^2}/{v_\rho^2|y_2|^2})$. Once we know
the sets of $(\alpha, \beta)$  that produces $\sin\theta_{13}$ 
in the 3$\sigma$ allowed range and  $r=0.03$, it is possible to determine $k$ 
from the best fit values of solar (or atmospheric) mass square differences,  
$m^2_2 - m^2_1 = 7.6\times 10^{-5}$ \hspace{.0cm} eV$^2$ 
($\vert \Delta m^2_{atm} \vert=2.48\times 10^{-3}$ 
eV$^2$)~\cite{Forero:2014bxa}. Hence corresponding to a set ($\alpha,\beta$), 
we can determine $k$. Doing so,  we find the allowed range for $k$ turns out to 
be $(1.82 - 1.84) \times 10^{-2}$ eV.
Using such a set of values of ($\alpha, ~\beta, ~k$)  we  plot  the sum of the 
light neutrino 
masses and effective mass parameter $|m_{ee}|$ for 
neutrinoless double beta decay  in the left and right panels of Fig. \ref{fig:meem00} respectively. 
Our findings are summarized in Table \ref{tab:000} in terms of allowed ranges 
for parameters and observables.

\subsection{Case B: [$\phi_{ba}=0$]}

With $\phi_{ba}=0$, Eqs. (\ref{tpsi}) and (\ref{tpsi2}) reduce into 
\begin{equation}
\tan{2\theta} = \frac{ \sqrt{3}\beta\cos\phi_{da}}{(\beta\cos\phi_{da}-2)\cos\psi}, ~~
\tan \delta = \frac{\tan\phi_{da}}{\alpha}.
\label{tanB}
\end{equation}
As before, $\sin\theta_{13}$ can be obtained from the relation 
$\sin\theta_{13}=\sqrt{\frac{2}{3}}\left|\sin\theta\right|$. Using 
Eqs.(\ref{m1}-\ref{m3}), the ratio of solar to atmospheric mass squared differences 
in this case can be written as
\begin{eqnarray}\label{r:ba0}
 r=\frac{1}{4\alpha P}\left[ 1+\beta^2+2\beta\cos\phi_{da} - (P-\alpha)^2-Q^2 \right],
\end{eqnarray}
where $P$ and $Q$ are same as given in Eqs. (\ref{pqab}). 
\begin{figure}[h!]
$$
\includegraphics[height=5.0cm]{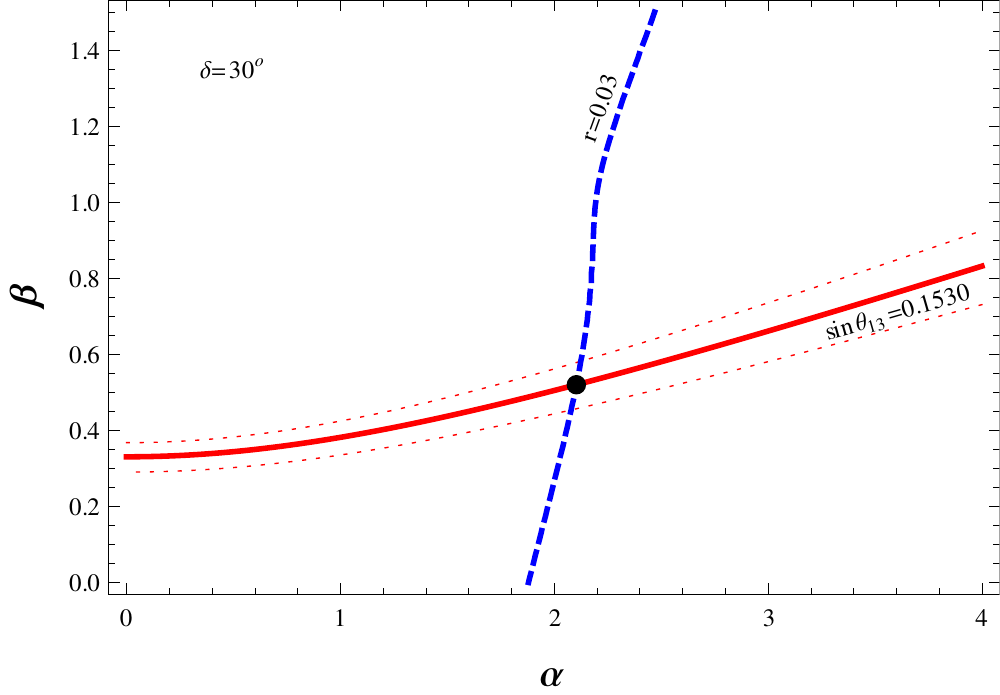}
\includegraphics[height=5.0cm]{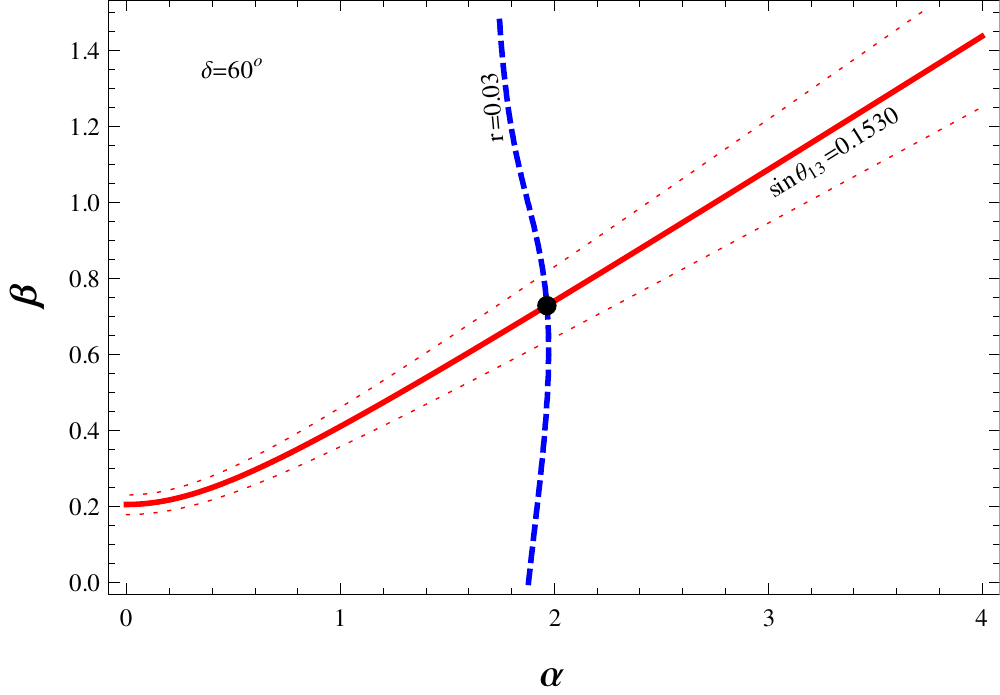}
$$
\caption{ Contour plot for  $r=0.03$ (dashed line) and 
$\sin\theta_{13}=0.153$ (continuous line) for $\delta=30^{\circ}$ (left panel) 
and $\delta=60^{\circ}$ (right panel) respectively. Red dotted lines represent 
a 3$\sigma$ variation of $\sin\theta_{13}$ while black dots stand for 
intersection (solution) points for best fit values of $\sin\theta_{13}$ and $r$ 
in both panels. }
\label{fig:s13rba0}
\end{figure}

\begin{table}[h!]
\centering
\begin{tabular}{cccccc}
\hline
$\delta$ & $\alpha$ & $\beta$ & $k$ (eV) & $\Sigma m_i$  (eV) & $|m_{ee}|$ 
(eV)\\
\hline
$0^{\circ}$  & 2.162 & 0.372 & 0.0183 & 0.1042& 0.0222 \\
$10^{\circ}$ & 2.155 & 0.393 & 0.0184 & 0.1047& 0.0225 \\
$20^{\circ}$ & 2.136 & 0.448 & 0.0188 & 0.1065& 0.0233 \\
$30^{\circ}$ & 2.103 & 0.521 & 0.0195 & 0.1093& 0.0245 \\
$40^{\circ}$ & 2.060 & 0.596 & 0.0204 & 0.1128& 0.0260 \\
$50^{\circ}$ & 2.011 & 0.666 & 0.0213 & 0.1162& 0.0274 \\
$60^{\circ}$ & 1.965 & 0.728 & 0.0220 & 0.1182& 0.0280 \\
$70^{\circ}$ & 1.928 & 0.782 & 0.0221 & 0.1179& 0.0275 \\
$80^{\circ}$ & 1.901 & 0.827 & 0.0217 & 0.1152& 0.0259 \\
$90^{\circ}$ & 1.879 & 0.859 & 0.0210 & 0.1109& 0.0270 \\
\hline\hline
\end{tabular}\caption{\label{tab:ba0} {\small Parameters satisfying neutrino 
oscillation data for various values of $\delta$ with $\phi_{ba}=0$ [Case B].}}	
\end{table}

The above expressions show that $\sin\theta_{13}$ and $r$ both are dependent 
on three parameters namely $\alpha,\beta$ and $\phi_{da}$ contrary to Case A where 
they depend only on two parameters  $\alpha$ and $\beta$. However if we choose a particular 
$\delta$,  we can replace $\phi_{da}$ dependence in terms of $\alpha$ by using the 
second relation from  Eq.(\ref{tanB}). 
Then if we draw contours of $r$ and $\sin\theta_{13}$ in the 
$\alpha, ~\beta$ plane where a simultaneous satisfaction of best fit values 
of $\sin\theta_{13}$ and $r$ provide solutions for $\alpha$ and $\beta$ with  
that specific choice of $\delta$.
As an example, we have drawn contour plots for $\sin\theta_{13}=0.153$ and 
$r=0.03$ in Fig. \ref{fig:s13rba0} for $\delta=30^{\circ}$ (left panel) and 
$\delta=60^{\circ}$ (right panel) in $\alpha-\beta$ plane. Intersecting points 
between the $\sin\theta_{13}$  
and $r$ contours in these plots, denoted by black dots represent the set of 
solutions
($\alpha, ~\beta$) satisfying neutrino oscillation data. $\theta_{12}$ and $\theta_{23}$
 fall in the right range for the entire 3$\sigma$ range of 
$\sin\theta_{13}$ considered. With each such set of 
solution points ($\alpha, ~\beta$) for a fixed $\delta$, we can compute the other 
parameter $k$ in order to obtain the correct solar (or atmospheric) mass splitting. 
Here in Table \ref{tab:ba0} we have provided sets of values for ($\alpha, 
\beta, k$) for various $\delta$ satisfying  $\sin\theta_{13}=0.153$ and 
$r=0.03$ obtained from neutrino oscillation experiments. 

It is to be noted that with a particular choice of $\delta$, contour plots for 
both $\sin\theta_{13}$ and $r$ are identical with the one  obtained from 
$|\pi-\delta|$. Here in this set-up, scanning over all values of $\delta$ (with 
3$\sigma$ variation of $\sin\theta_{13}$ taken into account), sum of the three 
light neutrino 
masses and effective mass parameter are predicted to be in the range :
$ 0.104 \hspace{.05cm}{~\rm eV} \hspace{.05cm} \lesssim \Sigma m_i 
\lesssim 0.118 \hspace{.05cm}{\rm eV}$ and 
$0.022 \hspace{.05cm}{~\rm eV} \hspace{.05cm} \lesssim |m_{ee}| \lesssim 
0.028 \hspace{.05cm}{\rm eV}$. These are mentioned in the two rightmost 
columns in Table \ref{tab:ba0}.

\subsection{Case C: [$\phi_{da}=0$]}
We consider here the other possibility of choosing one of the two phases as zero, $i.e. ~\phi_{da}=0$. 
Then we have relations $\tan{2\theta} =\frac{ 
                    \sqrt{3}\beta
                    }
                    {
                    (\beta-2)
                    },\hspace{.1cm}$
$\sin\theta_{13}=\sqrt{\frac{2}{3}}|\sin\theta|$ 
with $\tan\delta=0$. This coincides with Eq. (\ref{dd0}) of Case A. Hence we 
can use the outcome of Fig. \ref{fig:s13r00} (left panel) for specifying the 
range of $\alpha,\beta$ which reproduce the value of $\sin\theta_{13}$ (with in 
3$\sigma$ allowed range) and $r$ respectively. With $\phi_{da}=0$, the real and 
positive mass eigenvalues take the form
\begin{eqnarray}
m_{1}&=&k\left[(\sqrt{1+\beta^2-\beta}-\alpha\cos\phi_{ba})^2+(\alpha\sin\phi_{
ba})^2\right]^{1/2},\label{m111}\\
 m_{2}&=&k\left[1+\beta \right],\\
m_{3}&=&k\left[(\sqrt{1+\beta^2-\beta}+\alpha\cos\phi_{ba})^2+(\alpha\sin\phi_{
ba})^2\right]^{1/2}.
\label{m123}
\end{eqnarray}
\begin{figure}[h]
$$
\includegraphics[height=5.0cm]{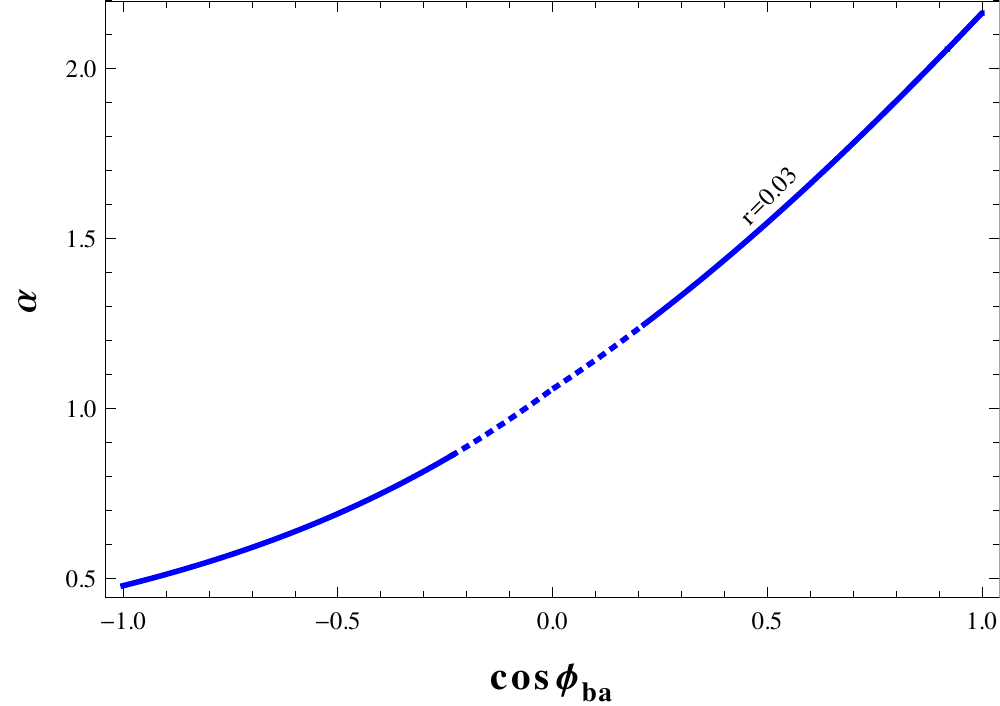}
$$
\caption{  Contour plot for $r=0.03$ in the $\alpha-\cos\phi_{ba}$ plane for 
           $\phi_{ba}=0$. The disallowed range of $\alpha,\cos\phi_{ba}$  is 
indicated by the dotted portion. }
\label{fig:rba0}
\end{figure}

In this case, the ratio of solar to atmospheric mass-squared differences $r$, 
is related to the parameters by the relation, 
\begin{equation}
r=\frac {3\beta-\alpha^2+2\alpha\cos\phi_{ba} 
\sqrt{1+\beta^2-\beta}}{4\alpha \sqrt{1+\beta^2-\beta} |\cos\phi_{ba}|}. 
\label{cosbaC}
\end{equation}

\begin{figure}[h!]
$$
\includegraphics[height=5.0cm]{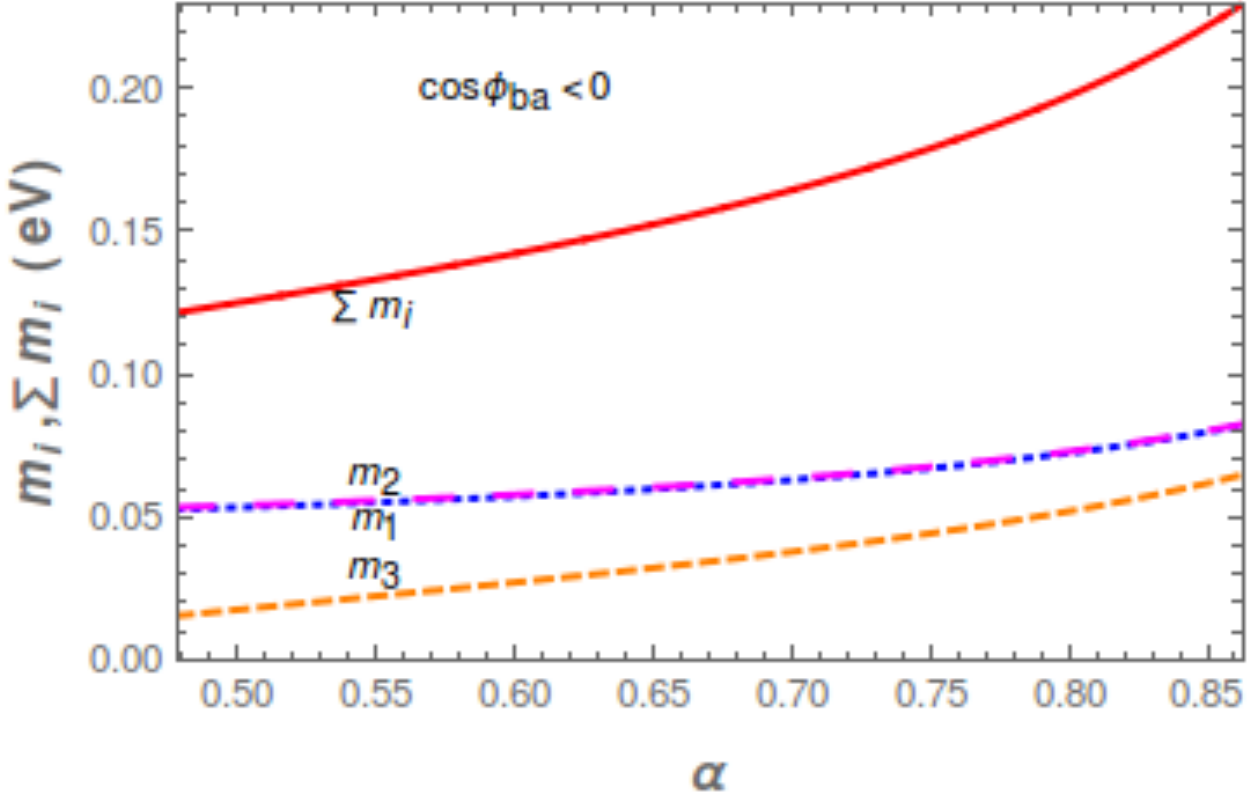}
\includegraphics[height=5.0cm]{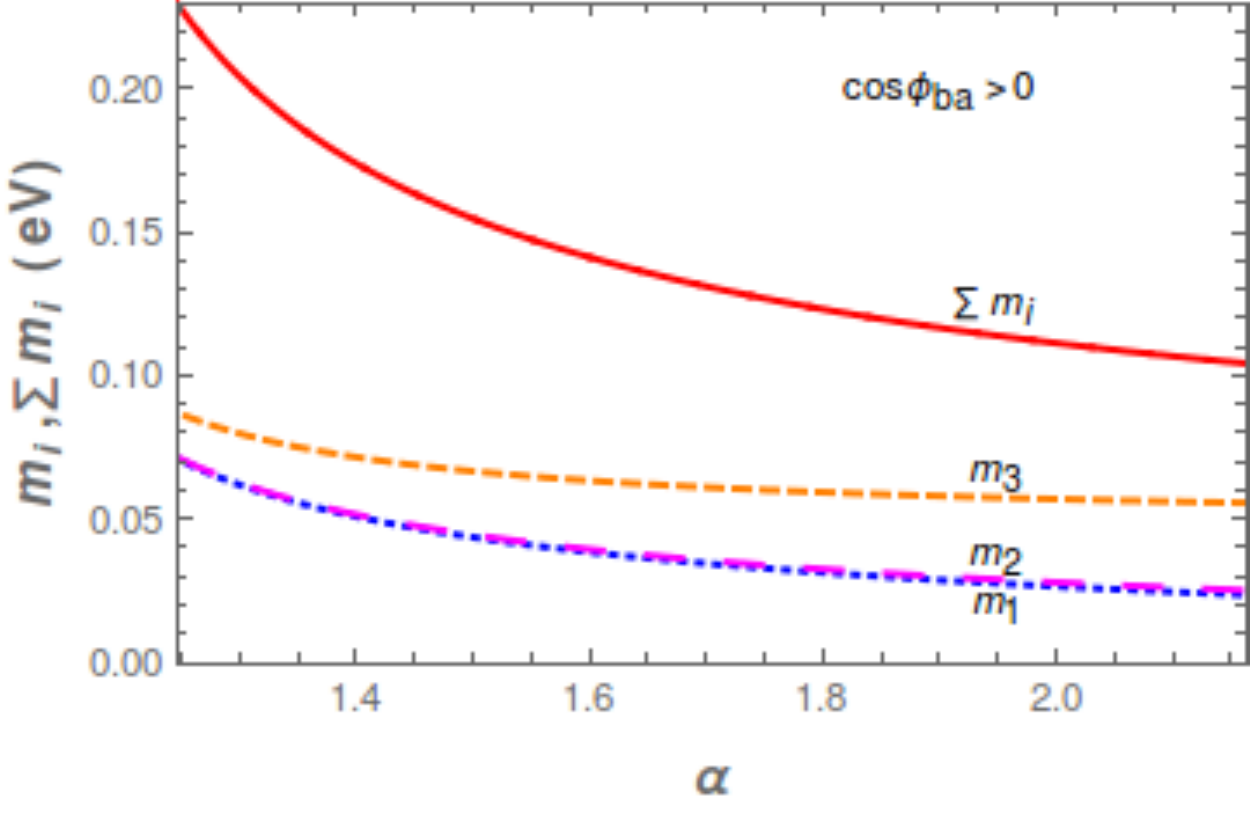}
$$
\caption{Absolute neutrino masses vs $\alpha$ (blue dotted,  magenta large 
dashed, orange dashed  and red continuous lines represent $m_1$, $m_2$, $m_3$ 
and $\sum m_{i}$ respectively). The left panel is for $\cos\phi_{ba}<0$ and  
right panel is for $\cos\phi_{ba}>0$.}
\label{fig:mimeeba0}
\end{figure}

The Eq. (\ref{cosbaC})  describes a relation between parameters 
$\alpha, \beta$ and $\phi_{ba}$. 
We fix  $\beta$ at 0.372 which corresponds to the best fit value of 
$\sin\theta_{13} = 0.153$ 
as seen from Fig. \ref{fig:s13r00} (left panel). Then 
$\cos\phi_{ba}$ and $\alpha$ correlation is addressed 
through a contour plot of $r = 0.03$ in Fig. \ref{fig:rba0} using  Eq. 
(\ref{cosbaC}). We find for $-1\leq\cos\phi_{ba}\leq 1$, 
$\alpha$ falls with the region $0.478\leq\alpha\leq 2.162$. This range is  
further constrained once 
we use cosmological constraint on sum of the light neutrino masses to be  below 
0.23 eV~\cite{Ade:2013zuv}. This exclusion part 
is indicated by the dotted portion of the $r$ contour in Fig. \ref{fig:rba0}. Now in order to have an estimate of absolute 
neutrino masses, first we need to know the other parameter $k$. Corresponding to the fixed value of 
$\beta = 0.372$, we have sets of values of ($\cos\phi_{ba}, ~\alpha$) which leads to $r=0.03$ from 
Fig.\ref{fig:rba0}. For each such set of ($\cos\phi_{ba}, ~\alpha$), we can have the corresponding $k$ 
value in order to get the best fit value for solar mass squared 
difference,  $m^2_2 - m^2_1 = 7.6\times 10^{-5}$ eV$^2$,  and obtain
\begin{equation}\label{k:ba0}
 k=\left[\frac{7.6\times 10^{-5}}{4\alpha r\sqrt{1+\beta^2-\beta} |\cos\phi_{ba}|}\right]^{1/2},
\end{equation}
where the Eqs. (\ref{m111}, \ref{cosbaC}) are employed and $\beta=0.372$ is taken.   In Fig. \ref{fig:mimeeba0} (left panel
and right panel) we have plotted 
absolute neutrino masses ($m_i$) against $\alpha$ (with $\beta = 0.372$) where one to one correspondence between 
$\alpha$ and $\cos\phi_{ba} (<0$ and $>0)$ from Fig. \ref{fig:rba0}  is taken 
into account.
Here 
$m_1, m_2, m_3$ and $\sum m_i$ are denoted by blue dotted, magenta large dashed
orange dashed and red continuous lines respectively. Note that 
$\cos\phi_{ba}<0$ indicates the inverted hierarchy while
$\cos\phi_{ba}>0$ corresponds to the normal hierarchy for light neutrinos. We 
have found the prediction for $|m_{ee}|$ to be within 
$0.016 \hspace{.1cm}{\rm eV} <|m_{ee}|<0.052 \hspace{.1cm}{\rm eV}$ for normal hierarchy
and $0.047 \hspace{.1cm}{\rm eV} <|m_{ee}|<0.066 \hspace{.1cm}{\rm eV}$ for 
inverted hierarchy considering the restricted variation of $\alpha$ 
 ($0.478\leqslant\alpha\leqslant 0.863$ for $\cos\phi_{ba}<0$ and 
$1.247 \leqslant\alpha\leqslant 2.162$ for $\cos\phi_{ba}>0$). Few of our 
findings are tabulated in Table \ref{tab:parac}.

\begin{table}[h]\centering
\resizebox{8.5cm}{!}{%
 \begin{tabular}{c|ccccc}
\hline
$\alpha$ & $\cos\phi_{ba}$ & $k$ & $\sum m_i$ & $|m_{ee}|$\\
\hline
1.904 & 0.8 &  0.0218 eV & 0.1164 eV & 0.0194 eV  \\
0.814 & -0.3 & 0.0544 eV & 0.0231 eV & 0.0604 eV \\
\hline\hline
\end{tabular}
}
\caption{\label{tab:parac} {\small Representative values of $k, \sum m_i$ 
and $|m_{ee}|$ in Case C. }}
\end{table}

\subsection{Case D: [$\phi_{ba}=\phi_{da}=\phi_{a}$]}
\begin{figure}[h!]
$$
\includegraphics[height=5.0cm]{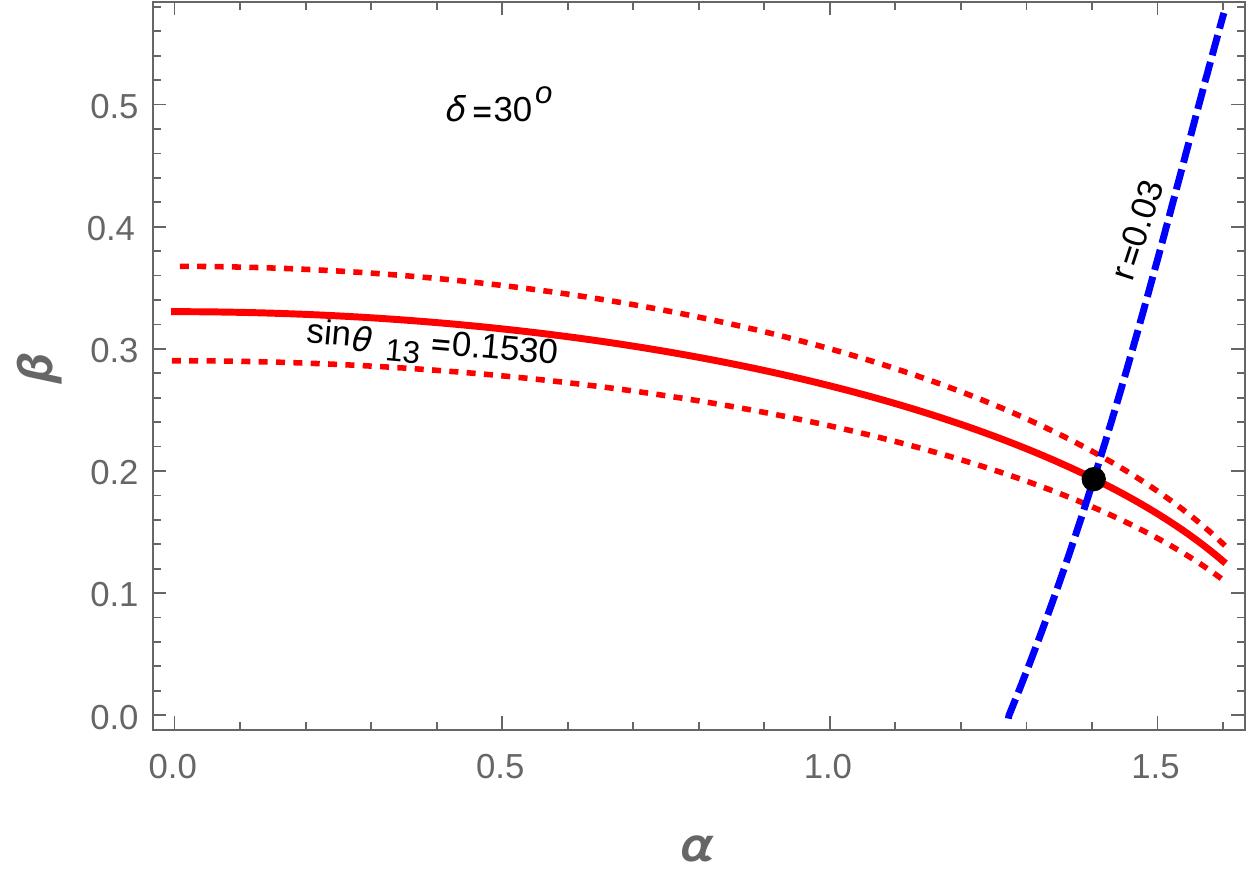}
\includegraphics[height=5.0cm]{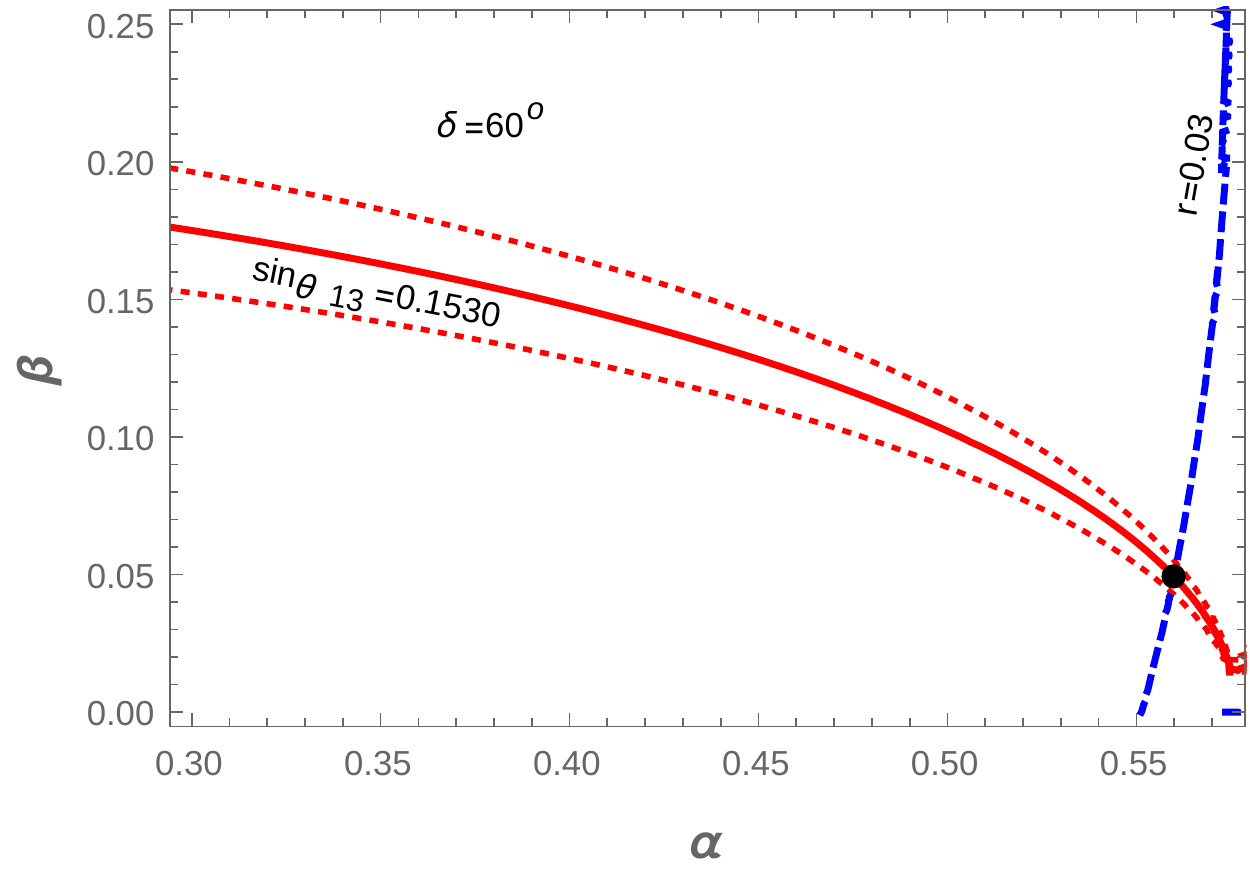}
$$
\caption{ Contour plot for  $r=0.03$ (dashed line) and 
$\sin\theta_{13}=0.153$ (continuous line) for $\delta=30^{\circ}$ (left panel) 
and 
$\delta=60^{\circ}$ (right panel) respectively. Red dotted lines represent a 
3$\sigma$ variation $\sin\theta_{13}$ and black dots stands for solution 
points for best fit values of $\sin\theta_{13}$ and $r$ in both panels.}
\label{fig:a}
\end{figure}

Now, if we consider $\phi_{ba}=\phi_{da}=\phi$, then Eqs. \ref{tpsi} and \ref{tpsi2} can be written as  
\begin{equation}
\tan{2\theta} = \frac{ \sqrt{3}\beta\cos\phi}{(\beta\cos\phi-2)\cos\psi+2\alpha\sin\phi\sin\psi}, ~~
\tan \delta =\tan\psi= \frac{\sin\phi}{\alpha}.
\label{}
\end{equation}
and hence $\sin\theta_{13}$ again can be computed using the relation  
$\sin\theta_{13}=\sqrt{\frac{2}{3}}|\sin\theta|$. 
The real and positive mass eigenvalues now take the form
\begin{eqnarray}
m_1&=&k\left[(P_D- \alpha\cos\phi)^2+(Q_D- \alpha\sin\phi)^2\right]^{1/2}\label{},\nonumber\\
m_{2}&=&k\left[1+\beta^2+2\beta\cos\phi\right]^{1/2}\label{},\nonumber \\
m_{3}&=&k\left[(P_D+\alpha\cos\phi)^2+(Q_D+\alpha\sin\phi)^2\right]^{1/2},\label
{}\nonumber
\label{}
\end{eqnarray}
with
\begin{eqnarray}  
 P_D&=&\left[\frac{1}{2}\left(A_D+\sqrt{A_D^2+B_D^2}\right)\right]^{1/2}, ~~Q_D ~=~\left[\frac{1}{2}\left(-A_D+\sqrt{A_D^2+B_D^2}\right)\right]^{1/2}, \\    
 A_D&=&1+\beta^2\cos 2\phi-\beta\cos \phi, ~~B_D~ = ~ \beta^2\sin 2\phi-\beta\sin 
 \phi.
\label{} 
\end{eqnarray}
Using above expressions for light neutrino masses we can write the ratio of solar to atmospheric mass squared difference as
\begin{equation}
 r=\frac{(1+\beta^2+2\beta\cos\phi)-(P_D- \alpha\cos\phi)^2-(Q_D- \alpha\sin\phi)^2}
       {4\alpha(P_D\cos\phi+Q_D\sin\phi)}. 
\end{equation}
Clearly just like Case B, here also both $\sin\theta_{13}$ and $r$ both depends on $\alpha,\beta$ and the common phase $\phi$. 
Following the  same prescription as in Case B, one can draw contours for best fit values of $\sin\theta_{13}$ and $r$ in 
the $\alpha,\beta$ plane. Intersecting points of these two contours then represent simultaneous solutions for both $\alpha$ and 
$\beta$ for a particular value of $\delta$. In Fig. \ref{fig:a} we have drawn such contours for $\delta=30^{\circ}$ (left panel)
and $\delta=60^{\circ}$ (right panel) for demonstrative purpose. In this plot, 
black dots represent the intersecting points for 
$\sin\theta_{13}=0.153$ and $r=0.03$ contours and hence the solutions for $\alpha$ and $\beta$.
Here we find that solutions satisfying neutrino oscillation data exist for all 
values of $\delta$ between $0^{\circ}$ and $90^{\circ}$ as given in Table 
\ref{tab:a}. 
We find that the contour
plots for both $\sin{\theta_{13}}=0.1530$ and $r=0.03$ with a specific $\delta$ value coincides
(and hence the solutions for $\alpha$, $\beta$) with
the one with other $\delta$ values (in the range 0 to $2\pi$) obtained from $|\pi - \delta|$.

\begin{table}[h!]
\centering
\begin{tabular}{cccccc}
\hline
$\delta$ & $\alpha$ & $\beta$ & $k$ (eV) & $\Sigma m_i$  (eV) & $|m_{ee}|$ (eV)\\
\hline
$0^{\circ}$  & 2.162 & 0.372 & 0.0183 & 0.1042& 0.0220 \\
$10^{\circ}$ & 2.039 & 0.343 & 0.0194 & 0.1057& 0.0221 \\
$20^{\circ}$ & 1.755 & 0.272 & 0.0223 & 0.1095& 0.0214 \\
$30^{\circ}$ & 1.403 & 0.194 & 0.0273 & 0.1187& 0.0225 \\
$40^{\circ}$ & 1.070 & 0.131 & 0.0354 & 0.1365& 0.0319 \\
$50^{\circ}$ & 0.792 & 0.084 & 0.0472 & 0.1659& 0.0447 \\
$60^{\circ}$ & 0.560 & 0.049 & 0.0658 & 0.2159& 0.0641 \\
$62^{\circ}$ & 0.518 & 0.043 & 0.0701 & 0.2301& 0.0694 \\
$70^{\circ}$ & 0.359 & 0.023 & 0.1011 & 0.3157& 0.1000 \\
$80^{\circ}$ & 0.175 & 0.006 & 0.2027 & 0.6144& 0.2022\\
\hline\hline 
\end{tabular}\caption{\label{tab:a} {\small Parameters satisfying neutrino 
oscillation data for various values of $\delta$ with $\phi_{ba}=\phi_{ba}=\phi$  
[Case D].
}}	
\end{table}
Following the same algorithm as described in Case B, in the last two column of Table \ref{tab:a} 
we have listed allowed values for sum of all three light neutrinos and effective mass parameter. Therefore
varying $\delta$ between $0$ to $2\pi$, we find range of few quantities as 
$ 0.1042 \hspace{.05cm}{~\rm eV} \hspace{.05cm} \lesssim \sum m_i 
\lesssim 0.6144\hspace{.05cm}{\rm eV}$ and 
$0.0220 \hspace{.05cm}{~\rm eV} \hspace{.05cm} \lesssim |m_{ee}| \lesssim 
0.2022 \hspace{.05cm}{\rm eV}$ respectively. Therefore imposing the constraint 
$\sum m_{i}<0.23$ eV on sum of all three light neutrinos
coming from Planck~\cite{Ade:2013zuv}, the allowed range for $\delta$ gets restricted and it finally lies 
in the range $0^{\circ}\leq \delta < 62^{\circ}$ (in terms of the full range 
$0^{\circ}-360^{\circ}$, other allowed ranges are
$128^{\circ}-180^{\circ}$ and $180^{\circ}-242^{\circ}$, $308^{\circ}-360^{\circ}$). In this case only the normal hierarchy results as in Case A
and B.


\subsection{General Case}

In the previous sub-sections, we have considered four different cases with specific choices for 
$\phi_{ba}$ and/or $\phi_{da}$ for our analysis on neutrino masses and mixing. Here we 
discuss the most general case where we allow the variation of $\phi_{ba}$ and 
$\phi_{da}$  for their entire range between 0 and $2\pi$. For this purpose, we employ 
Eqs. (\ref{tpsi}-\ref{ang0}) in order to analyze the mixing angles. On the other hand, the 
ratio of solar to the atmospheric mass-squared differences $r$, defined in Eq. (\ref{exp:r}),  
can also be computed once we use the general expressions for absolute neutrino masses 
given in Eq. (\ref{m1}-\ref{m3}). 

\begin{figure}[h]
$$
\includegraphics[height=4.0cm]{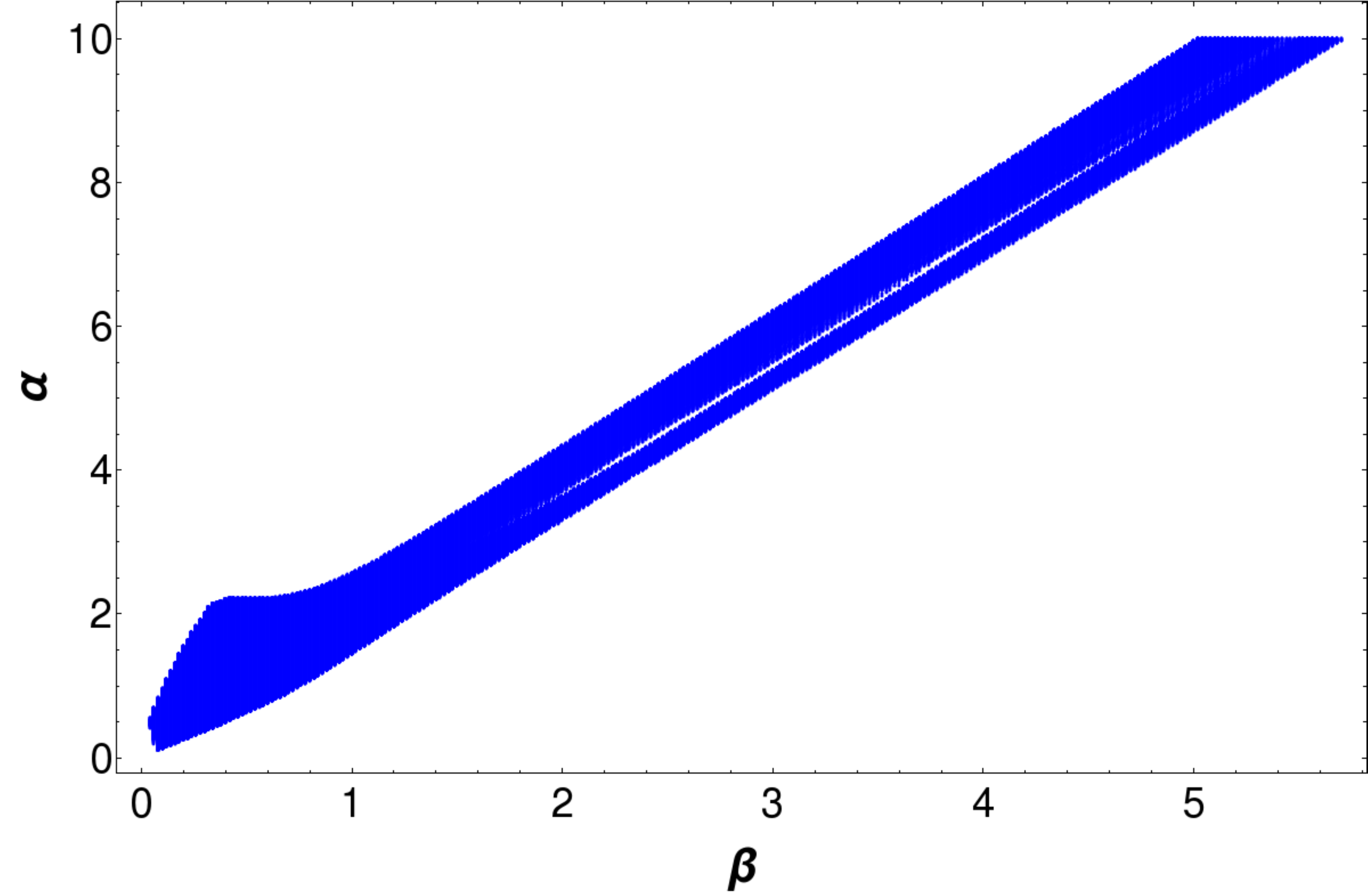}
$$
\caption{Allowed range (represented by the blue patch) of $\alpha$ and $\beta$ in order to 
satisfy 3$\sigma$ range of $\sin\theta_{13}$ and $r$. The phases $\phi_{ba,da}$ are 
allowed to vary within 0 to 2$\pi$.}
\label{fig:g-a-b}   
\end{figure}
\begin{figure}[h]
$$
\includegraphics[height=4.5cm]{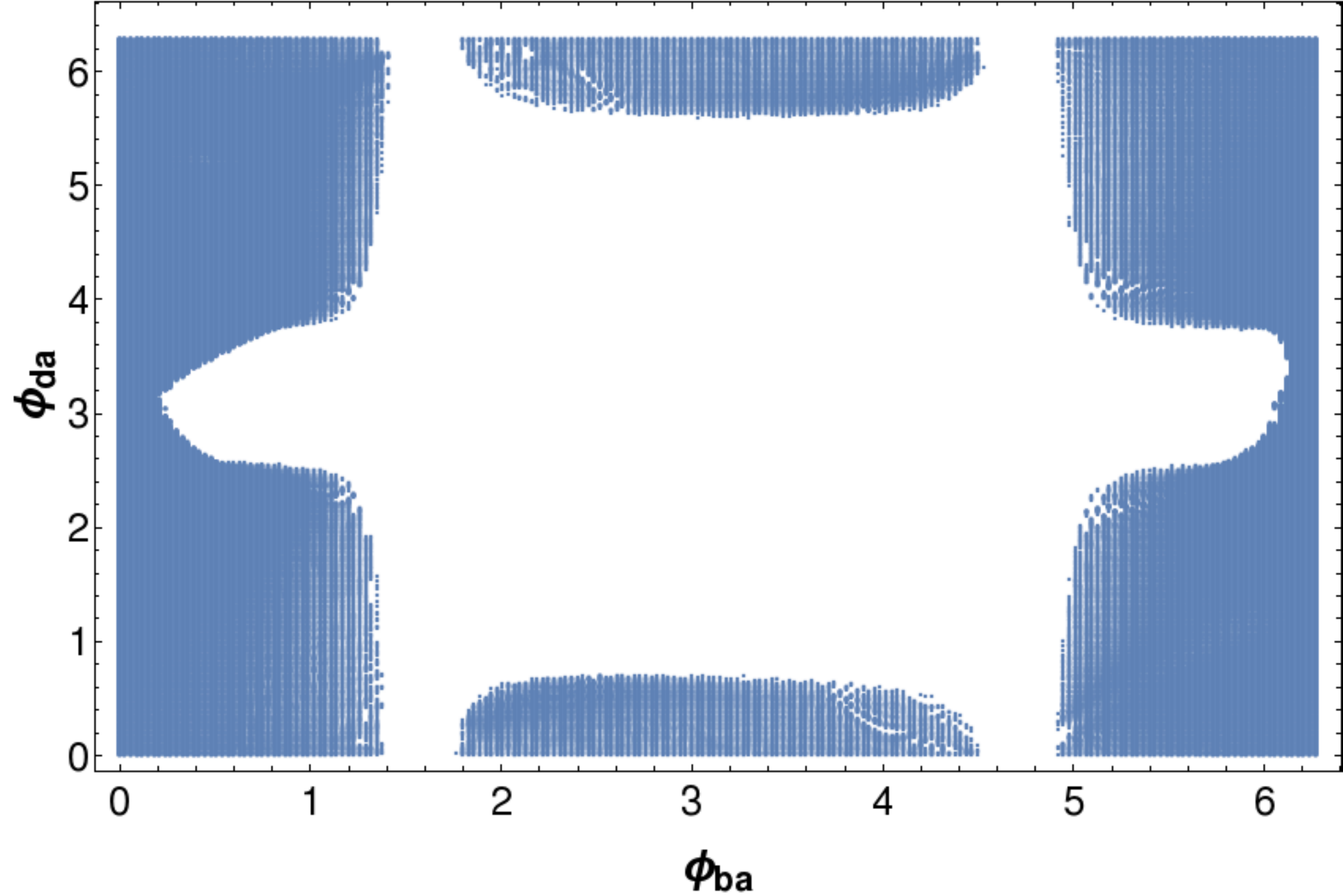}
$$
\caption{Allowed ranges of $\phi_{ba}$ and $\phi_{da}$ in order to produce $\sin\theta_{13}$ in the 
 3$\sigma$ range, correct $r$,  satisfying $\sum m_i<0.23$ eV.}
\label{fig:g-pba-pda} 
\end{figure}
\begin{figure}[h]
$$
\includegraphics[height=4.0cm]{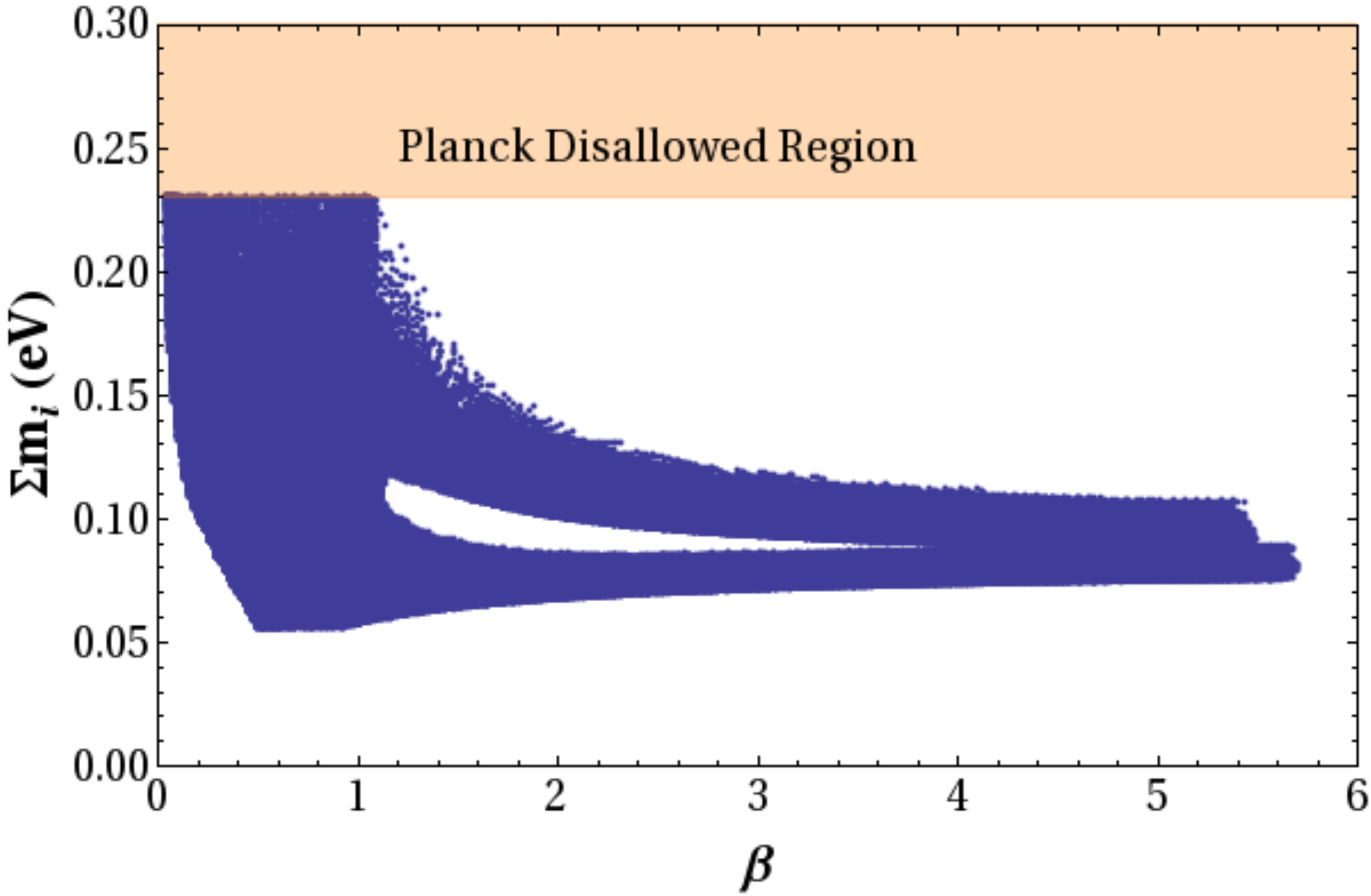}
\includegraphics[height=4.0cm]{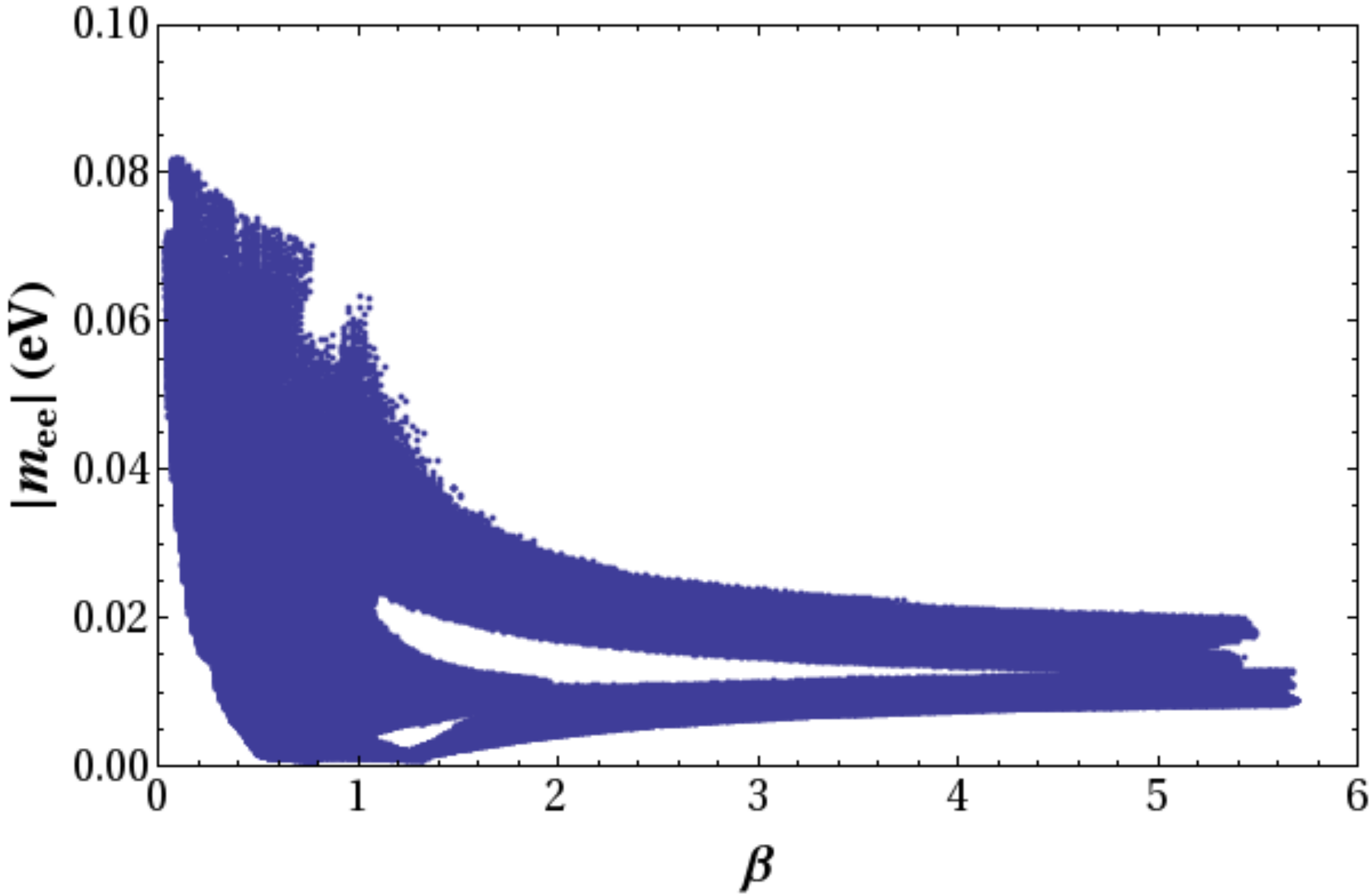}
$$
\caption{[Left panel] $\sum m_i$  vs $\beta$ which satisfy
         3$\sigma$ range of $\sin\theta_{13}$ and $r$.
        Horizontal orange patch represents the excluded region from the upper bound on sum of all the three light neutrino masses
        ($\sum m_i<0.23$ eV).
        [Right panel] $|m_{ee}|$ vs $\beta$ satisfying 
        $\sum m_i<0.23$ eV. In both the panels $\alpha, \phi_{ba}$ and $\phi_{da}$ vary according to Figs. 
        \ref{fig:g-a-b}-\ref{fig:g-pba-pda}.}
\label{fig:g-mi-mee}   
\end{figure}
\begin{figure}[h]
$$
\includegraphics[height=4.5cm]{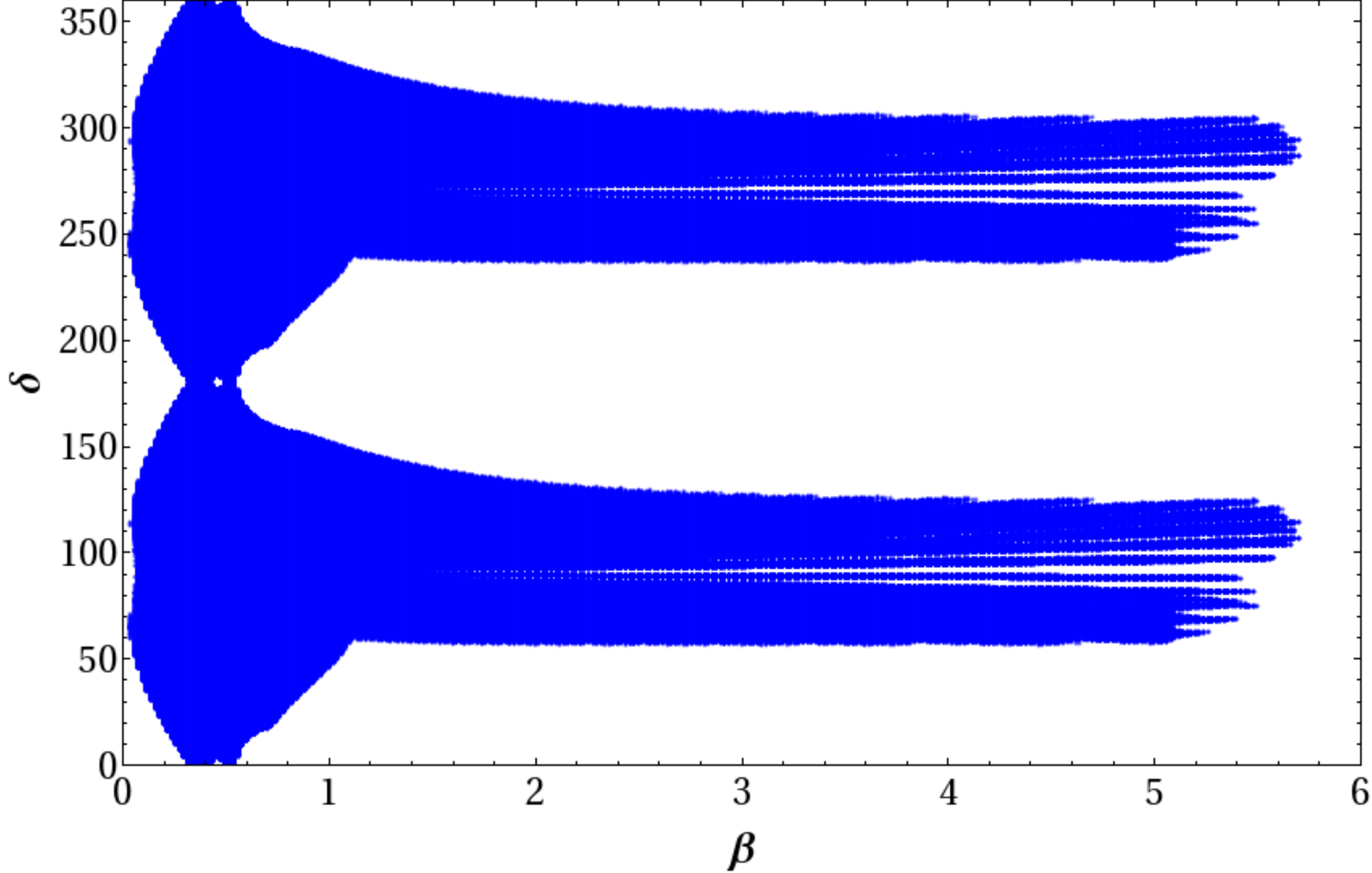}
$$
\caption{Dirac CP phase $\delta$ vs $\beta$  satisfying
         constraints from  3$\sigma$ range of $\sin\theta_{13}$ and $r$ (with $\sum m_i<0.23$ eV).  Here
        both $\phi_{ba}$ and $\phi_{da}$ varies between $0-2\pi$.}
\label{fig:g-delta-a}      
\end{figure}

 Now using the $3\sigma$ allowed ranges for $\theta_{13}$ and $r$ ~\cite{Forero:2014bxa}, we represent 
the allowed regions for $\alpha$ and $\beta$ in Fig. \ref{fig:g-a-b} represented by the blue patch. Here $\phi_{ba}$ 
and $\phi_{da}$ are allowed to vary within their full range ($0$ to $2\pi$). However it turns out that only a portion 
of this entire range can actually satisfy the required $\theta_{13}$ and $r$ through Eqs. (\ref{tpsi}-\ref{ang0})
along with the range of $\alpha-\beta$ depicted in Fig. \ref{fig:g-a-b}. This is shown  
in Fig. \ref{fig:g-pba-pda} in the $\phi_{ba} - \phi_{da}$ plane.
 Knowing the allowed range of $\alpha, ~\beta$ and their correlation with phases $\phi_{ba}, ~\phi_{da}$, 
we plot in Fig. \ref{fig:g-mi-mee} the prediction of the model in terms of sum of the three light neutrino masses 
($\sum m_i$) in the left panel and effective mass parameter for neutrinoless double beta decay ($|m_{ee}|$) in the 
right panel as functions of $\beta$.  In the left panel of Fig. \ref{fig:g-mi-mee}, the horizontal orange patch
represents the excluded region by the upper bound on sum of the absolute neutrino masses $\sum m_i\leq 0.23$ eV,  
whereas in the right panel 
of the same figure we have already included this additional constraint to plot $|m_{ee}|$.
 Finally in Fig. \ref{fig:g-delta-a}, we show the allowed range of the Dirac CP phase $\delta$ 
 against the range of $\beta$ (allowed) where we consider simultaneously the corresponding allowed range of 
 $\alpha, ~\phi_{ba}$ and $\phi_{da}$ following Figs. \ref{fig:g-a-b}-\ref{fig:g-mi-mee}.

\section{Non-unitary effect}\label{sec5}

In Section 3, we have determined the neutrino mixing matrix $U_{\nu}$ and 
identify 
it with the $U_{PMNS}$ (charged lepton mass matrix being diagonal) as it diagonalizes 
the effective light neutrino mass matrix $m_{\nu}$ through the unitary transformation
$ U_{\nu}^{T}m_{\nu}U_{\nu}={\rm diag}(m_1,m_2,m_3)$. However the
$U_{PMNS}$ should receive 
a correction over $U_{\nu}$ as the heavy states carries an admixture with the 
light 
neutrinos~\cite{Antusch:2006vwa}.
To clarify, suppose $V_{\nu}$ is the diagonalizing matrix which  makes
$M_{\nu}$ into the block diagonal form first,  $i.e.$
\begin{eqnarray}\label{w}
 V_{\nu}^T M_{\nu} V_{\nu} = \left(
\begin{array}{cc}\label{w-mat}
 {m_{\nu_{light}}}_{3\times 3}  & 0_{3\times 6}  \\
 0_{6\times 3}  &  {m_{\nu_{heavy}}}_{6\times 6}            
\end{array}
\right). 
\end{eqnarray}
At this point the light neutrino mass matrix $ {m_{\nu_{light}}} \simeq -m_{\nu} = -m_{D}M^{-1}\mu(M^{T})^{-1}m_D^T$ and 
the other one is given by 
\begin{eqnarray}\label{mheavy}
{m_{\nu_{heavy}}} \simeq \left(
\begin{array}{cc}\label{mheavy-mat}
0 & M^T  \\
M  & \mu           
\end{array}
\right),
\end{eqnarray} 
in the lowest order~\cite{GonzalezGarcia:1988rw}. Let $U$ be the matrix of the form 
\begin{eqnarray}\label{u}
U = \left(
\begin{array}{cc}\label{u-mat}
U_{\nu}  & 0  \\
 0  &  U_h       
\end{array}
\right), 
\end{eqnarray}
which will do the individual  diagonalization, $i.e. ~U_{\nu}$ and $U_h$ are 
expected to diagonalize  $m_{\nu_{light}}$ 
and $m_{\nu_{heavy}}$ respectively (remember that $U_{\nu}$ is the 
diagonalizing matrix of $m_{\nu}$ 
as already discussed in Section 3). So finally $W = V_{\nu}U$ diagonalizes the 
entire 
$9 \times 9$ matrix $M_{\nu}$ such that 
$W^T M_{\nu} W = {\rm {diag}} (m_{i=1,2,3}, m_{N_{k=1,2,..,6}})$. One can 
decompose $W$ as follows:
\begin{eqnarray}\label{mat:W}
 W=\left(
\begin{array}{cc}\label{W-mat}
 W_{3\times 3}  & W_{3\times 6}  \\
 W_{6\times 3}  & W_{6\times 6}            
\end{array}
\right), 
\end{eqnarray}
where the block $W_{3\times 3}$ is the leading order replacement of $U_{PMNS}$ matrix 
which is non-unitary \cite{Kanaya:1980cw, Altarelli:2008yr}. It is shown \cite{Kanaya:1980cw,Dev:2009aw} that $W_{3\times 3} \simeq 
(\mathds{I} - \frac{1}{2} F F^{\dagger}) U_{\nu}$, where the non unitary effect 
is parametrized by
\begin{equation}
 \eta = \frac{1}{2}FF^{\dagger},
\end{equation}
with $ F=m_D M^{-1}$ as defined before.
The present bound on $\eta$ (at 90$\%$  C.L.) can be summarized as \cite{Antusch:2008tz}
\begin{eqnarray}
|\eta|<\left(      \label{eta:expt}
\begin{array}{ccc}
 2.0\times 10^{-3}  & 3.5\times 10^{-5} & 8.0\times 10^{-3}  \\
 3.5\times 10^{-5}  & 8.0\times 10^{-4} & 5.1\times 10^{-3}  \\
 8.0\times 10^{-3}  & 5.1\times 10^{-3} & 2.7\times 10^{-3}
\end{array}
\right).
\end{eqnarray}

In our case $F$ is proportional to identity as mentioned before and so as $\eta$.
 In the present framework $\eta$ turns out to satisfy  
$|\eta|= \frac{v^2 |y_1|^2}{2 v_{\rho}^2 |y_2|^2} {\mathds{I}} = C_1 
{\mathds{I}}$ say, 
and hence the above bound on $\eta$ can be translated into 
\begin{equation}
C_1  ~<  ~8.0\times 10^{-4}, 
\label{c1}
\end{equation}
where $C_1=v^2|y_1|^2/2v_{\rho}^2|y_2|^2$. Using this bound, we can now 
estimate the scales involved in our scenario, $i.e.$ $v_\rho,\Lambda$ etc. 
For simplicity we assume all the flavons have the same vevs $v_{f}$. Then $C_1$  
is given by $\lambda v^2/2v_f^2$ where
 $\lambda = {|y_1|^2}/{|y_2|^2}$. Hence the common flavon vev $v_f$ is  bounded by 
\begin{equation}
 v_f ~> ~ 6.15 \sqrt{\lambda} {\rm{~~TeV}},
 \label{vflimit}
\end{equation}
which follows from Eq.(\ref{c1}).


\subsection{Determining the scales ($v_f,\Lambda$) involved in the set-up}
Note that the parameter $k$  defined in Section 3 can be written as 
\begin{equation}\label{kk}
 k=\frac{\lambda v^2}{v_f^2}|a|=2\lambda|\mu_1|\frac{v_fv^2}{\Lambda^2 },
\end{equation}
 once the common flavon vev $v_f$ is assumed and $a=2\mu_1 v^3_{f}/\Lambda^2$ 
is inserted. As we already have an estimate for the range of $k$ for all cases 
 (A, B, C and D), we can use that input on $k$ to study the correlation between 
$v_f$ and $\Lambda$ for various choices of $\lambda$ while $|\mu_1|$ is fixed, 
say at unity.
This correction  however satisfy Eq. (\ref{vflimit}) and we discuss it below 
case by case.

\subsubsection{Case A: [$\phi_{ba}=\phi_{da}=0$]} 
In this case, we have found $k=0.0183$ eV corresponding to the set of parameters 
$\alpha,\beta$ ($\alpha=2.16, \beta=0.372$) which produces the best fit value 
of 
$\sin \theta_{13}=0.1530$ and $r = 0.03$ so as to have the  solar and 
atmospheric mass squared splittings  $7.6\times 10^{-5}$ eV$^2$ and 
$2.48\times 10^{-3}$ eV$^2$ respectively via Eqs. (\ref{m100}-\ref{m300}). 
\begin{figure}[h]
$$
\includegraphics[height=6.0cm]{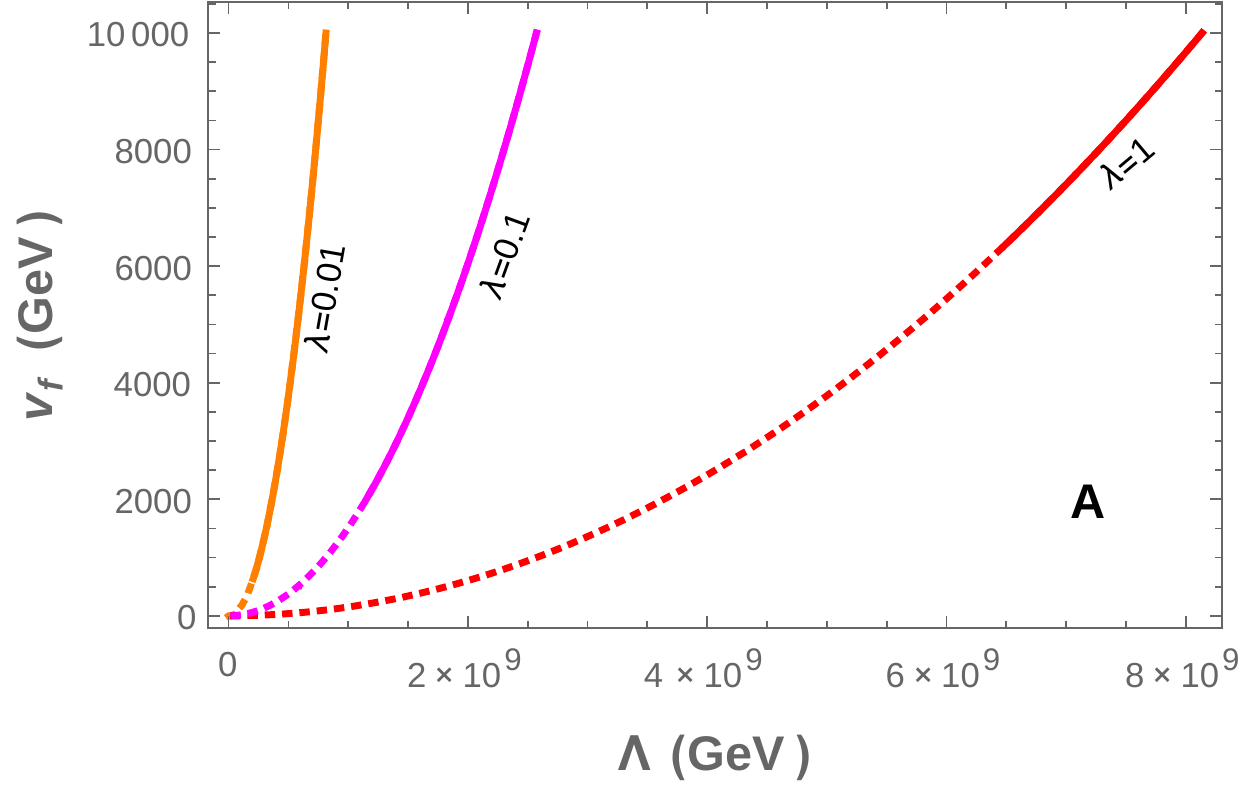}
$$
\caption{Contour plots for $k=0.0183 $ eV in the $v_{f}-\Lambda$ 
plane (using Eq. (\ref{kk})) for $\phi_{ba}=\phi_{da}=0$ (and $|\mu_1|=1$).
The dotted portion in each curve indicates the 
excluded part in view of Eq. (\ref{vflimit}).  Here the orange, 
magenta and red line stands for $\lambda$= 0.01, 0.1 and 1 respectively.}
\label{fig:flambda00}
\end{figure}
\begin{table}[h]\centering
\resizebox{10cm}{!}{%
 \begin{tabular}{c|ccccc}
\hline
 &$\lambda=0.01$ & $\lambda=0.1$ & $\lambda=1$ \\
\hline
 $\Lambda$ in GeV (for $C_1=7.5\times 10^{-4}$) &$2.06\times 10^{8}$  
&$1.16\times 10^{9}$ & $6.48\times 10^{9}$  \\
\hline
 $\Lambda$ in GeV (for $C_1=4\times 10^{-4}$) &$2.40\times 10^{8}$ &$1.35\times 
10^{9}$ & $7.59\times 10^{9}$ \\
\hline\hline
\end{tabular}
}
\caption{\label{tab:xxxx} {\small Cutoff scale  $\Lambda$ for different $C_1$  
(with $\phi_{da}=\phi_{ba}=0$)  when $\lambda=$ 0.01, 0.1 and 1.}}
\end{table}
Now using this particular value of $k$, we employ Eq.(\ref{kk}) to have an 
estimate of 
$v_f$ and $\Lambda$ once the couplings $\lambda$ and $|\mu_1|$ are fixed. In Fig. \ref{fig:flambda00},
we plot the contour lines for $k = 0.0183$ eV in the $v_f - \Lambda$ plane for different choices of $\lambda$. 
Here $|\mu_1|$ is assumed to be unity for simplicity. 
Following Eq. (\ref{vflimit}), the $v_f - \Lambda$ correlation gets further constrained. Depending on the 
specific choices of $\lambda$, the lower bound on $v_f$ is obtained through Eq.
(\ref{vflimit}). The portion of each 
$k$ contour line which does not satisfy Eq.(\ref{vflimit}) is indicated by the  
dotted segment. Note that corresponding to a specific choice of the 
non-unitarity parameter $\eta$, $v_f$ would fixed through 
$C_1=\frac{\lambda v^2}{2 v_f^2}$ (for fixed $\lambda$) which then  indicates a 
particular $\Lambda$. In Table \ref{tab:xxxx}, we 
provide some such specific choices of $\Lambda$ corresponding to different 
choice of  $\eta$. We find that with $\lambda$ small enough, the cut-off scale 
can also be lowered $\sim$ TeV.

\subsubsection{Case B: [$\phi_{ba}=0$]}

\begin{figure}[h]
$$
\includegraphics[height=5.0cm]{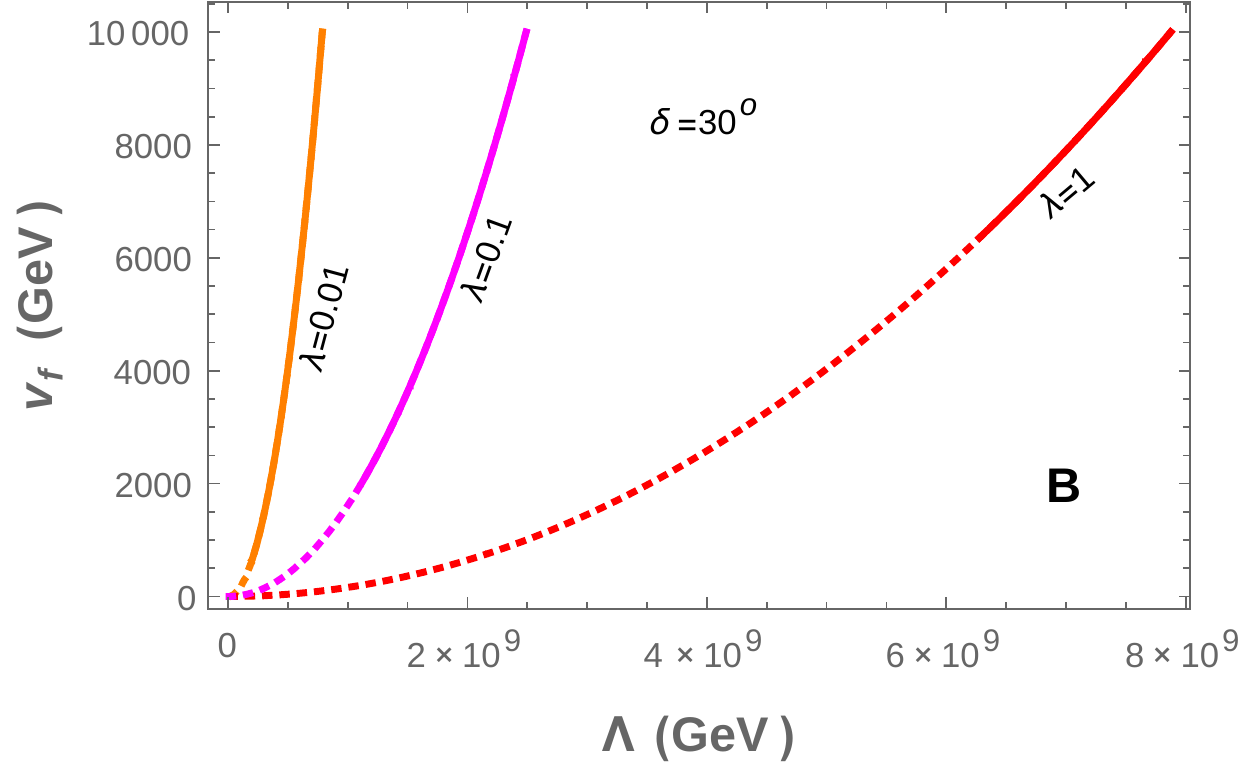}
\includegraphics[height=5.0cm]{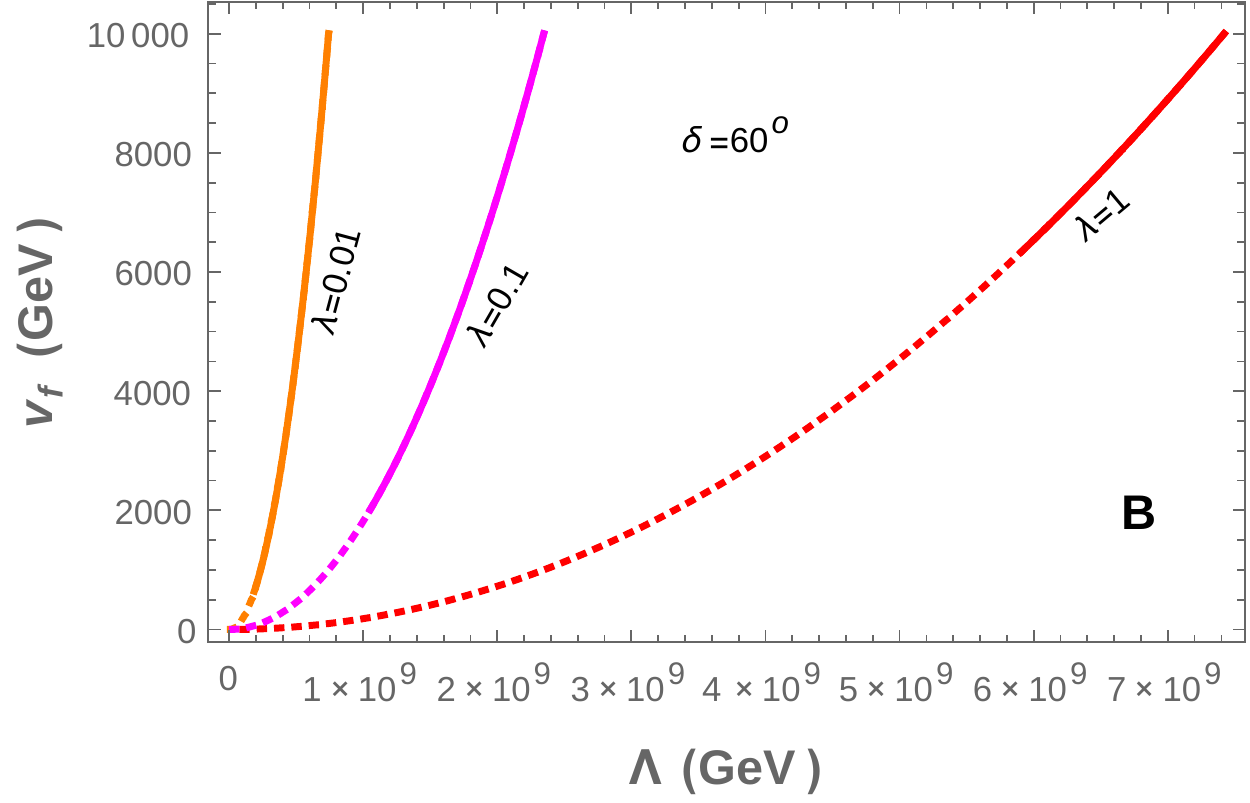}
$$
\caption{[Left panel] Contour plot for $k=0.0195$ eV in the $v_f-\Lambda$ plane
for $\phi_{ba}=0$ and $\delta=30^{\circ}$. [Right panel] Contour plot 
for $k=0.0220$ eV in the $v_f-\Lambda$ plane
for $\phi_{ba}=0$ and $\delta=60^{\circ}$. In both the panels orange, 
magenta and red lines stand for $\lambda$= 0.01, 0.1 and 1 respectively.   
}
\label{fig:flamba0}
\end{figure}

\begin{table}[h]\centering\resizebox{14cm}{!}{
\begin{tabular}{|c|c|c|c|c|c|c|}
\hline
\multicolumn{1}{|l|}{} & \multicolumn{3}{c|}{$C_1= 7.5\times 
10^{-4}$} & \multicolumn{3}{c|}{$C_1= 4\times 10^{-4}$} \\ 
\cline{2-7}
 &  $\lambda=0.01$     &   $\lambda=0.1$    &    $\lambda=1$   & 
$\lambda=0.01$      & $\lambda=0.1$      &   $\lambda=1$    \\ \hline \hline
$\Lambda$ in GeV $(\delta=30^{\circ})$ &   $1.98\times 10^{8}$    &  
$1.12\times 
10^{9}$     &  $6.3\times 10^{9}$     &     $2.32\times 10^{8}$  &  $1.31\times 
10^{9}$     &    $7.34\times 10^{9}$   \\ \hline
$\Lambda$ in GeV $(\delta=60^{\circ})$ &    $1.89\times 10^{8}$   & 
$1.05\times 
10^{9}$      &  $5.9\times 10^{9}$     &    $2.19\times 10^{8}$   &   
$1.23\times 10^{9}$    &     $6.92\times 10^{9}$  \\ \hline
\end{tabular}
}
\caption{\label{tab:c136} {\small Cutoff scale  $\Lambda$ for different $C_1$  
(with $\phi_{ba}=0$)  and $\lambda$ (= 0.01, 0.1 and 1).}}
\end{table}


In Section \ref{sec:analysis} we have seen that for $\phi_{ba}=0$, $r$ and 
$\sin\theta_{13}$ depend not only on $\alpha,\beta$, but also on the choice of 
Dirac CP phase $\delta$. We have already listed our finding toward this 
dependency in 
Table \ref{tab:ba0}. Corresponding to each $\delta$, we have sets of 
$(\alpha,\beta)$ and $k$ from Table \ref{tab:ba0}. Now for a fixed $\delta$ 
and $k$ we can study the correlation of $v_f$ and $\Lambda$ in a similar way as 
described in Case A above. In Fig. {\ref{fig:flamba0}}, we have studied this 
correlation for two different choices for $\delta=30^{\circ}$, $k=0.0195$ eV 
(left panel) and $\delta=60^{\circ}$, $k=0.0220$ eV (right panel). We
consider $|\mu_1|=1$ and  choices for $\lambda=|y_1|^2/|y_2|^2=0.01$ 
(orange line), 0.1 (magenta line) and 1 (red line) in both panels are 
shown. Since $k$ (see Table \ref{tab:ba0}) does not change much with the 
change of $\delta$, correlation between $v_f$ and $\Lambda$ remains 
almost unaltered as seen from the two panels of Fig. \ref{fig:flamba0}. 
The dotted section of each contour line in Fig. \ref{fig:flamba0} represents 
the excluded part in view of Eq. (\ref{vflimit}). With some specific choices 
of $C_1$ (satisfying Eq. (\ref{c1})) we have listed the corresponding scale 
$\Lambda$ in Table \ref{tab:c136}. 

\subsubsection{Case C: [$\phi_{da}=0$]}
\begin{figure}[h] 
$$
\includegraphics[height=5.0cm]{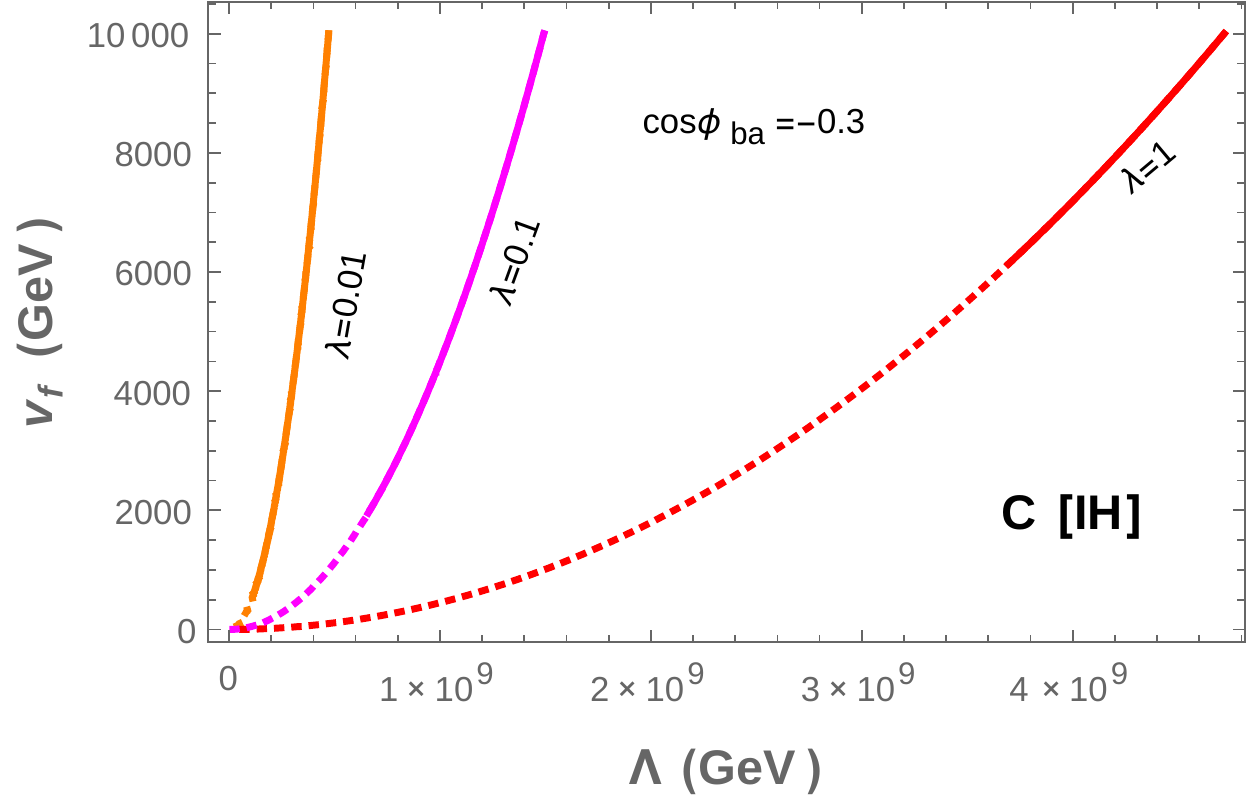}
\includegraphics[height=5.0cm]{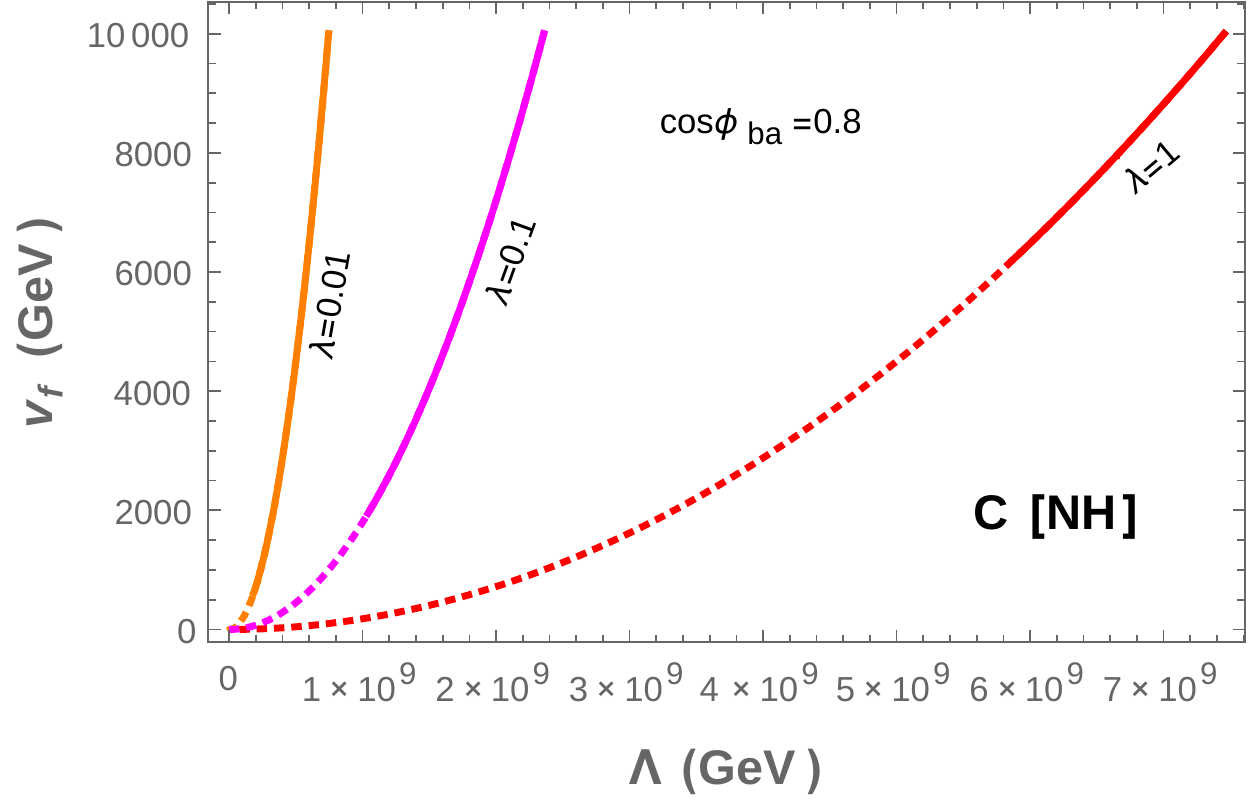}
$$
\caption{[Left panel] Contour plot for $k=0.0544$ eV in the $v_f-\Lambda$ plane
         for $\phi_{da}=0$ and $\cos\phi_{ba}=-0.3$ $(\alpha=0.814)$ 
[IH: Inverted 
hierarchy]. [Right panel] Contour plot for $k=0.0218$ GeV in the $v_f-\Lambda$ 
plane for $\phi_{ba}=0$ and $\cos\phi_{ba}=0.8$ $(\alpha=1.904)$ [NH: Normal 
hierarchy]. In both the panels orange, 
magenta and red lines stand for $\lambda$= 0.01, 0.1 and 1 respectively.}
\label{fig:flamda0}
\end{figure}

In this case, as we conclude from Fig. \ref{fig:rba0}, the range of $\alpha$ is 
restricted as $0.478<\alpha< 0.863$ for $\cos\phi_{ba}<0$ and 
$1.247 <\alpha < 2.162$ for $\cos\phi_{ba}>0$. We have found that 
$\cos\phi_{ba}<0$ represents inverted hierarchy while $\cos\phi_{ba}>0$ stands 
for normal hierarchy. Here $\delta$ turns out to be zero. Therefore for a 
specific value of $\alpha$ (and hence also for $\cos\phi_{ba}$) we obtain the 
corresponding value of $k$ as mentioned in Table \ref{tab:parac}. Using that 
particular $k$, we draw contour plot of $k$ in $v_f$-$\Lambda$ plane in Fig. 
\ref{fig:flamda0} where Eq. (\ref{kk}) is employed. Left panel of Fig. 
\ref{fig:flamda0} is for inverted hierarchy of light neutrinos and right panel 
represents the case of normal hierarchy. Using the non-unitarity constraints 
through Eq. (\ref{vflimit}), similar to Case A and Case B, here also we 
indicate 
the disallowed portion of $v_f$-$\Lambda$ correlation. Considering some 
specific choice of $C_1$, we provide sample values of $\Lambda$ in Table 
\ref{tab:xy}.

\begin{table}[h]
\centering\resizebox{14cm}{!}{
\begin{tabular}{|c|c|c|c|c|c|c|}
\hline
\multicolumn{1}{|l|}{} & \multicolumn{3}{c|}{$C_1= 7.5\times 
10^{-4}$} & \multicolumn{3}{c|}{$C_1= 4\times 10^{-4}$} \\ 
\cline{2-7}
 &  $\lambda=0.01$     &   $\lambda=0.1$    &    $\lambda=1$   & 
$\lambda=0.01$      & $\lambda=0.1$      &   $\lambda=1$    \\ \hline \hline
$\Lambda$ in GeV $(\cos\phi_{ba}=-0.3)$  &  $1.19\times 10^{8}$  &$6.69\times 
10^{8}$ & $3.76\times 
10^{9}$    &   $1.40\times 10^{8}$    &  $7.82\times 10^{8}$   &  $4.40\times 
10^{9}$   \\ \hline
$\Lambda$ in GeV $(\cos\phi_{ba}=0.8)$ &  $1.88\times 10^{8}$  
&$1.06\times 
10^{9}$ & $5.94\times 
10^{9}$      &   $2.20\times 10^{8}$  & $1.24\times 10^{9}$  &    $6.95\times 
10^{8}$  \\ \hline
\end{tabular}
}
\caption{\label{tab:xy}  {\small Cutoff scale  $\Lambda$ for different $C_1$  and $\cos\phi_{ba}$
(with $\phi_{da}=0$) when $\lambda$= 0.01, 0.1 and 1.}}
\end{table}


\subsubsection{Case D: [$\phi_{ba}=\phi_{da}=\phi$]}
\begin{figure}[h]
$$
\includegraphics[height=5.0cm]{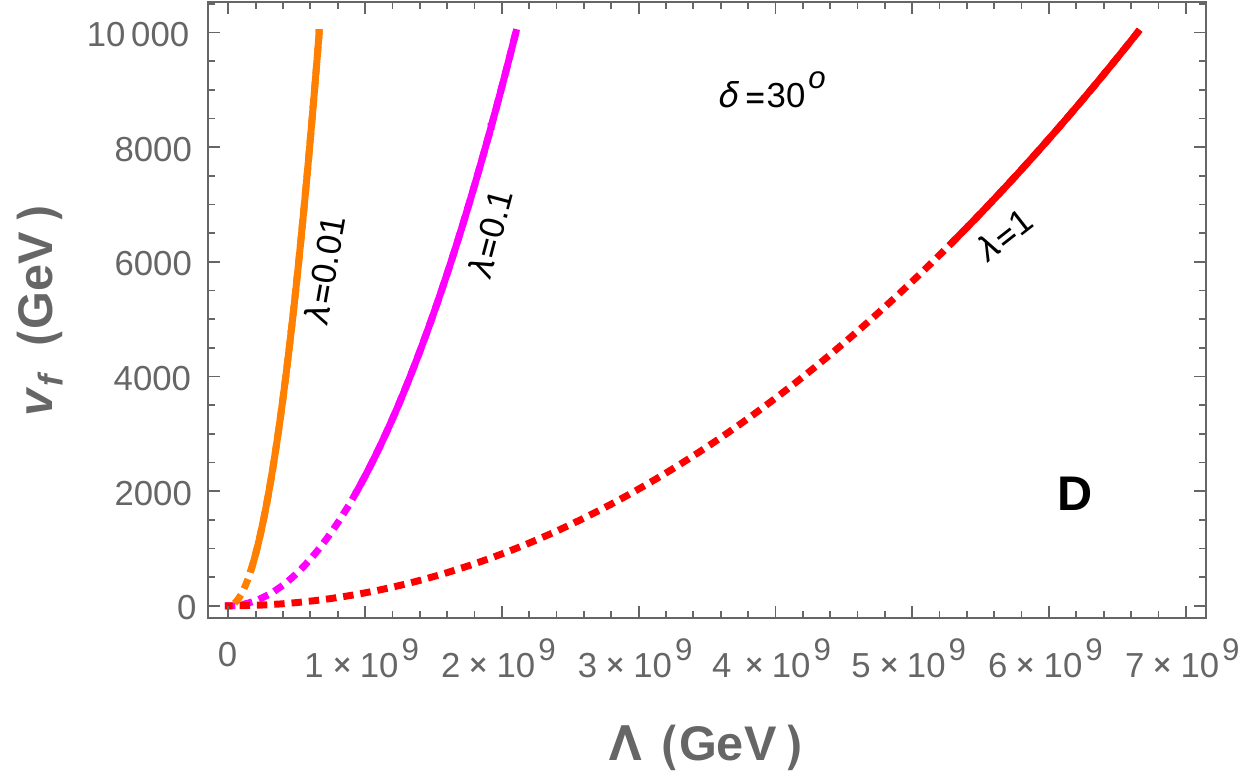}
\includegraphics[height=5.0cm]{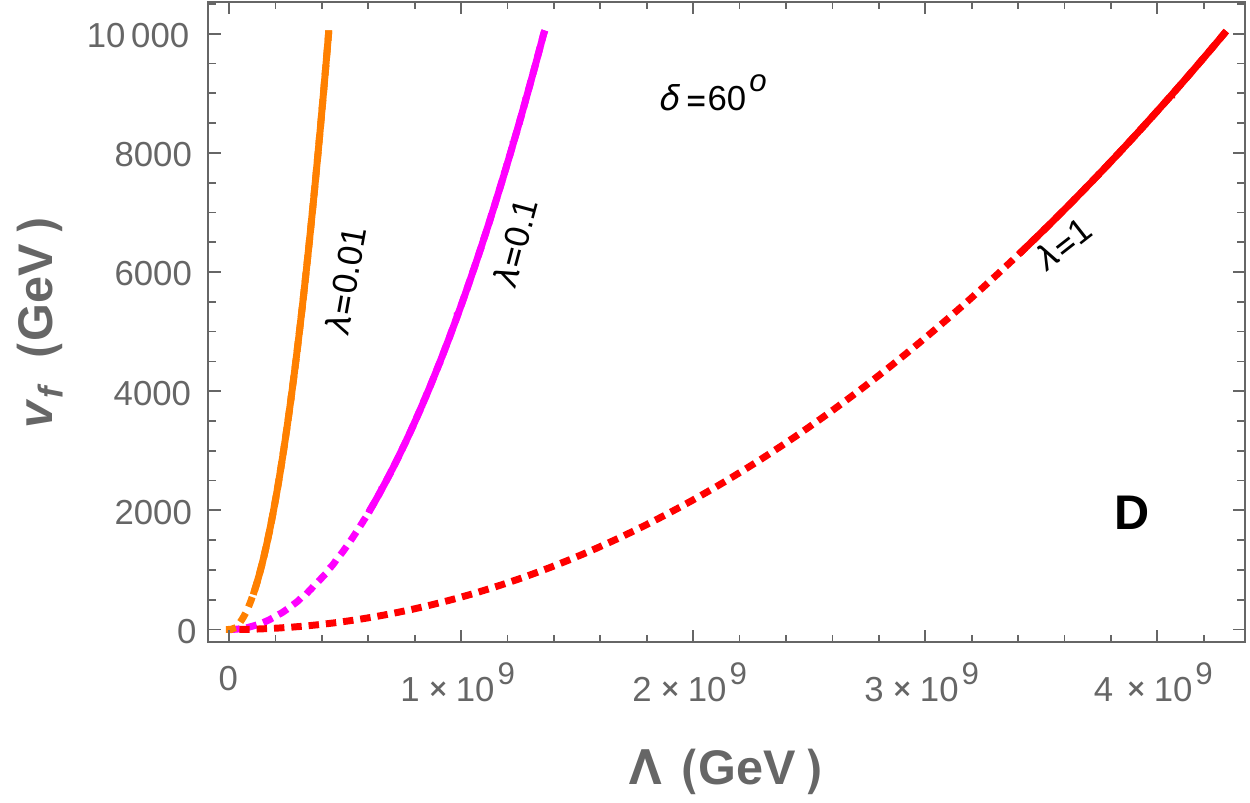}
$$
\caption{ [Left panel] Contour plot for $k=0.0274$ eV in the $v_f-\Lambda$ plane
for $\phi_{ba}=\phi_{da}=\phi$ and $\delta=30^{\circ}$. [Right panel] 
Contour plot for $k=0.0658$ eV in the $v_f-\Lambda$ plane
for $\phi_{ba}=\phi_{da}=\phi$ and $\delta=60^{\circ}$.
In both the panels orange, 
magenta and red lines stand for $\lambda$= 0.01, 0.1 and 1 respectively.}
\label{fla}
\end{figure}

With the consideration $\phi_{ba}=\phi_{da}=\phi$, we have already discussed in the previous 
section that $\alpha,\beta$ $\sin\theta_{13}$ and $r$ are correlated with the 
choice of $\delta$. We have listed $\alpha, \beta$ as well as $k$ for 
different allowed values of $\delta$ in 
Table \ref{tab:a}. As discussed before, here also we can plot the dependency of 
$v_{f}-\Lambda$
using Eq. (\ref{kk}) and estimate the allowed regions for $v_{f}$ and $\Lambda$ employing 
Eq. (\ref{vflimit}). In Fig. \ref{fla} we have plotted this dependency for various choice 
of $\lambda$ with $\delta=30^{\circ}$ (left panel) and $\delta=60^{\circ}$ (right panel). In 
both of these panels orange, magenta and red lines stand for $\lambda=0.01, 
0.1$  
and 1 respectively. 
Following this we have listed few representative values of $\Lambda$ in Table 
\ref{tab:c137}.
\begin{table}[h]\centering\resizebox{14cm}{!}{
\begin{tabular}{|c|c|c|c|c|c|c|}
\hline
\multicolumn{1}{|l|}{} & \multicolumn{3}{c|}{$C_1= 7.5\times 
10^{-4}$} & \multicolumn{3}{c|}{$C_1= 4\times 10^{-4}$} \\ 
\cline{2-7}
 &  $\lambda=0.01$     &   $\lambda=0.1$    &    $\lambda=1$   & 
$\lambda=0.01$      & $\lambda=0.1$      &   $\lambda=1$    \\ \hline \hline
$\Lambda$ in GeV $(\delta=30^{\circ})$  & $1.68\times 10^{8}$ 
&$9.48\times 10^{8}$
& $5.30\times 10^{9}$     &     $1.96\times 10^{8}$  &  $1.10\times 
10^{9}$     &    $6.20\times 10^{9}$   \\ \hline
$\Lambda$ in GeV $(\delta=60^{\circ})$ &   $1.08\times 10^{8}$ 
&$6.08\times 10^{8}$ 
& $3.42\times 10^{9}$      &    $1.26\times 10^{8}$   &   
$7.11\times 10^{8}$    &     $4.00\times 10^{9}$  \\ \hline
\end{tabular}
}
\caption{\label{tab:c137} {\small Cutoff scale  $\Lambda$ for different $C_1$  and $\delta$
(with $\phi_{ba}=\phi_{da}=\phi$) when $\lambda$= 0.01, 0.1 and 1.}}
\end{table}

\subsubsection{General Case} 

\begin{figure}[h]
$$
\includegraphics[height=5.0cm]{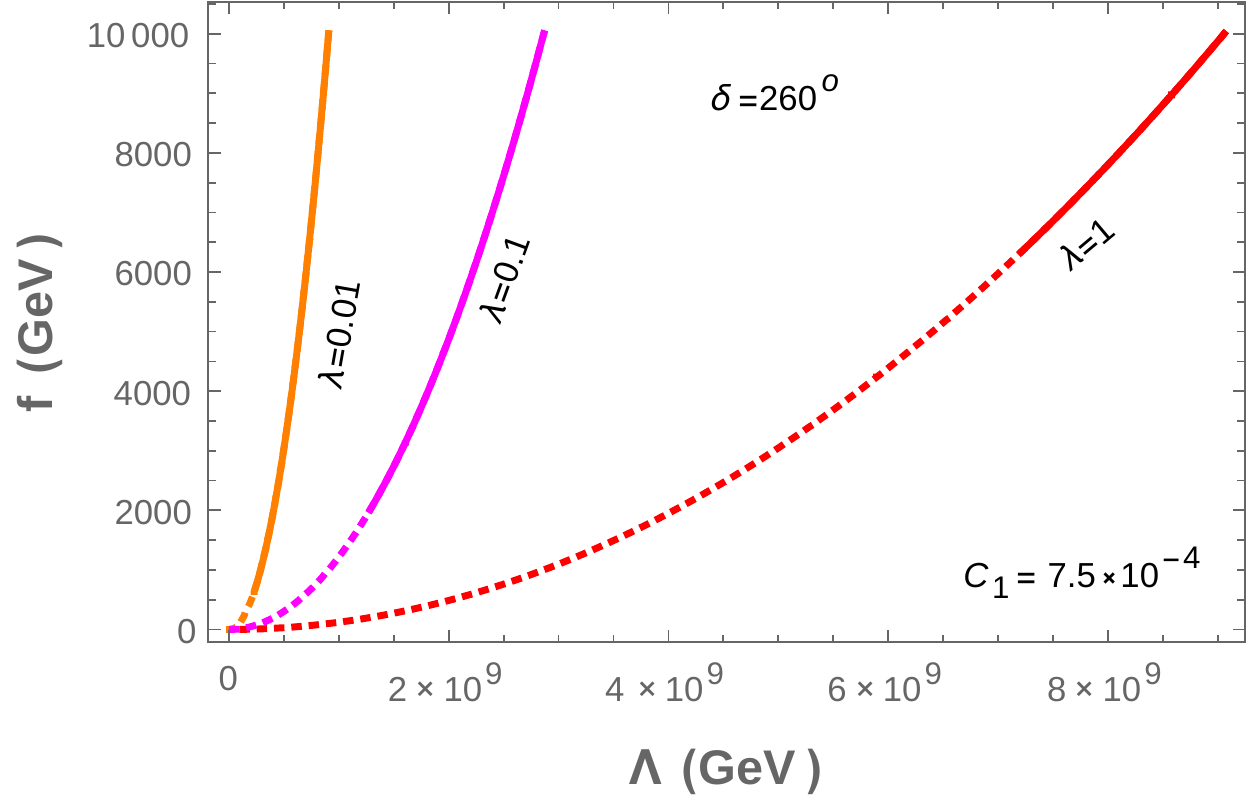}
\includegraphics[height=5.0cm]{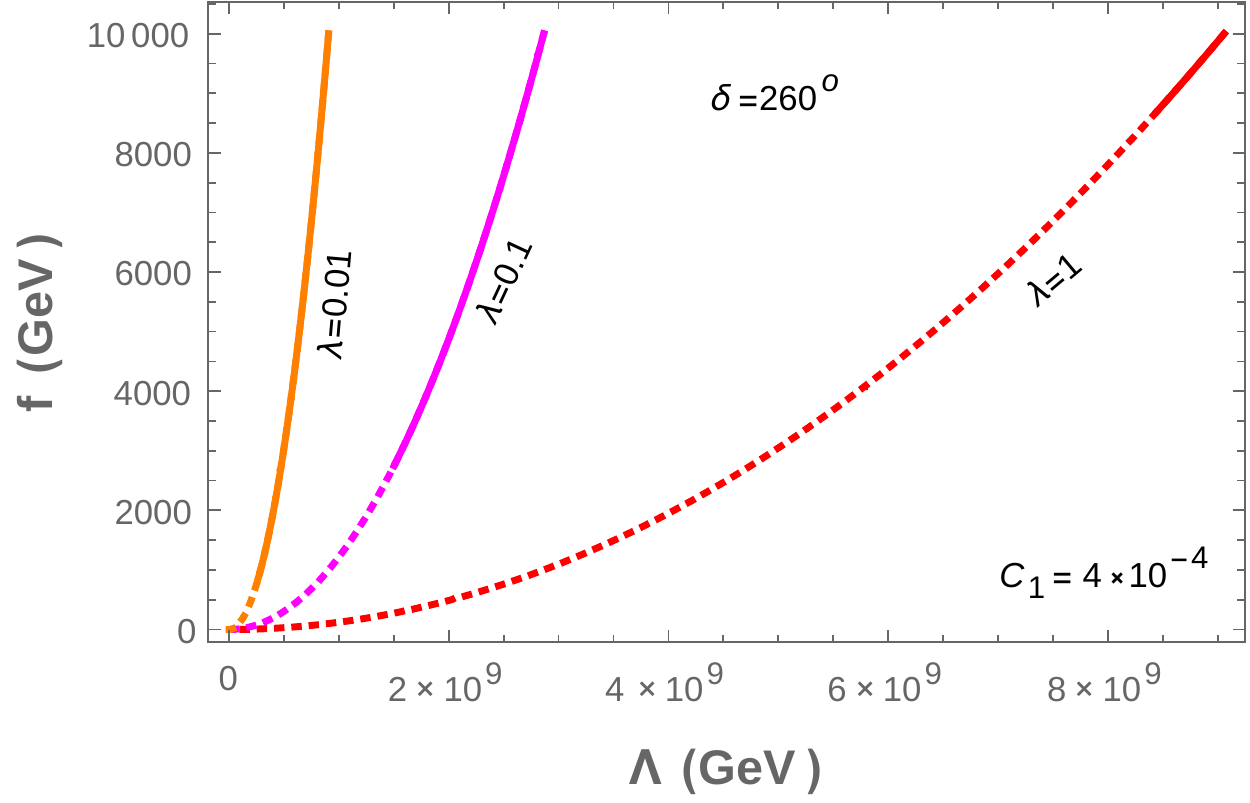}
$$
\caption{Contour plot for $k=0.0147 $ eV in the $v_{f}-\Lambda$ plane
where $\alpha=2.25, \beta=1, \phi_{ba}=0.5$ and $\phi_{da}=2$ and with $C_1=7.5\times 10^{-4}$ 
(left panel) and $C_1=4\times 10^{-4}$ (right panel). In both the panels orange, magenta and
red lines stand for $\lambda$= 0.01, 0.1 and 1 respectively.}
\label{flgen}
\end{figure}

 From our previous analysis in section \ref{sec:analysis}, we choose a particular value of $\delta=260^{\circ}$ for 
 this general case (a value close 
 to recent hint \cite{Abe:2013hdq, Forero:2014bxa, Capozzi:2013csa, Gonzalez-Garcia:2014bfa
}) to study the scales $v_f, \Lambda$. The set of parameters that would correspond to 
 this value of $\delta$ are found to be $\alpha=2.25, \beta=1, \phi_{ba}=0.5$ and $\phi_{da}=2$ which satisfy 
 constrains imposed from mixing angles, $r$ and $\sum m_i < 0.23$ eV. Here $k$ is found to be 0.0147 eV in order to 
 have adequate solar and atmospheric splittings. Using this 
 $k$ through Eq. (\ref{kk}),  we then obtain the contour plot of $v_f$ against $\lambda$ as shown in Fig. \ref{flgen} for 
 different choices of $\lambda$.  The dotted portion in each curve indicates the excluded part in view of Eq. (\ref{vflimit}) 
 with $C_1=7.5\times 10^{-4}$  (left panel) and $C_1=4\times 10^{-4}$ (right panel). These numerical estimates 
 are summarized in Table \ref{cons:gen}.
\begin{table}[h]\centering
\resizebox{10cm}{!}{%
 \begin{tabular}{c|ccccc}
\hline
 &$\lambda=0.01$ & $\lambda=0.1$ & $\lambda=1$ \\
\hline
 $\Lambda$ in GeV (for $C_1=7.5\times 10^{-4}$) &$2.29\times 10^{8}$  
&$1.28\times 10^{9}$ & $7.22\times 10^{9}$  \\
\hline
 $\Lambda$ in GeV (for $C_1=4\times 10^{-4}$) &$2.68\times 10^{8}$ &$1.50\times 
10^{9}$ & $8.45\times 10^{9}$ \\
\hline\hline
\end{tabular}
}
\caption{\label{cons:gen} {\small Cutoff scale  $\Lambda$ for different $C_1$  
with $\delta=260^{\circ}$ (with $\alpha=2.25, \beta=1, \phi_{ba}=0.5$ and
 $\phi_{da}=2$) and $\lambda$= 0.01, 0.1 and 1.}}
\end{table}

\subsection{Lepton flavor violation}

In view of the presence of this non-unitarity effect, 
the neutrino states ($\nu_{\alpha L}$ with $\alpha = e, \mu, \tau$)   
appearing in the SM charged current interaction Lagrangian now can be written as, 
\begin{equation}
 \nu_{\alpha L}=[\left( 1-C_1 \right)U_{\nu}]_{\alpha i} ~{\nu}_{i}+ [\mathcal{K}]_{\alpha j} ~N_{j},
\end{equation}
where the matrix $W_{3 \times 6}$ (see Eq. (\ref{mat:W})) is conventionally 
denoted by $\mathcal{K}$. $\nu_{i=1,2,3}$ and $N_{j =4,5,...9}$ are the light 
and heavy neutrino mass eigenstates respectively. 
Then in a basis where charged leptons are diagonal (as in our case), the 
charged current interactions have contributions involving three light neutrinos $\nu_i$ and six heavy neutrinos $N_j$ as
\begin{eqnarray}
 -\mathcal{L}_{CC}&
       =&  \frac{g}{\sqrt{2}}\bar{l}_{\alpha}\gamma^{\mu}\left\{ [\left( 1-C_1\right)U_{\nu}]_{\alpha i} ~{\nu}_{i}+ [\mathcal{K}]_{\alpha j} ~N_{j}\right\}
                     W^{-}_{\mu}+{\rm h.c.}. 
\end{eqnarray}
These nine neutrino states can therefore mediate lepton flavor violating decays like 
$l_{\alpha}\rightarrow l_{\beta} \gamma$ in one loop ($e.g$ $\mu \longrightarrow e \gamma $ ). 
Resulting branching ratio for such processes ( in the limit $m_{\beta} \rightarrow 0$) now can be written as
~\cite{Ilakovac:1994kj,Alonso:2012ji,  LalAwasthi:2011aa,Parida:2014dba,Parida:2016asc,Bilenky:1977du,
He:2002pva, Forero:2011pc,Rose:2015fua},
\begin{equation}\label{br}
 {\rm BR}(L_{\alpha}\rightarrow L_{\beta} \gamma) \simeq  \frac{\alpha^3_W \sin^2{{\theta}_W} m^5_{l_{\alpha}}}{256\pi^2m^4_{W}\Gamma_{l_{\alpha}}}
 \left|
 \sum\limits_{j=1}^3[\left( 1-C_1 \right) U_{\nu}]_{\alpha j}^*[\left( 1-C_1 \right)U_{\nu}]_{\beta k\j}I_{\gamma}\left(\frac{m^2_{\nu_j}}{m^2_W}\right)
+ 
\sum\limits_{l=4}^9 \mathcal{K}^*_{\alpha l}\mathcal{K}_{\beta l} I_{\gamma}\left(\frac{m^2_{N_l}}{m^2_W}
\right)
 \right|^2,
\end{equation}
where 
\begin{equation}
 I_{\gamma}\left(x \right)=\frac{10-43x+78x^2-49x^3+18x^3{\rm ln}x+4x^4}{12(1-x)^4}, 
 ~{\rm with}~x=\frac{m^2_{\nu, N}}{m^2_W}.
\end{equation}
Here $\alpha_W=g^2/4\pi$, with $g$ as the weak coupling, $\theta_W$ is electroweak mixing angle, $m_W$ is $W^{\pm}$ boson mass and 
$\Gamma_{l_{\alpha}}$ is the total decay width of the decaying charged lepton $l_{\alpha}$. Current upper bound for the branching ratio of the 
LFV decays are~\cite{Agashe:2014kda} (at 90\% CL)
\begin{eqnarray}
 {\rm BR}(\mu\longrightarrow e\gamma) & < &5.7\times 10^{-13},\\
 {\rm BR}(\tau\longrightarrow e\gamma) & < &3.3\times 10^{-8},\\
 {\rm BR}(\tau\longrightarrow \mu\gamma)& < &4.4\times 10^{-8}.
\end{eqnarray}
Another important lepton flavor violating decay $\mu\rightarrow eee$ is also  
worthy to mention and details 
of computation of branching ratio calculation can be found in~\cite{Ilakovac:1994kj,Alonso:2012ji}.  Current upper 
limit for this decay is ${\rm BR}(\mu\longrightarrow eee)  < 1.0\times 10^{-12}$ ($90\% $ CL)~\cite{Agashe:2014kda}.

Since the flavor structure of the neutrino mass matrix is already fixed in our  
present scenario (from the $A_4$ and additional symmetry consideration), it would provide some 
concrete understanding for the LFV processes  in this inverse 
seesaw model. Both the $W_{3\times3}=\left( 1-C_1 \right)U_{\nu}$ and $W_{3\times 6}=\mathcal{K}$ 
matrices play the instrumental role here. Remember that, $U_\nu$ is the diagonalizing matrix for the 
light neutrinos, defined by $U_\nu = U_{TB} U_1 U_m$ as discussed in section \ref{sec3}. Hence this can 
be obtained in terms of $\alpha, \beta, k, \phi_{ba}$ and $\phi_{da}$. The non-unitary parmeter 
$C_1$ is required to satisfy, $C_1=\lambda v^2 /2 v_f^2<8\times 10^{-4}$ as discussed earlier. 
Therefore we can completely evaluate $W_{3\times3}=\left( 1-C_1 \right)U_{\nu}$, once 
a specific value of $C_1$ is chosen.

On the other hand  the rectangular matrix $\mathcal{K}$ is approximately given 
by~\cite{Dev:2009aw,Dias:2012xp} 
\begin{equation}\label{eq:def-k-mat}
 \mathcal{K}\simeq (-F\mu M^{-1},F)U_h,
\end{equation}

\noindent
where $U_h$ is the diagonalizing matrix of $m_{\nu_{heavy}}$ given in Eq. 
(\ref{mheavy}). As previously mentioned, $F=m_{D}M^{-1}$ in our scenario is 
proportional to identity matrix of order $3\times 3$, $i.e.$ 
$F=\frac{y_1 v}{y_2 v_f}\mathds{I}_{3\times 3}=f_1 \mathds{I}_{3\times 3}$.
Hence, the matrix $\mathcal{K}$ turns out to be 
\begin{eqnarray}
 \mathcal{K}&=& (-f_1 \mu M^{-1},f_1\mathds{I}_{3\times 3})U_h.
\end{eqnarray}
Using Eqs. (\ref{mD-M}) and (\ref{mumatrix}), we find 
\begin{eqnarray} \label{muminv}
 \mu M^{-1}= \frac{a}{y_2v_f}\left(
\begin{array}{ccc}
 1-\frac{2}{3}\alpha e^{i\phi_{ba}}                     & \frac{1}{3}\alpha e^{i\phi_{ba}}+\beta e^{i\phi_{da}}  & \frac{1}{3}\alpha e^{i\phi_{ba}} \\
\frac{1}{3}\alpha e^{i\phi_{ba}}                        & 1+\frac{1}{3}\alpha e^{i\phi_{ba}}                     &-\frac{2}{3}\alpha e^{i\phi_{ba}}+\beta e^{i\phi_{da}}\\
\frac{1}{3}\alpha e^{i\phi_{ba}}+\beta e^{i\phi_{da}}   &-\frac{2}{3}\alpha e^{i\phi_{ba}}                       &  1+\frac{1}{3}\alpha e^{i\phi_{ba}}.
\end{array}
\right),
\end{eqnarray}
where we have used the explicit flavor structure of $\mu$ and $M$. 

 We now proceed to find out the 
form of $U_h$, the diagonalizing matrix of $m_{\nu_{heavy}}$. Note that the $m_{\nu_{heavy}}$ matrix can first be block diagonalized by $V_0$ as 
\begin{eqnarray}
{m'_{\nu_{heavy}}}=(V_0)^T  {m_{\nu_{heavy}}} V_0 \simeq\left(
\begin{array}{cc}
-M+\mu/2 & 0  \\
0  & M+\mu/2           
\end{array}
\right),
\end{eqnarray}
where $V_0$ is given by (in our scenario both $\mu$ and $M$ are symmetric 
matrices)
\begin{eqnarray}
 V_0 \simeq \frac{1}{\sqrt{2}}\left(
\begin{array}{cc}
\mathds{I}+\frac{\mu M^{-1}}{4} & \mathds{I}-\frac{\mu M^{-1}}{4}  \\
-\mathds{I}+\frac{\mu M^{-1}}{4} & \mathds{I}+\frac{\mu M^{-1}}{4}          
\end{array}
\right).
\end{eqnarray}
Here we have neglected the terms involving higher orders in $\mu M^{-1}$ as expected in 
inverse seesaw scenario in general. 
Now the upper $(-M+\mu/2)$ and lower $(M+\mu/2)$  block matrices of 
${m'_{\nu_{heavy}}}$ carry the form of $\mu$ matrix itself (or $m_{\nu}$). 
The presence of $M$ just redefines the previous parameter
$a$ by $a_{1,2}=a/2 \mp y_2v_f$ (see Eq. (\ref{mD-M}) and (\ref{mumatrix})). 
Therefore we can follow the similar prescription for diagonalizing these blocks 
as we did in case of $m_{\nu}$ diagonalization. Hence ${m'_{\nu_{heavy}}}$ can 
further be diagonalized by $V^T {m'_{\nu_{heavy}}} V$   with
\begin{eqnarray}\label{def:vmat}
V=\left(
\begin{array}{cc}
U_{TB}.V_1 (\theta_1, \psi_1)& 0  \\
0  & U_{TB}.V_2 (\theta_2, \psi_2)          
\end{array}
\right),
\end{eqnarray}
where $V_i$ has the form similar to $U_1$, $i.e.$
\begin{eqnarray}
V_{i}=\left(
\begin{array}{ccc}
 \cos\theta_{i}                   & 0 & \sin\theta_{i}{e^{-i\psi_{i}}} \\
     0                            & 1 &            0 \\
 -\sin\theta_{i}{e^{i\psi_{i}}} & 0 &        \cos\theta_{i}
\end{array}
\right).
\end{eqnarray}
Therefore the diagonalizing matrix of $m_{\nu_{heavy}}$ can be written as 
\begin{eqnarray}
 U_h\simeq \frac{1}{\sqrt{2}}\left(
\begin{array}{cc}
{\mathds{I}}+\frac{\mu M^{-1}}{4} & {\mathds{I}}-\frac{\mu M^{-1}}{4}  \\
-{\mathds{I}}+\frac{\mu M^{-1}}{4} & {\mathds{I}}+\frac{\mu M^{-1}}{4}          
\end{array}
\right)\left(
\begin{array}{cc}
U_{TB}.V_1 (\theta_1, \psi_1)& 0  \\
0  & U_{TB}.V_2 (\theta_2, \psi_1)          
\end{array}
\right). 
\end{eqnarray}
In order to find $U_h$, we use $\mu M^{-1}$ as obtained in Eq. (\ref{muminv}). Furthermore, 
we get $\theta_{1,2}$ and $\psi_{1,2}$ appearing in $V_{1,2}$ as discussed earlier. Hence 
following the same way as in Eq. (\ref{tpsi}) and Eq. (\ref{tpsi2}) we find 
\begin{eqnarray}
 \tan2\theta_i&=&\frac{\sqrt{3}\beta_i\cos\phi_{da}}{(\beta_i\cos\phi_{da}-2)\cos\psi_i
               +2\alpha_i\sin\phi_{ba}\sin\psi_i},\\
  \tan\psi_{i}&=&\frac{\sin\phi_{da}}{\alpha_i\cos({\phi_{ba}-\phi_{da}})},
\end{eqnarray}
with $i=1,2$ and we use the definition of $\alpha_i$ and $\beta_i$ as,  
\begin{eqnarray}\label{def:a1a2b1b2}
\alpha_{1,2}=\frac{|b|}{|a|\mp 2|y_2|v_f} ~ {\rm and} ~ \beta_{1,2}=\frac{|d|}{|a|\mp 2|y_2|v_f}.
\end{eqnarray}
For simplicity we discard phase difference between $y_2$ and $a$, and set $\phi_{y_2a}=0$. 

Note that from our understanding in Sections \ref{sec3}-\ref{sec:analysis}, we can have estimates over the parameters 
$\alpha, ~\beta, k$ along with the phases $\phi_{ba}, ~\phi_{da}$ in order to satisfy 
$\sin\theta_{13}$, other mixing angles, $r$, individual solar and atmospheric splittings, also to 
be consistent with the upper bound on sum of the light neutrino masses. Specific choice of 
$C_1$ enables us to compute magnitude of the flavon vev $v_f$ and hence $|a|$ from 
Eq. (\ref{kk}). With all these values in hand we can finally evaluate parameters $\theta_i, 
\psi_i, \alpha_i$ and $\beta_i$ appearing in $U_h$. Here we consider 
\footnote{A common phase $\phi_0$ as described in the discussion above Eq. (\ref{majo:eq}) in Section \ref{sec3}, is irrelevant 
for neutrino phenomenology and hence we put it at zero. We also set phases of $f_1$ and 
$a/y_2$ to zero.} $|y_2| = 1$. Now following the analytic expressions in Eqs. (\ref{eq:def-k-mat}-\ref{muminv}), 
(\ref{def:vmat}-\ref{def:a1a2b1b2}), we can estimate  $W_{3\times 3}$ and $\mathcal{K}$  and hence 
the corresponding contribution to the branching ratio (see Eq. (\ref{br})). 
Due to particular flavor structures of the matrices $\mu$ as well as $m_D$ and $M$, we find 
$W_{3\times 3}$ and $\mathcal{K}$ are such that this scenario predicts vanishingly small 
branching ratio  ($\sim 10^{-35}$) for LFV decays. 

In addition, we have performed the evaluation 
numerically also. In order to evaluate 
it, we need to diagonalize the entire $9 \times 9$ neutrino mass matrix $M_{\nu}$. Since the 
neutrino mixings are entirely dictated by the flavor structure of $\mu$ matrix, we could have 
find the entire $M_{\nu}$ numerically with the choices of $\alpha, ~\beta, k$ along with the 
phases $\phi_{ba}, ~\phi_{da}$ as done in cases A, B, C, D and the general case. 
However to compute $m_D$ and $M$, we need consider to $|y_1|$ and $|y_2|$ 
separately (for example, to have $\lambda = 1$, we assume $|y_1| = |y_2| = 1$). Then 
following Eq. (\ref{CMmu}), we can entirely construct the  $M_{\nu}$ matrix
numerically. Then with the help of Mathematica\footnote{We also use Takagi factorization 
\cite{Hahn:2006hr} to find $W$.}, we are able to find the diagonalizing matrix $W$ (and 
hence $\mathcal{K}$ matrix also) and have estimate over the LFV decays. It turns out that 
the numerical estimate coinsides with our analytical evaluation of vanishingly small branching 
ratios for LFV decays to a good extent.

\subsection{Neutrinoless double beta decay and contribution of heavy neutrinos}

We note that in addition to the standard contribution to the effective  mass 
parameter involved in neutrinoless double beta decay as described in 
Section \ref{sec3}, there will be additional contribution due the presence of 
mixing between light and heavy neutrinos ($i.e.$ with nonzero $W_{3\times 6}$). 
Hence the half life associated with neutrinoless double beta can be expressed 
as~\cite{Mitra:2011qr,Chakrabortty:2012mh,Tello:2010am,Khan:2012zw}
\begin{equation}\label{half-life}
(T^{0\nu}_{1/2})^{-1}= \mathcal{G}^{0\nu}\left|\frac{\mathcal{M}_{\nu}}{m_e}
\right|^2  \left|\sum^3_{i=1} (W_{3\times3})^2_{e i}m_{i} + <q^2> 
\sum^6_{i=1}(W_{3\times6})^2_{e i} m_{N}^{-1}\right|^2
\end{equation}
where $\mathcal{G}^{0\nu}$ is the phase space factor and  $<q^2>=-m_{e} 
m_{p}\mathcal{M}_{\nu}/\mathcal{M}_{N}=-(182$ MeV$^2)$~\cite{Mitra:2011qr}. Here 
$m_e$ is the 
mass of electron, $m_p$ is the mass of proton, $\mathcal{M}_{\nu}$  is the 
nuclear matrix element for light neutrino states and 
$\mathcal{M}_{N}$ is nuclear matrix element for heavy neutrino states.  Here the 
first and second contribution in Eq. (\ref{half-life})
is due to the light and heavy neutrinos respectively.
We already have an estimate for the first contribution (with $W_{3 \times 3} 
\simeq \left(1 - \eta \right) U_{\nu }$) as provided in several tables of 
Section 4, which turns out to be of order $\sim 10^{-2}$ eV. Now with 
some specific choice of $\lambda$ and $|y_2|$, we can determine the $W$ matrix 
numerically as discussed in the previous subsection where information on other 
parameters $\alpha, \beta, k ~etc.$ are taken from Section 
4 (different cases). Then we evaluate numerically the $\mathcal{K} ~( i.e.~ 
W_{3 
\times 6})$. In order to maximize this contribution, we consider lowest value 
of $v_f$ which is allowed from Eq. (\ref{vflimit}). It turns out then that the 
second 
contribution remains sub-dominant ($\sim 10^{-6}$ eV or less) compared to the 
first 
contribution of Eq. (\ref{half-life}). The smallness of the second term can also 
be understood from our finding for $\mathcal{K}$ as $\mathcal{K}_{ei} = f 
\left( U_h \right)_{4i}$. A naive estimate for this contribution (to 
$|m_{ee}|$) therefore is 
of order $\frac{\lambda}{y_2} v^2\langle q^2\rangle/v_f^3$. The using the 
lowest possible $v_f$ consistent with Eq. (\ref{vflimit}), the estimate 
indicates that this contribution is essentially small compared to the first 
contribution. So the effective mass involved in the neutrinoless double beta 
decay process is mostly unaffected with the presence of heavy neutrinos in the 
present set-up.

\section{Conclusion}\label{sec6}
 We have considered an inverse seesaw framework embedded in a flavor symmetric 
environment in order to study whether it can accommodate the neutrino masses 
and mixing as suggested from present experimental data, particularly in view of 
nonzero $\theta_{13}$. We employ an $A_4 \times  Z_4 \times Z_3$ discrete 
symmetry which is concocted with a global $B-L$ symmetry. We note that 
 the flavor structure of light neutrino mass matrix is essentially dictated by 
that of the $\mu$ matrix itself,  which is the matrix containing the lepton 
number breaking contribution in the inverse seesaw scenario. The flavor 
structure of $\mu$ matrix is generated when the flavons have vevs. We notice 
that the typical structure of this matrix can lead to a lepton mixing consistent 
with neutrino data where the charged lepton mass matrix is found to be 
diagonal in the framework. In doing this analysis, we have studied the 
correlation between different parameters of the model and their dependence on 
the neutrino parameters such as mass-squared differences, mixing angles etc., 
evaluated from experimental results. Dependency on the Dirac CP violating 
phase is also studied.

Since there exists a small mixing between light and heavy neutrino states in 
the framework, we have also checked the non-unitarity effects in our set-up 
which contribute to LFV processes, neutrinoless double beta decay etc.. We have 
found that owing to the typical flavor structure of the neutrino mass matrix 
here, the effective contribution of it to the LFV processes and neutrinoless 
double beta decays are vanishingly small. It can be noted that the $\mu$ matrix 
results from the breaking of a flavon which carries charge under the global 
$U(1)_{B-L}$. Hence we expect to have Goldstone boson or majoron ($J$) 
\cite{Joshipura:1992hp}. It may open Higgs boson decay channel ($H\rightarrow 
JJ$)
and demands extensive analysis in the context of current and future LHC data. 
Discussions in this direction can be found 
in~\cite{Bonilla:2015uwa, Bonilla:2015kna}. Particularly current 13 TeV run of 
LHC will be important for such analysis. However further discussion in this 
regard is beyond the scope of the present study. 
Since the new physics scale in the present set-up is
around few TeV, collider aspects of such a scenario turns out to be
 intersting and discussion 
in this direction can be found in~\cite{BhupalDev:2012zg,Das:2014jxa}.


\appendix
\numberwithin{equation}{section}
\section*{Appendix}
\section{VEV alignments of flavons}\label{apa}

The most general renormalizable potential involving all the flavons of our set-up which is invariant under 
$A_4 \times Z_4 \times Z_3$ and respecting $U(1)_{B-L}$ can be written as 
\begin{equation}
 V=V(H)+V(\phi_S)+V(\phi_T)+V(\xi)+V(\xi')+V(\rho)+V(H,\phi_S,\phi_T,\xi,\xi',\rho)+V(\phi_S,\phi_T,\xi,\xi',\rho)\nonumber
\end{equation}
where
\begin{eqnarray}
 V(H)&=&\mu_H^2 H^{\dagger}H + \lambda_H (H^{\dagger}H)(H^{\dagger}H)\\
 V(\phi_S)&=&\mu_S^2 ({\phi_S}^{\dagger}\phi_S)_1 + \lambda_1^S ({\phi_S}^{\dagger}\phi_S)_1({\phi_S}^{\dagger}\phi_S)_1
              +\lambda_2^S ({\phi_S}^{\dagger}\phi_S)_{1'}({\phi_S}^{\dagger}\phi_S)_{1''}\nonumber\\
          & & \lambda_3^S ({\phi_S}^{\dagger}\phi_S)_{3S}({\phi_S}^{\dagger}\phi_S)_{3S}  + 
              \lambda_4^S ({\phi_S}^{\dagger}\phi_S)_{3A}({\phi_S}^{\dagger}\phi_S)_{3A}\nonumber \\ 
          & & +\lambda_5^S ({\phi_S}^{\dagger}\phi_S)_{3S}({\phi_S}^{\dagger}\phi_S)_{3A} 
\end{eqnarray}

\begin{eqnarray}
 V(\phi_T)&=&\mu_T^2 ({\phi_T}^{\dagger}\phi_T)_1 + \lambda^1_T ({\phi_T}^{\dagger}\phi_T)_1({\phi_T}^{\dagger}\phi_T)_1
              +\lambda^2_T ({\phi_T}^{\dagger}\phi_T)_{1'}({\phi_T}^{\dagger}\phi_T)_{1''}\nonumber\\
          & & +\lambda^3_T ({\phi_T}^{\dagger}\phi_T)_{3S}({\phi_T}^{\dagger}\phi_T)_{3S}+
               \lambda^4_T ({\phi_T}^{\dagger}\phi_T)_{3A}({\phi_T}^{\dagger}\phi_T)_{3A}\nonumber\\
          & & +\lambda^4_T ({\phi_T}^{\dagger}\phi_T)_{3S}({\phi_T}^{\dagger}\phi_T)_{3A}\\
  V(\xi)&=&\mu_{\xi}^2 {\xi}^{\dagger}\xi + \lambda_{\xi} ({\xi}^{\dagger}\xi)({\xi}^{\dagger}\xi)\\
  V(\xi')&=&\mu_{\xi'}^2 {\xi'}^{\dagger}\xi' + \lambda_{\xi'} ({\xi'}^{\dagger}\xi')({\xi'}^{\dagger}\xi')\\
    V(\rho)&=&\mu_{\rho}^2 {\rho}^{\dagger}\rho + \lambda_{\rho} ({\rho}^{\dagger}\rho)({\rho}^{\dagger}\rho)\\
  V(H,\phi_S,\phi_T,\xi,\xi',\rho)&=&\lambda_{HS}(H^{\dagger}H)({\phi_S}^{\dagger}\phi_S)_1  + \lambda_{HT}(H^{\dagger}H)({\phi_T}^{\dagger}\phi_T)_1\nonumber  \\
                               & & +\lambda_{H\xi}(H^{\dagger}H)({\xi}^{\dagger}\xi) +\lambda_{H\rho}(H^{\dagger}H)({\rho}^{\dagger}\rho)  
                                   +\lambda_{H\xi'}(H^{\dagger}H)({\xi'}^{\dagger}\xi')
\end{eqnarray}

\begin{eqnarray}                          
V(\phi_S,\phi_T,\xi,\xi',\rho)&=&    k_{11}(\phi_T\phi_T)_{3S} \phi_T+k_{12}(\phi_T\phi_T)_{3A}\phi_T+k_{31} (\phi_S^{\dagger}\phi_S)_{3S}\phi_T 
                                     +k_{32} (\phi_S^{\dagger}\phi_S)_{3A}\phi_T \nonumber\\
                              & &    +k_4 (\phi_S\phi_T)_{1}\xi^{\dagger}
                                     +k_5 (\phi_S\phi_T)_{1'}\xi'^{\dagger}
                                     +k_6 (\phi_S^{\dagger}\phi_T)_{1}\xi
                                     +k_7 (\phi_S^{\dagger}\phi_T)_{1''}\xi'\nonumber\\
                              & &    +k_8 (\phi_S\phi_T^{\dagger})_{1}\xi^{\dagger}
                                     +k_9 (\phi_S\phi_T^{\dagger})_{1'}\xi'^{\dagger}
                                     +k_{10} (\phi_S^{\dagger}\phi_T^{\dagger})_{1}\xi
                                     +k'_{10} (\phi_S^{\dagger}\phi_T^{\dagger})_{1''}\xi'\nonumber\\       
                              & &    +\lambda^1_{ST}({\phi_S}^{\dagger}\phi_S)_1({\phi_T}^{\dagger}\phi_T)_1 
                                     +\lambda^2_{ST}({\phi_S}^{\dagger}\phi_S)_{1'}({\phi_T}^{\dagger}\phi_T)_{1''} \nonumber\\
                              & &    + \lambda^{22}_{ST}({\phi_S}^{\dagger}\phi_S)_{1''}({\phi_T}^{\dagger}\phi_T)_{1'}
                                     + \lambda^3_{ST}({\phi_S}^{\dagger}\phi_S)_{3S}({\phi_T}^{\dagger}\phi_T)_{3S}\nonumber\\
                              & &    + \lambda^4_{ST}({\phi_S}^{\dagger}\phi_S)_{3A}({\phi_T}^{\dagger}\phi_T)_{3A}
                                     +\lambda^5_{ST}({\phi_S}^{\dagger}\phi_S)_{3S}({\phi_T}^{\dagger}\phi_T)_{3A}\nonumber\\
                              & &    + \lambda^6_{ST}({\phi_S}^{\dagger}\phi_S)_{3A}({\phi_T}^{\dagger}\phi_T)_{3S}
                                     +\lambda'^{1}_{ST}({\phi_S}^{\dagger}\phi_T)_1({\phi_T}^{\dagger}\phi_S)_1\nonumber\\
                              & &    +\lambda'^{2}_{ST}({\phi_S}^{\dagger}\phi_T)_{1'}({\phi_T}^{\dagger}\phi_S)_{1''}
                                     +\lambda'^{22}_{ST}({\phi_S}^{\dagger}\phi_T)_{1''}({\phi_T}^{\dagger}\phi_S)_{1'}\nonumber\\
                              & &    +\lambda'^{3}_{ST}({\phi_S}^{\dagger}\phi_T)_{3S}({\phi_T}^{\dagger}\phi_S)_{3S}  
                                     +\lambda'^{4}_{ST}({\phi_S}^{\dagger}\phi_T)_{3A}({\phi_T}^{\dagger}\phi_S)_{3A} \nonumber     \\
                              & &      +\lambda'^{5}_{ST}({\phi_S}^{\dagger}\phi_T)_{3S}({\phi_T}^{\dagger}\phi_S)_{3A}
                                     +\lambda'^{6}_{ST}({\phi_S}^{\dagger}\phi_T)_{3A}({\phi_T}^{\dagger}\phi_S)_{3S} \nonumber     \\
                              & &    + \lambda_{S\xi}({\phi_S}^{\dagger}\phi_S)_1({\xi}^{\dagger}\xi) 
                                     + \lambda'_{S\xi}({\phi_S}^{\dagger}\xi)({\xi}^{\dagger}\phi_S)\nonumber\\
                    & & +\lambda_{S\xi'}({\phi_S}^{\dagger}\phi_S)_{1}({\xi'}^{\dagger}\xi')_{1} 
                        + \lambda'_{S\xi'}({\phi_S}^{\dagger}\xi')({\xi'}^{\dagger}\phi_S)\nonumber\\
                    & & +\lambda_{S\rho}({\phi_S}^{\dagger}\phi_S)_1({\rho}^{\dagger}\rho) 
                        + \lambda'_{S\rho}({\phi_S}^{\dagger}\rho)({\rho}^{\dagger}\phi_S)\nonumber\\
                    & & +\lambda_{T\xi}({\phi_T}^{\dagger}\phi_T)_1({\xi}^{\dagger}\xi) 
                        + \lambda'_{T\xi}({\phi_T}^{\dagger}\xi)({\xi}^{\dagger}\phi_T)\nonumber\\   
                    & & +\lambda_{T\xi'}({\phi_T}^{\dagger}\phi_T)_{1}({\xi'}^{\dagger}\xi')_{1} 
                        + \lambda'_{T\xi'}({\phi_T}^{\dagger}\xi')({\xi'}^{\dagger}\phi_T)\nonumber\\
                    & & +\lambda_{T\rho}({\phi_T}^{\dagger}\phi_T)_1({\rho}^{\dagger}\rho) 
                        + \lambda'_{T\rho}({\phi_T}^{\dagger}\rho)({\rho}^{\dagger}\phi_T)\nonumber\\
                    & & +\lambda_{\xi\rho}({\xi}^{\dagger}\xi)({\rho}^{\dagger}\rho) 
                        + \lambda'_{\xi\rho}({\xi}^{\dagger}\rho)({\rho}^{\dagger}\xi)
                        +\lambda_{\xi\xi'}({\xi}\xi^{\dagger})_{1}({\xi'}\xi'^{\dagger})_{1}\nonumber\\                      
                    & & +\lambda_{\xi'\rho}({\xi'}\xi'^{\dagger})({\rho}^{\dagger}\rho)
                        +\lambda'_{\xi'\rho}({\xi'}{\rho}^{\dagger})(\xi'^{\dagger}\rho)\nonumber\\   
                    & & +\lambda_{S\xi\xi'}({\phi_S}^{\dagger}\phi_S)_{1''}({\xi}^{\dagger}\xi')_{1'}
                        +\lambda_{T\xi\xi'}({\phi_T}^{\dagger}\phi_T)_{1''}({\xi}^{\dagger}\xi')_{1'}\nonumber
\end{eqnarray}                                     

\begin{eqnarray}                          
                    & & +\lambda^{2}_{S\xi\xi'}({\phi_S}^{\dagger}\phi_S)_{1'}({\xi}\xi'^{\dagger})_{1''}
                        +\lambda^{2}_{T\xi\xi'}({\phi_T}^{\dagger}\phi_T)_{1'}({\xi}\xi'^{\dagger})_{1''}\nonumber\\
                    & & +\lambda^1_{SS\xi}({\phi_S}^{\dagger}\phi_S)_{3S}({\phi_S}\xi^{\dagger})
                        +\lambda^2_{SS\xi}({\phi_S}^{\dagger}\phi_S)_{3A}({\phi_S}\xi^{\dagger})\nonumber\\
                    & & +\lambda^1_{SS\xi'}({\phi_S}^{\dagger}\phi_S)_{3S}({\phi_S}\xi'^{\dagger})  
                        +\lambda^2_{SS\xi'}({\phi_S}^{\dagger}\phi_S)_{3A}({\phi_S}\xi'^{\dagger})  \nonumber\\
                    & & +\lambda^1_{TS\xi}({\phi_T}^{\dagger}\phi_T)_{3S}({\phi_S}\xi^{\dagger})
                        +\lambda^2_{TS\xi}({\phi_T}^{\dagger}\phi_T)_{3A}({\phi_S}\xi^{\dagger}) \nonumber\\
                    & & +\lambda^1_{TS\xi'}({\phi_T}^{\dagger}\phi_T)_{3S}({\phi_S}\xi'^{\dagger})
                        +\lambda^2_{TS\xi'}({\phi_T}^{\dagger}\phi_T)_{3A}({\phi_S}\xi'^{\dagger})\nonumber \\
                    & & +\lambda^{11}_{SS\xi}({\phi_S}^{\dagger}\phi_S)_{3S}({\phi_S}^{\dagger}\xi)
                        +\lambda^{22}_{SS\xi}({\phi_S}^{\dagger}\phi_S)_{3A}({\phi_S}^{\dagger}\xi)\nonumber\\ 
                    & & +\lambda^{11}_{SS\xi'}({\phi_S}^{\dagger}\phi_S)_{3S}({\phi_S}^{\dagger}\xi')  
                        +\lambda^{22}_{SS\xi'}({\phi_S}^{\dagger}\phi_S)_{3A}({\phi_S}^{\dagger}\xi')  \nonumber\\    
                    & & +\lambda^{11}_{TS\xi}({\phi_T}^{\dagger}\phi_T)_{3S}({\phi_S}^{\dagger}\xi)
                        +\lambda^{22}_{TS\xi}({\phi_T}^{\dagger}\phi_T)_{3A}({\phi_S}^{\dagger}\xi) \nonumber\\    
                    & & +\lambda^{11}_{TS\xi'}({\phi_T}^{\dagger}\phi_T)_{3S}({\phi_S}^{\dagger}\xi')
                        +\lambda^{22}_{TS\xi'}({\phi_T}^{\dagger}\phi_T)_{3A}({\phi_S}^{\dagger}\xi').   
\end{eqnarray}
Here the explicit multiplication of $A_4$ are taken into account. In general this potential involves several free parameters. 
These plenty free parameters should naturally allow therefore the required vev alignment of the flavons we considered, 
$\langle \phi_S \rangle=v_S(1,1,1), ~ \langle \phi_T \rangle =v_T(1,0,0),~  
\langle \xi  \rangle=v_{\xi},~ \langle\xi' \rangle=v_{\xi'}, ~ \langle \rho  \rangle =v_{\rho}$.
For example, with a particular choice like  the followings\footnote{Along the specified vev directions, terms
involving the couplings $\lambda_{3,4,5}^S$, $\lambda_{2,4,5}^T$, 
       $k_{12,31,32}$ $\lambda^2_{ST}$, $\lambda^{22}_{ST}$, $\lambda^{3}_{ST}$, $\lambda^{4}_{ST}$,
       $\lambda^{5}_{ST}$,  $\lambda^{6}_{ST}$, $\lambda'^{5}_{ST}$, $\lambda'^{6}_{ST}$
       , $\lambda'_{T\xi'}$,$\lambda_{T\xi\xi'}$, $\lambda^{2}_{T\xi\xi'}$, $\lambda^1_{SS\xi}$, $\lambda^2_{SS\xi}$,
       $\lambda^1_{SS\xi'}$, $\lambda^2_{SS\xi'}$, $\lambda^2_{TS\xi}$, $\lambda^1_{TS\xi'}$, $\lambda^2_{TS\xi'}$
       $\lambda^{11}_{SS\xi}$, $\lambda^{22}_{SS\xi}$, $\lambda^{11}_{SS\xi'}$, $\lambda^{22}_{SS\xi'}$,
       $\lambda^{22}_{TS\xi}$, $\lambda^{11}_{TS\xi'}$ and 
       $\lambda^{22}_{TS\xi'}$ do not contribute.}   
\begin{align}
 [3\lambda_{HS}v^2+3\lambda_{HT}v^2+\lambda_{H\xi}v^2+\lambda_{H\xi'}v^2+\lambda_{H\rho}v^2-3\mu^2_S-3\mu^2_T-\mu^2_{\xi}-\mu^2_{\xi'}-\mu^2_{\rho}]\sim&~ 1~ {\rm GeV}^2\nonumber\\
 [ 2 k_{11}+k_4+k_5+k_6+k_7+k_8+k_9+k_{10}+k'_{10}]\sim& -1~{\rm GeV}\nonumber\\
\left[9(\lambda_1^S+\lambda_2^S) + (3\lambda^1_{ST}+ \lambda^{22}_{ST}+\lambda'^{1}_{ST}+\lambda'^{2}_{ST}
       +\lambda'^{22}_{ST}+6\lambda'^{3}_{ST}+2\lambda'^{4}_{ST})+
\right. \nonumber\\
  \left.
  (3\lambda_{S\xi}+3\lambda'_{S\xi})+(3\lambda_{S\xi'}+3\lambda'_{S\xi'})+(9\lambda_1^T+4\lambda_3^T)
       +(3\lambda_{S\rho}+3\lambda'_{S\rho})
  \right.\nonumber\\
  \left.
  (\lambda_{\xi\rho}+\lambda'_{\xi\rho})+\lambda_{\rho}+
       (\lambda_{T\xi'}+\lambda'_{T\xi'})+(\lambda_{\xi'\rho}+\lambda'_{\xi'\rho})
       +2\lambda^{11}_{TS\xi}
       +3\lambda^{2}_{S\xi\xi'}
              \right]& \sim~0.00075\nonumber
    \,.
\end{align}      
can actually lead to a common vev $v_S = v_T = v_{\xi} = v_{\xi'} = v_{\rho} \sim$ 1 TeV along the required direction.



\begin{thebibliography}{100}
\bibitem{Minkowski:1977sc} 
  P.~Minkowski,
  Phys.\ Lett.\ B {\bf 67}, 421 (1977).

\bibitem{GellMann:1980vs} 
  M.~Gell-Mann, P.~Ramond and R.~Slansky,
  Conf.\ Proc.\ C {\bf 790927}, 315 (1979)
  [arXiv:1306.4669 [hep-th]].
  
\bibitem{Mohapatra:1979ia} 
  R.~N.~Mohapatra and G.~Senjanovic,
  Phys.\ Rev.\ Lett.\  {\bf 44}, 912 (1980).

\bibitem{Mohapatra:1986aw} 
  R.~N.~Mohapatra,
  Phys.\ Rev.\ Lett.\  {\bf 56}, 561 (1986).
  
\bibitem{Mohapatra:1986bd} 
  R.~N.~Mohapatra and J.~W.~F.~Valle,
  Phys.\ Rev.\ D {\bf 34}, 1642 (1986).
  
  

  

  
  
\bibitem{Abe:2011fz} 
  Y.~Abe {\it et al.} [Double Chooz Collaboration],
  Phys.\ Rev.\ Lett.\  {\bf 108}, 131801 (2012)
  [arXiv:1112.6353 [hep-ex]].

\bibitem{An:2012eh} 
  F.~P.~An {\it et al.} [Daya Bay Collaboration],
  Phys.\ Rev.\ Lett.\  {\bf 108}, 171803 (2012)
  [arXiv:1203.1669 [hep-ex]].

\bibitem{Ahn:2012nd} 
  J.~K.~Ahn {\it et al.} [RENO Collaboration],
  Phys.\ Rev.\ Lett.\  {\bf 108}, 191802 (2012)
  [arXiv:1204.0626 [hep-ex]].

\bibitem{Abe:2013hdq} 
  K.~Abe {\it et al.} [T2K Collaboration],
  Phys.\ Rev.\ Lett.\  {\bf 112}, 061802 (2014)
  [arXiv:1311.4750 [hep-ex]].  
  
  
\bibitem{Forero:2014bxa} 
  D.~V.~Forero, M.~Tortola and J.~W.~F.~Valle,
  Phys.\ Rev.\ D {\bf 90}, no. 9, 093006 (2014)
  [arXiv:1405.7540 [hep-ph]].
  
  
\bibitem{Capozzi:2013csa} 
  F.~Capozzi, G.~L.~Fogli, E.~Lisi, A.~Marrone, D.~Montanino and A.~Palazzo,
  Phys.\ Rev.\ D {\bf 89}, no. 9, 093018 (2014)
  [arXiv:1312.2878 [hep-ph]].

\bibitem{Gonzalez-Garcia:2014bfa} 
  M.~C.~Gonzalez-Garcia, M.~Maltoni and T.~Schwetz,
  JHEP {\bf 1411}, 052 (2014)
  [arXiv:1409.5439 [hep-ph]].
    
  
  
\bibitem{NuExpt} 
  S.~Fukuda {\it et al.}  [Super-Kamiokande Collaboration],
  Phys.\ Lett.\ B {\bf 539}, 179 (2002)
  [hep-ex/0205075];
  Y.~Ashie {\it et al.}  [Super-Kamiokande Collaboration],
  Phys.\ Rev.\ D {\bf 71}, 112005 (2005)
  [hep-ex/0501064].
  P.~Adamson {\it et al.}  [MINOS Collaboration],
  Phys.\ Rev.\ Lett.\  {\bf 106}, 181801 (2011)
  [arXiv:1103.0340 [hep-ex]].
  T.~Araki {\it et al.}  [KamLAND Collaboration],
  Phys.\ Rev.\ Lett.\  {\bf 94}, 081801 (2005)
  [hep-ex/0406035].
  
  
  
\bibitem{Harrison:1999cf} 
  P.~F.~Harrison, D.~H.~Perkins and W.~G.~Scott,
  Phys.\ Lett.\ B {\bf 458}, 79 (1999)
  [hep-ph/9904297].  
  
\bibitem{King:2013eh}
  S.~F.~King and C.~Luhn,
  Rept.\ Prog.\ Phys.\  {\bf 76}, 056201 (2013)
  [arXiv:1301.1340 [hep-ph]]. 
  
  
\bibitem{Ma:2001dn}
  E.~Ma and G.~Rajasekaran,
  Phys.\ Rev.\ D {\bf 64} (2001) 113012
  [hep-ph/0106291].

\bibitem{Altarelli:2005yx} 
  G.~Altarelli and F.~Feruglio,
  Nucl.\ Phys.\ B {\bf 741}, 215 (2006)
  [hep-ph/0512103].
  
\bibitem{Barry:2010zk} 
  J.~Barry and W.~Rodejohann,
  Phys.\ Rev.\ D {\bf 81}, 093002 (2010)
  Erratum: [Phys.\ Rev.\ D {\bf 81}, 119901 (2010)]
  [arXiv:1003.2385 [hep-ph]].
  
\bibitem{Karmakar:2014dva} 
  B.~Karmakar and A.~Sil,
  Phys.\ Rev.\ D {\bf 91}, 013004 (2015)
  [arXiv:1407.5826 [hep-ph]] and references therein.
  

  
  
  

  
\bibitem{Karmakar:2015jza} 
  B.~Karmakar and A.~Sil,
  Phys.\ Rev.\ D {\bf 93}, no. 1, 013006 (2016)
  [arXiv:1509.07090 [hep-ph]] and references therein.
  

    
\bibitem{Shimizu:2011xg} 
  Y.~Shimizu, M.~Tanimoto and A.~Watanabe,
  Prog.\ Theor.\ Phys.\  {\bf 126}, 81 (2011)
  [arXiv:1105.2929 [hep-ph]].
  
  
  
\bibitem{King:2011zj} 
  S.~F.~King and C.~Luhn,
  JHEP {\bf 1109}, 042 (2011)
  [arXiv:1107.5332 [hep-ph]].
  
\bibitem{Hirsch:2009mx} 
  M.~Hirsch, S.~Morisi and J.~W.~F.~Valle,
  Phys.\ Lett.\ B {\bf 679}, 454 (2009)
  [arXiv:0905.3056 [hep-ph]].
  
  
\bibitem{Altarelli:2010gt} 
  G.~Altarelli and F.~Feruglio,
  Rev.\ Mod.\ Phys.\  {\bf 82}, 2701 (2010)
  [arXiv:1002.0211 [hep-ph]].
  
  
  
\bibitem{Dorame:2012zv} 
  L.~Dorame, S.~Morisi, E.~Peinado, J.~W.~F.~Valle and A.~D.~Rojas,
  Phys.\ Rev.\ D {\bf 86}, 056001 (2012)
  [arXiv:1203.0155 [hep-ph]].
  
\bibitem{Abbas:2015zna} 
  M.~Abbas, S.~Khalil, A.~Rashed and A.~Sil,
  Phys.\ Rev.\ D {\bf 93}, no. 1, 013018 (2016)
  [arXiv:1508.03727 [hep-ph]];
  S.~Fraser, E.~Ma and O.~Popov,
  Phys.\ Lett.\ B {\bf 737}, 280 (2014)
  [arXiv:1408.4785 [hep-ph]];
  E.~Ma and R.~Srivastava,
  Mod.\ Phys.\ Lett.\ A {\bf 30}, no. 26, 1530020 (2015)
  [arXiv:1504.00111 [hep-ph]];
  A.~Mukherjee and M.~K.~Das,
  Nucl.\ Phys.\ B {\bf 913}, 643 (2016)
  [arXiv:1512.02384 [hep-ph]].
  
 
 
\bibitem{Asakura:2014lma} 
  K.~Asakura {\it et al.} [KamLAND-Zen Collaboration],
  AIP Conf.\ Proc.\  {\bf 1666}, 170003 (2015)
  [arXiv:1409.0077 [physics.ins-det]].
  
\bibitem{Albert:2014awa} 
  J.~B.~Albert {\it et al.} [EXO-200 Collaboration],
  Nature {\bf 510}, 229 (2014)
  [arXiv:1402.6956 [nucl-ex]].
 
  
\bibitem{Lindner:2005pk} 
  M.~Lindner, M.~A.~Schmidt and A.~Y.~Smirnov,
  JHEP {\bf 0507}, 048 (2005)
  [hep-ph/0505067].
  
  
\bibitem{Agashe:2014kda} 
  K.~A.~Olive {\it et al.}  [Particle Data Group Collaboration],
  Chin.\ Phys.\ C {\bf 38}, 090001 (2014).
  
  
\bibitem{He:2006qd} 
  X.~G.~He and A.~Zee,
  Phys.\ Lett.\ B {\bf 645}, 427 (2007)
  [hep-ph/0607163].
  
  
\bibitem{Albright:2008rp} 
  C.~H.~Albright and W.~Rodejohann,
  Eur.\ Phys.\ J.\ C {\bf 62}, 599 (2009)
  [arXiv:0812.0436 [hep-ph]].
  
  

  
  
\bibitem{Altarelli:2012bn} 
  G.~Altarelli, F.~Feruglio, L.~Merlo and E.~Stamou,
  JHEP {\bf 1208}, 021 (2012)
  [arXiv:1205.4670 [hep-ph]].
  
  
  
  
  
  
  
  
  
  
  
  
  
  
  
  

  
\bibitem{Ade:2013zuv} 
  P.~A.~R.~Ade {\it et al.}  [Planck Collaboration],
  arXiv:1303.5076 [astro-ph.CO].
  

    
  

  
  
 
  
  



  

  
  
\bibitem{Antusch:2006vwa} 
  S.~Antusch, C.~Biggio, E.~Fernandez-Martinez, M.~B.~Gavela and J.~Lopez-Pavon,
  JHEP {\bf 0610}, 084 (2006)
  [hep-ph/0607020].
  
  
\bibitem{GonzalezGarcia:1988rw} 
  M.~C.~Gonzalez-Garcia and J.~W.~F.~Valle,
  Phys.\ Lett.\ B {\bf 216}, 360 (1989).
  
  
  

\bibitem{Kanaya:1980cw} 
  K.~Kanaya,
  Prog.\ Theor.\ Phys.\  {\bf 64}, 2278 (1980).
  
\bibitem{Altarelli:2008yr} 
  G.~Altarelli and D.~Meloni,
  Nucl.\ Phys.\ B {\bf 809}, 158 (2009)
  [arXiv:0809.1041 [hep-ph]].
  
\bibitem{Antusch:2008tz} 
  S.~Antusch, J.~P.~Baumann and E.~Fernandez-Martinez,
  Nucl.\ Phys.\ B {\bf 810}, 369 (2009)
  [arXiv:0807.1003 [hep-ph]].
  
  
\bibitem{Dev:2009aw} 
  P.~S.~B.~Dev and R.~N.~Mohapatra,
  Phys.\ Rev.\ D {\bf 81}, 013001 (2010)
  [arXiv:0910.3924 [hep-ph]].
  
  
\bibitem{Dias:2012xp} 
  A.~G.~Dias, C.~A.~de S.Pires, P.~S.~Rodrigues da Silva and A.~Sampieri,
  Phys.\ Rev.\ D {\bf 86}, 035007 (2012)
  [arXiv:1206.2590].
  
\bibitem{Ilakovac:1994kj} 
  A.~Ilakovac and A.~Pilaftsis,
  Nucl.\ Phys.\ B {\bf 437}, 491 (1995)
  [hep-ph/9403398].
  
\bibitem{Alonso:2012ji} 
  R.~Alonso, M.~Dhen, M.~B.~Gavela and T.~Hambye,
  JHEP {\bf 1301}, 118 (2013)
  [arXiv:1209.2679 [hep-ph]].
  
  
  
 \bibitem{Hahn:2006hr} 
   T.~Hahn,
   physics/0607103.
  
  
\bibitem{LalAwasthi:2011aa} 
  R.~Lal Awasthi and M.~K.~Parida,
  Phys.\ Rev.\ D {\bf 86}, 093004 (2012)
  [arXiv:1112.1826 [hep-ph]].
  
  
  
  
  
  
\bibitem{Parida:2014dba} 
  M.~K.~Parida, R.~L.~Awasthi and P.~K.~Sahu,
  JHEP {\bf 1501}, 045 (2015)
  [arXiv:1401.1412 [hep-ph]].
  
  
  
  
\bibitem{Parida:2016asc} 
  M.~K.~Parida and B.~P.~Nayak,
  arXiv:1607.07236 [hep-ph].
  
\bibitem{Bilenky:1977du} 
  S.~M.~Bilenky, S.~T.~Petcov and B.~Pontecorvo,
  Phys.\ Lett.\  {\bf 67B}, 309 (1977).
 
 
\bibitem{He:2002pva} 
  B.~He, T.~P.~Cheng and L.~F.~Li,
  Phys.\ Lett.\ B {\bf 553}, 277 (2003)
  [hep-ph/0209175].
 
\bibitem{Forero:2011pc} 
  D.~V.~Forero, S.~Morisi, M.~Tortola and J.~W.~F.~Valle,
  JHEP {\bf 1109}, 142 (2011)
  [arXiv:1107.6009 [hep-ph]].
 
\bibitem{Rose:2015fua} 
  L.~Delle Rose, C.~Marzo and A.~Urbano,
  JHEP {\bf 1512}, 050 (2015)
  [arXiv:1506.03360 [hep-ph]].
  
  
\bibitem{Bonilla:2015uwa} 
  C.~Bonilla, J.~W.~F.~Valle and J.~C.~Romão,
  Phys.\ Rev.\ D {\bf 91}, no. 11, 113015 (2015)
  [arXiv:1502.01649 [hep-ph]].
  
  
  
\bibitem{Bonilla:2015kna} 
  C.~Bonilla, R.~M.~Fonseca and J.~W.~F.~Valle,
  Phys.\ Lett.\ B {\bf 756}, 345 (2016)
  [arXiv:1506.04031 [hep-ph]].
  

  
\bibitem{Mitra:2011qr} 
  M.~Mitra, G.~Senjanovic and F.~Vissani,
  Nucl.\ Phys.\ B {\bf 856}, 26 (2012)
  [arXiv:1108.0004 [hep-ph]].
  
  
\bibitem{Chakrabortty:2012mh} 
  J.~Chakrabortty, H.~Z.~Devi, S.~Goswami and S.~Patra,
  JHEP {\bf 1208}, 008 (2012)
  [arXiv:1204.2527 [hep-ph]].
  
  

  
  
\bibitem{Joshipura:1992hp} 
  A.~S.~Joshipura and J.~W.~F.~Valle,
  Nucl.\ Phys.\ B {\bf 397}, 105 (1993).
  A.~S.~Joshipura and S.~D.~Rindani,
  Phys.\ Rev.\ Lett.\  {\bf 69}, 3269 (1992).
  J.~C.~Romao, F.~de Campos and J.~W.~F.~Valle,
  Phys.\ Lett.\ B {\bf 292}, 329 (1992)
  [hep-ph/9207269].
  
  
  
\bibitem{Tello:2010am} 
  V.~Tello, M.~Nemevsek, F.~Nesti, G.~Senjanovic and F.~Vissani,
  Phys.\ Rev.\ Lett.\  {\bf 106}, 151801 (2011)
  [arXiv:1011.3522 [hep-ph]].
  
\bibitem{Khan:2012zw} 
  S.~Khan, S.~Goswami and S.~Roy,
  Phys.\ Rev.\ D {\bf 89}, no. 7, 073021 (2014)
  [arXiv:1212.3694 [hep-ph]].
  
  
\bibitem{BhupalDev:2012zg} 
  P.~S.~Bhupal Dev, R.~Franceschini and R.~N.~Mohapatra,
  Phys.\ Rev.\ D {\bf 86}, 093010 (2012)
  [arXiv:1207.2756 [hep-ph]].
  
  
\bibitem{Das:2014jxa} 
  A.~Das, P.~S.~Bhupal Dev and N.~Okada,
  Phys.\ Lett.\ B {\bf 735}, 364 (2014)
  [arXiv:1405.0177 [hep-ph]].
  
  
  
  
\end{thebibliography}
\end{document}